\documentclass[12pt]{article}
\pdfoutput=1

\usepackage[a4paper,text={16.8cm,22.4cm}]{geometry}
\usepackage{amsmath,amsfonts,slashed,amssymb,tikz,bm,psfrag,graphicx,color,dsfont}
\usepackage{multicol,multirow}
\usepackage{float}

\RequirePackage[sort&compress,square,comma,numbers]{natbib}
\allowdisplaybreaks
\renewcommand{\arraystretch}{1.2}

\newcommand{\SLASH}{\not  \! \!}

\begin{document}

\begin{titlepage}

\begin{flushright}
\normalsize
September 6, 2020
\end{flushright}

\vspace{0.1cm}
\begin{center}
\Large\bf
Precision calculations of the double radiative bottom-meson decays
in soft-collinear effective theory
\end{center}

\vspace{0.5cm}
\begin{center}
{\bf Yue-Long Shen$^{a}$,
Yu-Ming Wang$^{b}$,
Yan-Bing Wei$^{b}$}\\
\vspace{0.7cm}
{\sl   ${}^a$ \,College of Information Science and Engineering,
Ocean University of China,
\\
Songling Road 238, Qingdao, 266100 Shandong, China
\\
${}^b$ School of Physics, Nankai University, Weijin Road 94, 300071 Tianjin, China \,
}
\end{center}

\vspace{0.2cm}
\begin{abstract}

Employing the systematic framework of soft-collinear effective theory (SCET)
we perform an improved calculation of the leading-power contributions to
the double radiative $B_{d, \, s}$-meson decay amplitudes in the heavy quark expansion
by including the perturbative resummation of enhanced logarithms of $m_b/\Lambda_{\rm QCD}$
at the next-to-leading-logarithmic accuracy.
We then construct the QCD factorization formulae for the subleading power contributions
arising from the energetic photon radiation off the constituent light-flavour quark of
the bottom meson at tree level.
Furthermore, we explore the factorization properties of the subleading power correction
from the effective SCET current
$\left ( \bar \xi \,  W_c \right ) \, \Gamma \, \left [ i \, \SLASH  D_{s} / (2 m_b) \right ] \, h_v$
at ${\cal O} (\alpha_s^0)$ by virtue of the operator identities due to the classical  equations of motion.
The higher-twist contributions to the $B_{d, \, s} \to \gamma \gamma $ helicity form factors
from the two-particle and three-particle bottom-meson distribution amplitudes are evaluated
with the perturbative factorization technique, up to the twist-six accuracy.
In addition, the subleading power weak-annihilation contributions from both the current-current
and QCD penguin operators are taken into account at the one-loop accuracy.
We proceed to apply the operator-production-expansion-controlled dispersion relation
for estimating the power-suppressed soft contributions to the double radiative
$B_{d, s}$-meson decay form factors, which cannot be factorized into
the light-cone distribution amplitudes of the heavy-meson and the resolved photon
as well as the hard-scattering kernel calculable in perturbation theory canonically.
Phenomenological explorations of the radiative $B_{d, \, s} \to \gamma \, \gamma$ decay observables
in the presence of the neutral-meson mixing, including the CP-averaged branching fractions,
the polarization fractions and the time-dependent CP asymmetries,
are carried out subsequently with an emphasis on the numerical impacts of the newly computed
ingredients together with the theory uncertainties from the shape parameters of
the HQET bottom-meson distribution amplitudes.

\end{abstract}

\vfil

\end{titlepage}

\tableofcontents

\newpage

\section{Introduction}

The exclusive radiative penguin $B$-meson decays are evidently of particular importance
to explore the delicate quark-flavour mixing mechanism in the Standard Model (SM)
and to probe the new dynamics of flavour-changing neutral currents above the electroweak scale.
Consequently, intensive efforts have been devoted to developing the systematic approaches for
computing the radiative $B \to V \gamma$ decay amplitudes in QCD beyond the leading-order approximation
based upon the diagrammatic factorization technique
\cite{Beneke:2001at,Beneke:2004dp,Bosch:2001gv,DescotesGenon:2004hd,Ali:2004hn}
 and soft-collinear effective theory \cite{Becher:2005fg,Ali:2007sj}
(see \cite{Ball:2006eu,Khodjamirian:2010vf} for further discussions on the non-factorizable charm-loop effects).
It turns out that the hadronic matrix elements of the renormalized operators in the  effective weak Hamiltonian
responsible for the exclusive $B \to V \gamma$ decays are more complex than the heavy-to-light
transition form factors even at leading power in the heavy quark expansion,
let alone to anatomize  the factorization properties of such radiative $B$-meson decay amplitudes
at subleading power where the resolved photon corrections parameterized by the corresponding
light-cone distribution amplitudes (LCDAs) will be indispensable for the complete theoretical description
(see \cite{Wang:2017ijn,Li:2013xna,Wang:2018wfj,Li:2020rcg} for expanded discussions in different contexts).
Pinning down the theoretical uncertainties for QCD calculations of the radiative $B \to V \gamma$ decays
therefore necessitates an improved understanding of both experimentally and theoretically cleaner
exclusive  $b \to s (d) \, \gamma$ transitions in advance.

The double radiative $B_{d, \, s} \to \gamma \gamma$ decays with non-hadronic final states
are in this respect  the valuable touchstones for investigating the non-trivial strong interaction dynamics
of the heavy-meson systems, in analogy to the radiative leptonic $B \to \gamma \, \ell \, \bar \nu_{\ell}$ decays
generated  by the flavour-changing charged current.
In particular, the direct CP asymmetries of $B_{d, \, s} \to \gamma \gamma$ with
the distinct polarization configurations of two energetic photons allow for the direct extraction
of the CKM phase angle $\gamma$ \cite{Bosch:2002bv}.
In despite of their relatively small branching fractions in the SM
and the challenging detections of such exclusive reactions with two photons in the final states
at the LHCb  experiment, the radiative penguin decays $B_{d, \, s} \to \gamma \gamma$ are demonstrated
to be among the Golden/Silver channels for the Belle II program \cite{Kou:2018nap},
where the event reconstruction becomes straightforward with the electromagnetic calorimeter.

Employing the QCD factorization formalism, the double radiative $B_{d, \, s} \to \gamma \gamma$ decay
amplitudes have been  computed at next-to-leading order (NLO) in the strong coupling
and at leading power in the heavy quark expansion \cite{DescotesGenon:2002ja},
where the two-loop $b \to (s, d) \, \gamma$ matrix elements of QCD penguin operators
evaluated in \cite{Buras:2002tp,Asatrian:2004et} were unfortunately not taken into account
and the factorization-scale independence for the obtained  matrix elements
$\langle \gamma \gamma | H_{\rm eff}| B_{d, \, s}\rangle $ were not completely
achieved at ${\cal O}(\alpha_s)$.
The yielding factorization formulae for the radiative  $B_{d, \, s} \to \gamma \gamma$
decay form factors can be expressed in terms of the short-distance matching coefficients
and the leading-twist $B_{d, \, s}$-meson distribution amplitudes in the heavy-quark effective theory (HQET).
Accordingly, precision measurements of the branching fractions for the double radiative $B_{d, \, s}$-meson decays
can further provide meaningful constraints on the two crucial shape parameters $\lambda_{B_d}(\mu)$
and $\lambda_{B_s}(\mu)$, which serve as the fundamental ingredients for the model-independent calculations
of many exclusive bottom-meson decay observables in QCD.
The weak-annihilation contributions from the current-current operators,
which are suppressed by one power of $\Lambda_{\rm QCD}/m_b$ in the heavy quark limit,
have been shown to factorize into the perturbative hard-scattering kernels
and the $B_{d, \, s}$-meson decay constants at two loops \cite{Bosch:2002bv},
with the complex  short-distance functions computed explicitly  at the one-loop accuracy.

Previous calculations of the double radiative $B_{s} \to \gamma \gamma$ decay amplitude
with the aid of the static bottom- and strange-quark approximation
\cite{Lin:1989vj,Lin:1990kw,Reina:1997my,Herrlich:1991bq,Chang:1997fs,Devidze:1996np,Dincer:2001hu}
apparently introduce  conceptual and technical limitations for the theory description of
the corresponding hadronic matrix elements.
In addition, the long-distance hadronic contributions from the charm-meson rescattering mechanism
\cite{Choudhury:1998rb,Liu:1999qz}
and from the vector-meson coupling to the on-shell photon \cite{Hiller:1997ie,Hiller:1997kp}
have been estimated with model-dependent phenomenological approaches,
which prevent us from improving the current theory predictions systematically in the long run.

Advancing our understanding of the radiative penguin $B_{d, \, s} \to \gamma \gamma$
decay amplitudes further  by  taking advantage of the perturbative  factorization
approach and the hadronic dispersion relation will be of wide interest for providing
the helpful insights of evaluating the subleading power corrections to exclusive
heavy-to-light $B_q$-meson decays (with $q = d, \, s$) in QCD and for probing the dynamical information
of the bottom-meson distribution amplitudes in HQET.
The major new ingredients of the present paper  at the technical level can be summarized
in the following.

\begin{itemize}

\item{The next-to-leading logarithmic (NLL) resummmation for the enhanced logarithms of $m_b/\Lambda_{\rm QCD}$
entering the leading-power factorization formula of $B_{d, s} \to \gamma \gamma$
will be accomplished by solving the renormalization group equations for the hard matching coefficients
and the leading-twist $B$-meson LCDA at two loops.
As a consequence, our result goes  beyond the leading logarithmic (LL) approximation
implemented in \cite{DescotesGenon:2002ja}.
Furthermore, the complete scale-independent factorization formulae
for the double radiative decay amplitudes at ${\cal O} (\alpha_s^2)$ will be achieved
by including the two-loop $b \to (s, \, d) \, \gamma$ matrix elements of the QCD penguin operators. }

\item{
Applying the equations of motion for the soft quark field and the effective heavy quark,
we will derive the QCD factorization formulae of the subleading power corrections to
the energetic photon radiation off the (massless)-light  quark at tree level
on the basis of the two-particle and three-particle $B_q$-meson distribution amplitudes.
Moreover, it will be verified explicitly that the convolution integrals
of the hard-collinear  matching coefficients and $B_q$-meson LCDAs appearing in the constructed
factorization formulae are convergent.
By contrast,  the na\"{\i}ve soft-collinear factorization formula of the  power suppressed contribution
from the light-quark mass effect suffers from the rapidity divergences already at tree level. }

\item{We will derive the perturbative factorization formula for the subleading power
soft-collinear effective theory (SCET) matrix element
$\langle 0 | \left ( \bar \xi \,  W_c \right ) \, \Gamma \,
\left [ i \, \SLASH  D_{s} / (2 m_b) \right ] \, h_v | \bar B_{q} \rangle $
by employing the well-established relations between the non-local  operators at leading order (LO) in $\alpha_s$.
In light of the canonical behaviours of the appearing $B$-meson distribution amplitudes at small quark and gluon momenta,
the obtained convolution integrals of the leading- and higher-twist $B_q$-meson distribution amplitudes
with the  short-distance functions are both convergent.}

\item{Applying the QCD factorization approach, the higher-twist corrections
to the double radiative decay form factors  from both the two-particle and
three-particle $B_q$-meson distribution amplitudes will be evaluate at ${\cal O}(\alpha_s^0)$
up to the twist-six accuracy with the systematic parametrization of the HQET matrix element
for the three-body light-ray operator
$\bar q_s(\tau_1 \, n) \, g_s \, G_{\mu \nu}(\tau_2 \, n) \,\, \Gamma \, h_{v \, \beta}(0)$
\cite{Braun:2017liq}. The resulting factorization formulae expressed in terms of
the convolution integrals of the twist-three and twist-four LCDAs with
the corresponding hard-collinear functions appear to converge for the phenomenologically
acceptable models of $\phi_B^{-}(\omega, \mu)$ and $\Phi_4(\omega_1, \omega_2, \mu)$. }

\item{The subleading power soft contributions to the radiative  $B_{d, \, s} \to \gamma \gamma$ decay amplitudes
characterizing the resolved photon corrections will be computed with the dispersion approach motivated from
the operator-product-expansion (OPE) technique and the parton-hadron duality ansatz,
which has been successfully applied to estimate the  non-perturbative correction
to the pion-photon transition form factor at experimentally accessible momentum transfer
\cite{Khodjamirian:1997tk,Agaev:2010aq,Agaev:2012tm}.
To this end, the spectral representations of the soft-collinear factorization formulae for
the two transverse $B \to \gamma^{\ast}$ form factors  at NLO in QCD obtained in
\cite{Wang:2016qii,Beneke:2018wjp} will be in demand.
}

\end{itemize}

The remainder of this paper is structured as follows.
We set up the computational framework in Section \ref{section:theory summary}  by introducing
the  effective weak Hamiltonian governing the $b \to (s, d) \, \gamma$ transitions
and by expressing the decay amplitude of  $B_q \to \gamma \gamma$ to the lowest
non-vanishing order in the electromagnetic interactions in terms of
the appropriate QCD correlation functions,
where the two equivalent bases of the Lorenz-invariant amplitudes are also discussed
for the purpose of the phenomenological explorations.
We proceed to determine the ${\rm SCET_{I}}$ representations of the above-mentioned
QCD correlation functions at leading power in the heavy quark expansion
by integrating out the short-distance modes with virtualities
of order $m_b^2$ and then present the analytical expressions of the resulting hard functions
for an arbitrary  quark-mass ratio $m_c/m_b$ at ${\cal O}(\alpha_s)$ in Section \ref{section:LP}.
Subsequently, the soft-collinear factorization formulae for the yielding ${\rm SCET_{I}}$
correlation functions will be constructed by integrating out the hard-collinear fluctuations.
The radiative jet function from matching ${\rm SCET_{I}} \to {\rm SCET_{II}}$ at one loop
will be also collected here.
The NLL resummation improved factorization formulae for the helicity form factors
will be further derived with the standard renormalization-group formalism.
In Section \ref{section:NLP} we turn to compute
the subleading power corrections  generated by the energetic photon emission from the light soft quark
and to derive their tree-level  factorization formulae in terms of the HQET $B$-meson distribution amplitudes.
We then evaluate the subleading power  matrix element of the effective heavy-to-light current
$\left ( \bar \xi \,  W_c \right ) \, \Gamma \, \left [ i \, \SLASH  D_{s} / (2 m_b) \right ] \, h_v$
from matching ${\rm QCD} \to {\rm SCET_{I}}$ at LO in $\alpha_s$.
Drawing upon the QCD factorization technique, we  further calculate the subleading power contributions
from the higher-twist $B$-meson distribution amplitudes due to both the off light-cone corrections
and the higher Fock-state effects.
The power suppressed weak-annihilation contributions from all the four-quark operators at one loop
and the (anti)-collinear photon radiation off the bottom quark  at tree level will be presented here
as well by matching ${\rm QCD} \to {\rm SCET_{II}}$  directly.
Section \ref{section:soft NLP} is devoted to an estimate of the long-distance photon contributions to the
helicity amplitudes of  $B_{d, \, s} \to \gamma \gamma$  based upon the OPE-controlled dispersion technique
at ${\cal O}(\alpha_s)$.
Phenomenological implications of the newly derived expressions for the double radiative
decay amplitudes including the renormalization-group improvement for the
leading power contributions and the updated analysis of the  subleading power corrections
will be investigated in Section \ref{section:numerics} with an emphasis on the shape-parameter dependence
of the achieved theory predictions on the leading-twist $B$-meson LCDA.
Having at our disposal the numerical results for the helicity amplitudes of $B_{d, \, s} \to \gamma \gamma$,
we  predict a number of observables of experimental interest including the  CP-averaged branching fractions,
the polarization fractions and the time-dependent CP asymmetries in the presence of the neutral-meson mixing.
Our main observations and the concluding discussions will be presented
in Section \ref{section:conclusions}.
We further present the detailed expressions of the complete NLO QCD corrections
to the  $b \to s \gamma$ matrix elements obtained in \cite{Buras:2002tp}
and the renormalization-group evolution functions emerged in the NLL resummation
improved factorization formulae of the double radiative $B_q$-meson
decay  amplitudes at leading power in  Appendices \ref{appendix: hard functions} 
and \ref{appendix: B-meson LCDA}, respectively.

\section{Theory summary of  the double radiative $B_q$-meson decays}
\label{section:theory summary}

\subsection{The effective weak Hamiltonian}

Applying the classical equations of motion \cite{Politzer:1980me}, the effective weak Hamiltonian
for  $b \to q \gamma \gamma$ can be  reduced to the one for  the radiative $b \to q \gamma$
transitions at leading order in the Fermi coupling $G_F$ \cite{Grinstein:1990tj}.
Employing the unitarity relations of the CKM matrix elements, the relevant effective weak Hamiltonian
can be written as
\begin{eqnarray}
{\cal H}_{\rm eff} =  {4 \, G_F \over \sqrt{2}} \, \sum_{p=u, c} \, V_{p b} V_{p q}^{\ast}  \,
\left [ C_1(\nu) \, P_1^{p}(\nu) + C_2(\nu) \, P_2^{p}(\nu)
+ \sum_{i=3}^{8} C_i(\nu) \, P_i(\nu) \right ] + {\rm h.c.}   \,,
\label{weak effective  Hamiltonian}
\end{eqnarray}
where the complete set of physical operators are given by \cite{Chetyrkin:1996vx}
\begin{eqnarray}
P_1^{p} &=& (\bar q_L \gamma_{\mu} T^a p_L ) \, (\bar p_L \gamma^{\mu} T^a b_L )  \,,
 \qquad \hspace{2.0 cm}
P_2^{p} = (\bar q_L \gamma_{\mu}  p_L ) \, (\bar p_L \gamma^{\mu} b_L )  \,,   \nonumber \\
P_3 &=&   (\bar q_L \gamma_{\mu}  b_L ) \, \sum_{q^{\prime}} (\bar q^{\prime} \gamma^{\mu} q^{\prime} ) \,,
\qquad   \hspace{2.5 cm}
P_4 =   (\bar q_L \gamma_{\mu} T^a  b_L ) \, \sum_{q^{\prime}} (\bar q^{\prime} \gamma^{\mu} T^a q^{\prime} ) \,, \nonumber \\
P_5 &=&   (\bar q_L \gamma_{\mu_1} \gamma_{\mu_2} \gamma_{\mu_3}  b_L ) \,
\sum_{q^{\prime}} (\bar q^{\prime} \gamma^{\mu_1} \gamma^{\mu_2} \gamma^{\mu_3} q^{\prime} ) \,,
\qquad
P_6 =   (\bar q_L \gamma_{\mu_1} \gamma_{\mu_2} \gamma_{\mu_3}  T^a  b_L ) \,
\sum_{q^{\prime}} (\bar q^{\prime} \gamma^{\mu_1} \gamma^{\mu_2} \gamma^{\mu_3} T^a q^{\prime} ) \,, \nonumber \\
P_7 &=& - {g_{\rm em} \, \overline{m}_b(\nu)\over 16 \, \pi^2} \, (\bar q_L \sigma^{\mu \nu}  b_R ) F_{\mu \nu} \,,
\qquad \hspace{1.6 cm}
P_8 = - {g_{s} \, \overline{m}_b(\nu)\over 16 \, \pi^2} \, (\bar q_L \sigma^{\mu \nu} T^a b_R ) G^{a}_{\mu \nu} \,.
\label{weak operator basis}
\end{eqnarray}
The sign convention of the dipole operators $P_{7, 8}$ corresponds to the definition of the gauge covariant derivative
$D_{\mu}= \partial_{\mu} - i g_s \, T^a A^a_{\mu} - i e_f \, g_{\rm em} A_{\mu}^{\rm em}$
with $e_f=-1$ for the lepton fields \cite{Beneke:2001at}.
In addition, $\overline{m}_b(\nu)$ stands for the bottom-quark mass at the renormalization scale $\nu$
in the ${\rm \overline{MS}}$ scheme.
The four-quark operator basis introduced in (\ref{weak operator basis}) is more advantageous than
the conventional Buchalla-Buras-Lautenbacher scheme \cite{Buchalla:1995vs}
due to the disappearance of the closed fermion loop
with an odd number of $\gamma_5$ matrices in the interaction vertices.
The renormalized effective Wilson coefficients $C_i(\nu)$ at $\nu \simeq m_b$
at NLL in QCD will be achieved by solving the renormalization-group equations
including the three-loop mixing of the four-quark operators into $P_{7, 8}$
and the two-loop self-mixing of the magnetic moment operators \cite{Chetyrkin:1996vx,Gambino:2003zm}.

\subsection{The helicity and transversity form factors}

The exclusive radiative $\bar B_q \to \gamma \gamma$ decay amplitude can be expressed as
\begin{eqnarray}
{\cal \bar A}(\bar B_q \to \gamma \gamma)  =
- \langle \gamma(p, \epsilon_1^{\ast}) \gamma(q, \epsilon_2^{\ast}) |{\cal H}_{\rm eff} | \bar B_q(p+q) \rangle  \,,
\label{definition: the hadronic amplitude}
\end{eqnarray}
where $\epsilon_1^{\ast}$ and $\epsilon_2^{\ast}$ correspond to the polarization vectors of
the collinear and anti-collinear photons, respectively.
We work in the rest frame of the $B_q$-meson and define the two light-cone momenta $n_{\mu}$
and $\bar n_{\mu}$ satisfying the relations $n^2=\bar n^2=0$ and $n \cdot \bar n=2$ by
\begin{eqnarray}
p_{\mu} = {n \cdot p \over 2} \, \bar n_{\mu} \equiv  {m_{B_q} \over 2} \, \bar n_{\mu} \,, \qquad
q_{\mu} = {\bar n \cdot q \over 2} \, n_{\mu}  \equiv  {m_{B_q} \over 2} \, n_{\mu}  \,,
\end{eqnarray}
which  allows us to write down the four-velocity vector
$v_{\mu}=(p+q)_{\mu}/m_{B_q}=({n_{\mu} + \bar n_{\mu}})/2$.
In analogy to the radiative leptonic $B_q \to \gamma \ell \bar \ell$ decays \cite{Beneke:2020},
the hadronic matrix element (\ref{definition: the hadronic amplitude})
at the lowest non-vanishing order in the electromagnetic interactions
can be further brought into the following form
\begin{eqnarray}
{\cal \bar A}(\bar B_q \to \gamma \gamma)  =
- {4 \, G_F \over \sqrt{2}} \, {\alpha_{\rm em} \over 4 \pi}  \,
\epsilon_1^{\ast \alpha}(p)  \, \epsilon_2^{\ast \beta}(q)  \,
\sum_{p=u, c} \, V_{p b} V_{p q}^{\ast}  \,
\sum_{i=1}^{8} C_i \, T_{i, \, \alpha \beta}^{(p)}\,.
\label{definition:the hadronic amplitude at LO in QED}
\end{eqnarray}
The primary objective of the present paper consists in performing  improved QCD
calculations of  the hadronic tensors appearing in (\ref{definition:the hadronic amplitude at LO in QED})
with the QCD factorization and the dispersion relation techniques.
It is straightforward to derive the explicit expressions of $T_{i, \, \alpha \beta}^{(p)}$
in the following
\begin{eqnarray}
T_{7, \, \alpha \beta}&=& 2 \, \overline{m}_b(\nu) \,
\int d^4 x \, e^{i q \cdot x} \, \langle 0 |
{\rm T} \left  \{ j^{\rm em}_{\beta}(x),
\bar q_L(0) \sigma_{\mu \alpha} \, p^{\mu}  b_R(0) \right \}| \bar B_q(p+q) \rangle \nonumber \\
&& +  \left [ p \leftrightarrow q, \alpha \leftrightarrow \beta \right ] \,,
\label{QCD correlator: 7}\\
T_{i, \, \alpha \beta}^{(p)} &=& - (4 \, \pi)^2 \, \int d^4 x \,  \int d^4 y \,
e^{i p \cdot x} \, e^{i q \cdot y}  \, \langle 0 |  {\rm T} \left  \{ j^{\rm em}_{\alpha}(x), j^{\rm em}_{\beta}(y),
P_i^{(p)}(0) \right \}| \bar B_q(p+q) \rangle \,, \nonumber \\
&& 
(i=1,...6, \, 8,)
\label{QCD correlator: 1-6 and 8}
\end{eqnarray}
where the electromagnetic current of the active quark is given by
\begin{eqnarray}
j^{\rm em}_{\alpha}(x) = \sum_{q} e_q \, \bar q(x) \,  \gamma_{\alpha} \, q(x) \,.
\label{QED current}
\end{eqnarray}

Applying the transversality conditions for the photon polarization vectors
and the QED Ward-Takahashi identities,
we can parameterize  the obtained hadronic tensors in terms of the Lorenz-invariant helicity
form factors
\begin{eqnarray}
T_{i, \, \alpha \beta}^{(p)} = i \, m_{B_q}^3 \,
\left [ \left (g_{\alpha \beta}^{\perp} - i \, \varepsilon_{\alpha \beta}^{\perp} \right ) \, F_{i, L}^{ (p)}
-  \left (g_{\alpha \beta}^{\perp} + i \, \varepsilon_{\alpha \beta}^{\perp} \right ) \, F_{i, R}^{ (p)}  \right ] \,,
\end{eqnarray}
where we have introduced the shorthand notations and conventions
\begin{eqnarray}
g_{\alpha \beta}^{\perp} \equiv g_{\alpha \beta}-{n_{\alpha}  \bar n_{\beta} \over 2}
-{\bar n_{\alpha}   n_{\beta} \over 2},  \qquad
\varepsilon_{\alpha \beta}^{\perp} \equiv   \varepsilon_{\alpha \beta \rho \tau} \bar n^{\rho} v^{\tau}  \,,
\qquad  \varepsilon_{0123} = -1.
\end{eqnarray}
The transversity decay amplitudes of $B_q \to \gamma \gamma$ corresponding to the linearly polarized
(anti)-collinear photon states can be constructed along the line of the analogous analysis for
the charmless non-leptonic $B \to V V$ decays \cite{Beneke:2006hg}
\begin{eqnarray}
F_{i, \|}^{ (p)} =  F_{i, L}^{ (p)} -  F_{i, R}^{ (p)} \,,   \qquad
F_{i, \perp}^{ (p)} =  F_{i, L}^{ (p)} +  F_{i, R}^{ (p)}  \,.
\end{eqnarray}
It is evident that the two-photon final states contributing to
$F_{i, \|}^{ (p)}$ and $F_{i, \perp}^{ (p)}$ are also the CP eigenstates
with the eigenvalues $+1$ and $-1$, respectively.
Furthermore,  the left-handedness of the weak interaction Lagrangian and
the helicity conservation for the strong interaction at high energy
implies the well-known hierarchy structure
\begin{eqnarray}
F_{i, L}^{ (p)} : F_{i, R}^{ (p)}
=  1 : \left ( {\Lambda_{\rm QCD}\over m_b} \right )  \,.
\end{eqnarray}

\section{Factorization of the helicity form factors at leading power}
\label{section:LP}

In this section we aim at constructing the perturbative factorization
formulae for the helicity form factors at leading power in the heavy quark expansion
in the SCET framework and accomplishing the complete NLL resummation of
the parametrically large logarithms $\ln \left (m_b/\Lambda_{\rm QCD} \right )$
due to the renormalization-group evolutions of the hard matching coefficients
and the leading-twist $B$-meson distribution amplitude.
The short-distance Wilson coefficients from integrating out the hard
and hard-collinear fluctuations of the QCD correlation functions
(\ref{QCD correlator: 7}) and (\ref{QCD correlator: 1-6 and 8}) will be
determined by implementing the two-step matching program
${\rm QCD} \to {\rm SCET_{I}} \to {\rm SCET_{II}}$  in sequence.


We start with constructing the ${\rm SCET_{I}}$ representations of
the hadronic tensors $T_{i, \, \alpha \beta}^{(p)}$
in accordance with the position-space formalism \cite{Beneke:2002ph,Beneke:2002ni}
and then providing the detailed expressions of the hard matching coefficients
at NLO in QCD.
In contrast to the SCET factorization for the heavy-to-light $B$-meson form factors
\cite{Beneke:2004rc,Beneke:2005gs},
only the ${\rm A0}$-type  effective weak currents $O_{j}^{(\rm A0)}$ will contribute to
the double radiative decay amplitudes at leading power in the $\Lambda_{\rm QCD}/m_b$
expansion \cite{Lunghi:2002ju}.
The obtained matching equation from ${\rm QCD} \to {\rm SCET_{I}}$ can be written as
\begin{eqnarray}
\sum_{i=1}^{8} C_i \, T_{i, \, \alpha \beta}^{(p)}
&=& \sum_{i=1}^{8} C_i \, H_i^{(p)} \,
\bigg \{  \int d^4 x \, e^{i q \cdot x } \,
\langle 0 | {\rm T} \left \{ j_{\beta, \, {\rm SCET_I}}^{\rm em}(x), \,\,
\left [ (\bar \xi_{\rm \overline{hc}} W_{\rm \overline{hc}}) \gamma_{\alpha}^{\perp} \,
P_L \,  h_v  \right ] (0) \right \} |  \bar B_q \rangle  \nonumber \\
&& \hspace{2.0 cm}   +  \left [ p \leftrightarrow q, \alpha \leftrightarrow \beta \right ] \bigg \} \,,
\label{matching condition at LP}
\end{eqnarray}
with the chiral projection operator $P_L = (1-\gamma_5)/2$.
The desired ${\rm SCET_I}$ representation of the electromagnetic current
$j_{\beta, \, {\rm SCET_I}}^{\rm em}$ of our interest has been constructed in \cite{Lunghi:2002ju}
\begin{eqnarray}
j_{\beta, \, {\rm SCET_I}}^{\rm em}(x)
&=& \sum_q \, e_q \, \left [  \left (\bar \xi_{\rm \overline{hc}} \,  W_{\rm \overline{hc}} \right ) \,
\left ( \gamma_{\beta}^{\perp} {1 \over i \bar n \cdot D_{\rm \overline{hc}}} i \slashed D_{\rm \overline{hc} \, \perp}
+ i \, \slashed D_{\rm \overline{hc} \, \perp}   {1 \over i \bar n \cdot D_{\rm \overline{hc}}} \gamma_{\beta}^{\perp} \right )
{\slashed {\bar n} \over 2 }  \left ( W^{\dag}_{\rm \overline{hc}}  \, \xi_{\rm \overline{hc}} \right ) \right ](x) \nonumber \\
&& +  \sum_q \, e_q \,  \left [ \left ( \bar q_s  Y_s  \right )(x_{+}) \, \gamma_{\beta}^{\perp}  \,
 \left ( W^{\dag}_{\rm \overline{hc}}  \, \xi_{\rm \overline{hc}} \right )(x) \right ]  \,,
 \label{QED current in SCETI}
\end{eqnarray}
where the soft and (anti)-collinear Wilson lines are defined in the standard way
\begin{eqnarray}
W_{\rm \overline{hc}}(x) &=& {\rm P} {\rm exp} \,
\left [ i \, g_s \, \int_{-\infty}^{0} ds \,\, \bar n \cdot A_{\rm \overline{hc}}(x + s \, \bar n) \right ] \,,
\nonumber \\
Y_s(x)  &=& {\rm P} {\rm exp} \,
\left [ i \, g_s \, \int_{-\infty}^{0} dt \,\,  n \cdot A_{s}(x + t \, n) \right ]  \,.
\end{eqnarray}
The position argument of the soft-quark field $x_{+} = (\bar n \cdot x / 2) \, n$
arises from the multipole expansion.
Apparently, the representation for an electromagnetic current carrying the collinear momentum
can be obtained from (\ref{QED current in SCETI}) with the replacement rules
${\rm \overline{hc}} \leftrightarrow {\rm hc}$ and $n \leftrightarrow \bar n$.

The hard matching coefficients from the four-quark and choromomagnetic penguin operators
at ${\cal O}(\alpha_s)$ can be extracted from perturbative corrections to the QCD
$b \to s \gamma$ matrix elements, which have been computed independently in
\cite{Buras:2002tp,Asatrian:2004et} with different techniques for
evaluating the non-trivial two-loop diagrams.
Furthermore, we need the one-loop hard function from matching  the electromagnetic dipole
operator onto ${\rm SCET_{I}}$ at leading power in the heavy quark expansion
\cite{Beneke:2004rc,Beneke:2005gs}
\begin{eqnarray}
\bar q_L  \, i \, \sigma_{\mu \alpha} \, p^{\mu}\, b_R
= n \cdot p \, C_{T_1}^{\rm (A0)} \,
\left [ (\bar \xi_{\rm \overline{hc}} W_{\rm \overline{hc}}) \,
\gamma_{\alpha}^{\perp} \, P_L \,  h_v  \right ]
+ ...,
\end{eqnarray}
with
\begin{eqnarray}
C_{T_1}^{\rm (A0)} = 1 - {\alpha_s(\mu) C_F \over 4 \pi} \,
\left [ 2\, \ln{\nu \over m_{b}} + 2 \,\ln^2 {\mu \over m_{b}}
+ 5 \, \ln{\mu \over  m_{b}} + {\pi^2 \over 12} + 6 \right ].
\label{1-loop hard function}
\end{eqnarray}
Here $\mu$ stands for the renormalization scale of the ${\rm A0}$-type
${\rm SCET_{I}}$ current, whereas $\nu$ characterizes the renormalization
scale of the QCD tensor current.
We then derive the NLO short-distance matching coefficients appearing
in (\ref{matching condition at LP}) as follows
\begin{eqnarray}
\sum_{i=1}^{8} C_i \, H_i^{(p)}
&=&  (-2 \, i) \,\, \overline{m}_b(\nu) \, m_{B_q}
\, V_{7, \, \rm{eff}}^{(p)} \,,
\\
V_{7, \, \rm{eff}}^{(p)}  &=&  C_7^{\rm eff} \,C_{T_1}^{\rm (A0)}
+ \sum_{i=1,...,6, \, 8} {\alpha_s(\mu) \over 4 \pi} \,  C_{i}^{\rm eff} \, F_{i, 7} ^{(p)} \,,
\label{combined hard function at NLO}
\end{eqnarray}
where the effective Wilson coefficients $C_{i}^{\rm eff}$
in the ${\rm \overline{MS}}$ renormalization scheme and in the naive dimensional scheme
for $\gamma_5$ are given by \cite{Chetyrkin:1996vx}
\begin{eqnarray}
C_{i}^{\rm eff} =   \left\{
\begin{array}{l}
C_i  \,, \qquad  \hspace{7.0 cm}
({\rm for} \,\, i=1,..,6) \vspace{0.5 cm} \\
C_7 - {1 \over 3} \, C_3  - {4 \over 9} \, C_4
- {20 \over 3} \, C_5 - {80 \over 9} \, C_6  \,,
 \qquad  \hspace{1.5 cm}
({\rm for} \,\, i=7)  \vspace{0.5 cm} \\
 C_8 + \, C_3  - {1 \over 6} \, C_4
+ 20 \, C_5 - {10 \over 3} \, C_6   \,.
 \qquad  \hspace{1.8 cm}
({\rm for} \,\, i=8)
\end{array}
 \hspace{0.5 cm} \right.
\end{eqnarray}
The complete expressions of the two-loop functions $F_{i, 7}^{(p)} \, (i = 1,...,6)$
and the one-loop function $F_{8, 7}$ will be presented in Appendix \ref{appendix: hard functions}
with no series expansion in the quark-mass ratio $m_c/m_b$.

We are now in a position to perform the perturbative matching of the ${\rm SCET_{I}}$
correlation function in (\ref{matching condition at LP}) onto ${\rm SCET_{II}}$
by integrating out the hard-collinear fluctuations
\begin{eqnarray}
{\cal T}_{\alpha \beta} &=& \int d^4 x \, e^{i q \cdot x } \,
\langle 0 | {\rm T} \left \{ j_{\beta, \, {\rm SCET_I}}^{\rm em}(x), \,\,
\left [ (\bar \xi_{\rm \overline{hc}} W_{\rm \overline{hc}}) \,
\gamma_{\alpha}^{\perp} \, P_L \,  h_v  \right ] (0) \right \} |  \bar B_q \rangle \,
\nonumber \\
&=& \int d t \,\,  {\cal J} \left ({\bar n \cdot q \over \mu^2 \, t} \right ) \,
\langle 0 | (\bar q_s Y_s)(t n) \gamma_{\beta}^{\perp} \, \slashed n  \,
\gamma_{\alpha}^{\perp} \, P_L \, (Y_s^{\dag} h_v)(0) |   \bar B_q \rangle \,.
\end{eqnarray}
According to the definition of the HQET $B$-meson distribution amplitudes
\cite{Grozin:1996pq,Beneke:2000wa}
\begin{eqnarray}
&& \langle 0 | (\bar q_s Y_s)_{\beta}(t n) \,\,\,  (Y_s^{\dag} h_v)_{\alpha}(0) |   \bar B_q \rangle
= -  {i \, \tilde{f}_{B_q}(\mu) \,  m_{B_q} \over 4} \,
\int_0^{\infty} d \omega \, e^{-i \omega t} \,  \nonumber \\
&& \hspace{4.0 cm} \times \left \{  {1+ \slashed v \over 2}   \,
\left [2 \, \phi_B^{+}(\omega, \mu)
+ (\phi_B^{-}(\omega, \mu)-\phi_B^{+}(\omega, \mu)) \,
\slashed {\bar n}  \right ] \, \gamma_5  \right \}_{\alpha \beta} \,,
\label{definition: 2P B-meson LCDA}
\end{eqnarray}
we can readily derive the soft-collinear factorization formula for
the effective non-local matrix element ${\cal T}_{\alpha \beta}$ in momentum space
\begin{eqnarray}
{\cal T}_{\alpha \beta} =  {e_q \, \tilde{f}_{B_q}(\mu) \, m_{B_q} \over 4} \,
\left [g_{\alpha \beta}^{\perp} - i \varepsilon_{\alpha \beta}^{\perp} \right ] \,
\int_0^{\infty} \, { d \omega \over \omega} \, \phi_B^{+}(\omega, \mu) \,
J (\bar n \cdot q, \omega, \mu)  \,.
\end{eqnarray}
The hard-collinear function from the matching procedure ${\rm SCET_{I}} \to {\rm SCET_{II}}$
has been already computed with distinct techniques at the one-loop accuracy
\cite{Lunghi:2002ju,Bosch:2003fc} (see also \cite{Liu:2020ydl}
for the newly derived expression at two loops)
\begin{eqnarray}
J =  1 + {\alpha_s(\mu) \, C_F \over 4 \pi} \,
\left [ \ln^2 \left ( {\mu^2 \over m_b \, \omega} \right )
- {\pi^2 \over 6} - 1 \right ]  + {\cal O}(\alpha_s^2) \,.
\label{1-loop jet function}
\end{eqnarray}
Moreover,  the HQET $B$-meson decay constant $\tilde{f}_{B_q}(\mu)$ can be expressed in terms of
the QCD decay constant $f_{B_q}$ by integrating out the strong interaction dynamics at the hard scale
\begin{eqnarray}
\tilde{f}_{B_q}(\mu) = \left [  1 - {\alpha_s(\mu) \, C_F \over 4 \pi}
\left (3 \, \ln{\mu \over m_b} + 2 \right ) \right ]^{-1}  \, f_{B_q}
\equiv  K^{-1}(m_b, \mu)  \, f_{B_q} \,.
\label{1-loop K function}
\end{eqnarray}

The resulting factorization formula for the double radiative $B_q \to \gamma \gamma$
decay amplitude at leading power in the large energy expansion then  reads
\begin{eqnarray}
{\cal \bar A}_{\rm LP}(\bar B_q \to \gamma \gamma)  &=&
i \, {4 \, G_F \over \sqrt{2}} \, {\alpha_{\rm em} \over 4 \pi}  \,
\epsilon_1^{\ast \alpha}(p)  \, \epsilon_2^{\ast \beta}(q)  \,
\left [g_{\alpha \beta}^{\perp} - i \varepsilon_{\alpha \beta}^{\perp} \right ] \,
e_q \, f_{B_q} \, m_{B_q}^2 \,  K^{-1}(m_b, \mu)   \, \nonumber \\
&& \left [  \sum_{p=u, c} \, V_{p b} V_{p q}^{\ast}   \,\,
\overline{m}_b(\nu)  \, V_{7, \, \rm{eff}}^{(p)}(m_b, \mu, \nu) \right ] \,
\int_0^{\infty} \, { d \omega \over \omega} \, \phi_B^{+}(\omega, \mu) \,
J (m_{b}, \omega, \mu) \,. \hspace{1.0 cm}
\label{factorized amplitude at LP}
\end{eqnarray}
Evidently, there is no single choice of the factorization scale $\mu$
to get rid of the enhanced logarithms of $m_b/\Lambda_{\rm QCD}$
in the factorized amplitude.
Taking the factorization scale of order $\sqrt{m_b \, \Lambda_{\rm QCD}}$,
implementing the resummation of $\ln (m_b/\Lambda_{\rm QCD})$
for both the hard coefficient functions and the leading-twist $B_q$-meson
distribution amplitude to all orders in perturbation theory will be required
in the NLL approximation.
To achieve this goal,  we will first employ the renormalization-group
evolution equations for $V_{7, \, \rm{eff}}^{(p)}$ and $K^{-1}$
in momentum space, whose general solutions can be derived in the following form
\begin{eqnarray}
V_{7, \, \rm{eff}}^{(p)}(m_b, \mu, \nu) &=& \hat{U}_1(m_b, \mu_{\rm h}, \mu) \,
V_{7, \, \rm{eff}}^{(p)}(m_b, \mu_{\rm h}, \nu) \,,  \nonumber \\
K^{-1}(m_b, \mu) &=&   \hat{U}_2(m_b, \mu_{\rm h}, \mu) \, K^{-1}(m_b, \mu_{\rm h})  \,.
\end{eqnarray}
The explicit expression of the evolution function $\hat{U}_1$ can be obtained from
$U_1(E_{\gamma}, \mu_{\rm h}, \mu)$ displayed in \cite{Beneke:2011nf}
with the replacement $E_{\gamma} \to m_{B_q}/2$.
Furthermore, the evolution kernel $\hat{U}_2$ can be deduced from $\hat{U}_1$
with the proper substitutions of the anomalous dimensions.
The NLL resummation of large logarithms due to the scale evolution of the $B$-meson LCDA
can be accomplished by virtue of the corresponding renormalization group equation
at two loops \cite{Braun:2019wyx,Liu:2020ydl}
\begin{eqnarray}
{d \phi_B^{+}(\omega, \mu)  \over d \ln \mu}  &=&
\left [ \Gamma_{\rm cusp}(\alpha_s) \, \ln {\omega \over \mu}
- \gamma_{\eta}(\alpha_s)  \right ] \, \phi_B^{+}(\omega, \mu)
+ \Gamma_{\rm cusp}(\alpha_s) \, \int_0^{\infty}  d x \,
\Gamma(1, x) \, \phi_B^{+}(\omega/x, \mu)  \nonumber \\
&& + \, \left ( {\alpha_s \over 2 \pi} \right )^2  \, C_F \,
 \int_0^{\infty}  {d x \over 1 - x}  \, h(x) \,  \phi_B^{+}(\omega/x, \mu)  \,.
 \label{Two-loop RGE in momentum space}
\end{eqnarray}
The upper incomplete gamma function $\Gamma(s, x)$
and the perturbative kernel $h(x)$ due to conformal symmetry breaking
are given by
\begin{eqnarray}
\Gamma(s, x) = \int_{x}^{\infty} \, d t \, t^{s-1}  \, e^{-t} \,,
\qquad
h(x)= \ln x \, \left  \{  \beta_0 + 2 \, C_F \,
\left [ \ln x  - {1+x  \over x} \, \ln (1-x) - {3 \over 2} \right ]   \right  \}   \,.
\hspace{0.5 cm}
\end{eqnarray}
The perturbative expansions for the cusp anomalous dimension $\Gamma_{\rm cusp}(\alpha_s)$
and the non-logarithmic term $\gamma_{\eta}(\alpha_s)$ are defined by \cite{Liu:2020ydl,Beneke:2011nf}
\begin{eqnarray}
\Gamma_{\rm cusp}(\alpha_s)
= \sum_{n=0}^{\infty}  \, \left ( {\alpha_s \over 4 \pi} \right )^{n+1} \, \Gamma_{\rm cusp}^{(n)} \,,
\qquad
\gamma_{\eta}(\alpha_s)
= \sum_{n=0}^{\infty}  \, \left ( {\alpha_s \over 4 \pi} \right )^{n+1} \, \gamma_{\eta}^{(n)} \,,
\end{eqnarray}
where the series coefficients of particular relevance for the NLL resummation are
\begin{eqnarray}
\Gamma_{\rm cusp}^{(0)} &=& 4 \, C_F \,, \qquad
\Gamma_{\rm cusp}^{(1)} =  C_F \left [ {268 \over 3} - 4 \, \pi^2  - {40 \over 9} \, n_f \right ]  \,,
\nonumber \\
\Gamma_{\rm cusp}^{(2)} &=&  C_F \left \{ 1470 - {536 \pi^2 \over 3} + {44 \pi^4 \over 5}
+ 264 \, \zeta(3)  + \left [-{1276 \over 9} + {80 \pi^4 \over 9}  -{208 \over 3} \, \zeta(3) \right ] \, n_f
- {16 \over 27} \, n_f^2   \right \}  \,, \nonumber \\
\gamma_{\eta}^{(0)} &=& -2 \, C_F   \,,   \\
\gamma_{\eta}^{(1)} &=&  C_F \, \left \{ C_F \,
\left [ - 4  +  {14 \pi^4 \over 3}  - 24 \,  \zeta(3) \right ]
+  \left [ {254 \over 9}  -  {55 \pi^4 \over 6}   - 18 \,  \zeta(3)  \right ]
+ \left [ - {32 \over 27}  +  {5 \pi^2 \over 9}   \right ] \, n_f  \right \}   \,. \nonumber
\end{eqnarray}
Applying the Mellin transformation with respect to the variable $\omega$
\begin{eqnarray}
\tilde{\phi}_B^{+}(\eta, \mu) = \int_0^{\infty} \, {d \omega \over \omega}\,
\phi_B^{+}(\omega, \mu) \, \left ({\omega \over \bar \omega} \right )^{-\eta}
\end{eqnarray}
with a fixed reference scale $\bar \omega$,
the  general solution to the non-linear partial differential equation (\ref{Two-loop RGE in momentum space})
in Mellin representation can be constructed explicitly \cite{Galda:2020epp}
\begin{eqnarray}
\tilde{\phi}_B^{+}(\eta, \mu) &=& {\rm exp} \left  [ S(\mu_0, \mu) + a_{\gamma}(\mu_0, \mu)
+ 2 \, \gamma_{E} \, a_{\Gamma}(\mu_0, \mu)  \right ] \,
\left ({\bar{\omega} \over \mu_0} \right )^{-a_{\Gamma}(\mu_0, \mu)}  \nonumber \\
&&  \times \, \frac{\Gamma(1 + \eta + a_{\Gamma}(\mu_0, \mu)) \, \Gamma(1-\eta)}
{\Gamma(1 - \eta - a_{\Gamma}(\mu_0, \mu)) \, \Gamma(1+\eta)}  \,\,
{\rm exp} \left [  \int_{\alpha_s(\mu_0)}^{\alpha_s(\mu)} \, {d \alpha \over \beta(\alpha)}  \,
\mathcal{G}(\eta + a_{\Gamma}(\mu_{\alpha}, \mu), \alpha) \right ] \nonumber \\
&& \times \, \tilde{\phi}_B^{+}(\eta + a_{\Gamma}(\mu_0, \mu), \mu_0)  \,.
\end{eqnarray}
It proves helpful to introduce four dimensionless functions
\begin{eqnarray}
S(\mu_0, \mu) &=& -  \int_{\alpha_s(\mu_0)}^{\alpha_s(\mu)} \, d \alpha \,
{\Gamma_{\rm cusp}(\alpha)  \over \beta(\alpha)} \,
\int_{\alpha_s(\mu_0)}^{\alpha} \, {d \alpha^{\prime}  \over \beta(\alpha^{\prime}) }  \,,
 \qquad
\mathcal{G}(\eta, \alpha_s) =  \left ( {\alpha_s \over 2 \pi} \right )^{2} \, C_F \,
\int_0^1 d y   \, {y^{\eta}  \over 1 - y}  \, h(y)  \,,
\nonumber \\
a_{\Gamma}(\mu_0, \mu) &=&   -  \int_{\alpha_s(\mu_0)}^{\alpha_s(\mu)} \, d \alpha \,
{\Gamma_{\rm cusp}(\alpha)  \over \beta(\alpha)}  \,,
\qquad  \hspace{2.4 cm}
a_{\gamma}(\mu_0, \mu) =   -  \int_{\alpha_s(\mu_0)}^{\alpha_s(\mu)} \, d \alpha \,
{\gamma_{\eta}(\alpha)  \over \beta(\alpha)}   \,,
\label{solution to 2-loop RGE}
\end{eqnarray}
where the definition of the beta function reads \cite{Baikov:2016tgj}
\begin{eqnarray}
\beta(\alpha_s) &=& {d \, \alpha_s(\mu) \over d \ln \mu}
= -2 \,\alpha_s \, \sum_{n=0}^{\infty} \beta_n  \,
\left ( {\alpha_s \over 4 \pi} \right )^{n+1}   \,, \nonumber  \\
\beta_0 &=&  11 - {2 \over 3} \, n_f  \,,
\qquad
\beta_1 =  102 - {38 \over 3} \, n_f  \,,
\qquad
\beta_2 = {2857 \over 2} - {5033 \over 18} \, n_f + {325 \over 54} \, n_f^2   \,.
\end{eqnarray}

Equivalently, one can perform an alternative integral transformation
following \cite{Braun:2019zhp} \footnote{Implementing the one-loop approximation
for the general eigenfunctions of the Lange-Neubert evolution kernel \cite{Lange:2003ff}
in the $j$-space displayed in (3.12)  of \cite{Braun:2019zhp} yields
$\widehat {\cal Q}_{\lambda}(j)=\Gamma(j+2) \, {\rm exp} \, \left [ - j (\lambda - \gamma_E +1)  \right ]$,
which precisely correspond to the momentum-space eigenfunctions
${\cal Q}_{s}(\omega)= {1 \over \sqrt{\omega \, s}}\, J_1(2 \,\sqrt{\omega \, s})$
originally derived in \cite{Bell:2013tfa} (see \cite{Braun:2014owa} for an independent construction
with the aid of the collinear conformal symmetry) with the desired relation
$\left (\mu \, e^{\gamma_E} \, s \right )^{- j}
= {\rm exp} \left [ - \,  j \, \left ( \lambda - \gamma_E + 1 \right ) \right ]$
 at ${\cal O}(\alpha_s)$ and the integral transformation (\ref{new integral transformation}).}
\begin{eqnarray}
\varphi_B^{+}(j, \mu)  =  - {i \over 2 \, \pi} \, \Gamma(-j) \,
\int_{0}^{\infty} \, d \omega \, \phi_B^{+}(\omega, \mu) \,
\left (  {\omega  \over \mu \, e^{\gamma_E}}  \right )^{j}  \,,
\label{new integral transformation}
\end{eqnarray}
to reduce the momentum-space evolution equation (\ref{Two-loop RGE in momentum space})
to an elegant form
\begin{eqnarray}
\left [  {d \over d \ln \mu} + V(j, \alpha_s) \right ] \, \varphi_B^{+}(j, \mu) = 0 \,.
\label{Two-loop RGE in j space}
\end{eqnarray}
The general expression of the renormalization kernel $V(j, \alpha_s)$ is given by
\begin{eqnarray}
V(j, \alpha_s) = \Gamma_{\rm cusp}(\alpha_s) \, \left [ \psi(j+2) - \psi(2) + \vartheta(j) \right ]
+ \gamma_{+}(\alpha_s) + j  \,,
\end{eqnarray}
where $j(\mu)$ is determined by the solution to the following differential equation
\begin{eqnarray}
{d \, j(\mu) \over d \ln \mu} = - \Gamma_{\rm cusp}(\alpha_s)  \,.
\end{eqnarray}
The coefficient functions $\gamma_{+}(\alpha_s)$ and $\vartheta(j)$
at the NLL accuracy take the form
\begin{eqnarray}
\gamma_{+}(\alpha_s) &=& \gamma_{\eta}(\alpha_s) 
+ \Gamma_{\rm cusp}(\alpha_s) \,
\left \{ 1 - \left ( {\alpha_s \over 4 \pi} \right ) \, \left [ \beta_0 \, \left (1- {\pi^2  \over 6} \right )
- C_F \,  \left (3- {\pi^2  \over 6} \right ) \right ]  \right \}  \,,
\nonumber \\
\vartheta(j) &=& \left ( {\alpha_s \over 4 \pi} \right )  \,
\bigg \{  2 \, C_F  \bigg [  \psi^{\prime}(j+2) \, \left( \psi(j+2) - \psi(1)  \right )
+ \psi^{\prime}(j+1) \, \left( \psi(j+1) - \psi(1)  \right )    \nonumber \\
&& \hspace{1.5 cm}  + {1 \over (j+1)^3}  - {\pi^2 \over 6}  \bigg ]
+ (\beta_0 - 3 \, C_F) \, \left [\psi^{\prime}(j+2) - \psi^{\prime}(2) \right ] \bigg \}   \,.
\end{eqnarray}
It is straightforward to demonstrate the equivalence of the two renormalization
group equations (\ref{Two-loop RGE in momentum space})
and (\ref{Two-loop RGE in j space}) with the aid of the transformation relation
\begin{eqnarray}
\varphi_B^{+}(j, \mu) =   - {i \over 2 \, \pi} \, \Gamma(-j) \,
\bar \omega  \,  \left (  {\bar \omega  \over \mu \, e^{\gamma_E}}  \right )^{j}  \,
\tilde{\phi}_B^{+}(-j-1, \mu)  \,.
\end{eqnarray}

The obtained solution (\ref{solution to 2-loop RGE}) to the Mellin-space evolution equation
of the $B$-meson distribution amplitude allows us to derive the NLL resummation improved
expression for  the leading-power factorization formula (\ref{factorized amplitude at LP})
exactly
\begin{eqnarray}
{\cal \bar A}_{\rm LP}(\bar B_q \to \gamma \gamma)  &=&
i \, {4 \, G_F \over \sqrt{2}} \, {\alpha_{\rm em} \over 4 \pi}  \,
\epsilon_1^{\ast \alpha}(p)  \, \epsilon_2^{\ast \beta}(q)  \,
\left [g_{\alpha \beta}^{\perp} - i \varepsilon_{\alpha \beta}^{\perp} \right ] \,
e_q \, f_{B_q} \, m_{B_q}^2 \, \hat{U}_1(m_b, \mu_{\rm h}, \mu) \,
\hat{U}_2(m_b, \mu_{\rm h}, \mu) \, \nonumber \\
&&  K^{-1}( m_b, \, \mu_{\rm h})   \,
\left [  \sum_{p=u, c} \, V_{p b} V_{p q}^{\ast}   \,\,
\overline{m}_b(\nu)  \, V_{7, \, \rm{eff}}^{(p)}(m_b, \mu_{\rm h}, \nu) \right ] \,
\nonumber \\
&&   {\rm exp} \left  [ S(\mu_0, \mu) + a_{\gamma}(\mu_0, \mu)
+ 2 \, \gamma_{E} \, a_{\Gamma}(\mu_0, \mu)  \right ] \,
{\hat{\mathcal{J}}} \left ({\partial \over\partial \eta}, \mu \right )  \,
\left ( {m_{b} \, \bar{\omega} \over \mu^2} \right )^{-\eta}    \nonumber \\
&&  \frac{\Gamma(1 + \eta + a_{\Gamma}(\mu_0, \mu)) \, \Gamma(1-\eta)}
{\Gamma(1 - \eta - a_{\Gamma}(\mu_0, \mu)) \, \Gamma(1+\eta)} \,\,
{\rm exp} \left [  \int_{\alpha_s(\mu_0)}^{\alpha_s(\mu)} \, {d \alpha \over \beta(\alpha)}  \,
\mathcal{G}(\eta + a_{\Gamma}(\mu_{\alpha}, \mu), \alpha) \right ] \, \nonumber \\
&&  \left ({\bar{\omega} \over \mu_0} \right )^{-a_{\Gamma}(\mu_0, \mu)}
\, \tilde{\phi}_B^{+}(\eta + a_{\Gamma}(\mu_0, \mu), \mu_0) \,\, \bigg |_{\eta \to 0} \,,
\label{factorized amplitude at NLL}
\end{eqnarray}
where the operator-valued function ${\hat{\mathcal{J}}}$ is defined in terms of
the hard-collinear function
\begin{eqnarray}
{\hat{\mathcal{J}}} \left ( \ln{\mu^2 \over m_{b} \, \omega}, \, \mu \right )
\equiv J(m_{b},  \omega, \mu) \,.
\end{eqnarray}
Implementing perturbative expansions for the anomalous dimensions
as well as the QCD $\beta$-function at NLL leads to the approximate expressions
of the renormalization group evolution functions displayed in
Appendix \ref{appendix: B-meson LCDA}.
Consequently, we can readily write down the factorized expressions
for the  helicity form factors at leading order
in the power expansion of $\Lambda_{\rm QCD}/m_b$
\begin{eqnarray}
\sum_{i=1}^{8} C_i \, F_{i, L}^{(p), \, {\rm LP}} &=&
-  {e_q \, f_{B_q}  \over m_{B_q}} \, \hat{U}_1(m_b, \mu_{\rm h}, \mu) \,
\hat{U}_2(m_b, \mu_{\rm h}, \mu) \,  K^{-1}( m_b, \mu_{\rm h})   \nonumber \\
&& \, \overline{m}_b(\nu)  \, V_{7, \, \rm{eff}}^{(p)}(m_b, \mu_{\rm h}, \nu)
\, {\cal R}(m_b, \mu_0, \mu) \,,
\nonumber \\
\sum_{i=1}^{8}  C_i \, F_{i, R}^{(p), \, {\rm LP}} &=& 0 \,,
\label{NLL factorization formula of the helicity FFs}
\end{eqnarray}
where the explicit form of the dimensionful function ${\cal R}(m_b, \mu_0, \mu)$
can be found in (\ref{R function: RGE}).



\section{Factorization of the helicity form factors at subleading power}
\label{section:NLP}

In this section we proceed to construct the tree-level factorization formulae
for the subleading power corrections to the exclusive radiative $B_q \to \gamma \gamma$
decay amplitude from distinct sources in the framework of QCD factorization.
To achieve this goal, we will categorize the  power suppressed mechanisms of interest
according to the effective weak operators defining the QCD correlation functions
(\ref{QCD correlator: 7}) and (\ref{QCD correlator: 1-6 and 8})
and further take advantage of the non-trivial relations for the light-cone HQET operators
due to the classical equations of motion extensively.

\begin{figure}
\begin{center}
\includegraphics[width=1.0 \columnwidth]{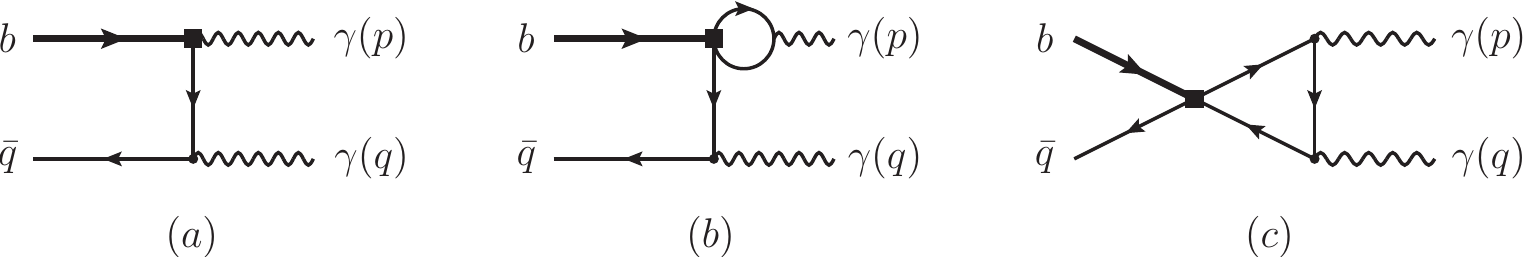}
\vspace*{-0.5 cm}
\caption{Diagrammatical representations of the QCD correlation functions
(\ref{QCD correlator: 7}) and (\ref{QCD correlator: 1-6 and 8})
contributing to the double radiative $B_q \to \gamma \gamma$ decay amplitudes
at LO in the strong coupling constant, where the accompanying  diagrams due to the
exchanges of two energetic photons in the final states are not presented. }
\label{fig: Feynman diagrams}
\end{center}
\end{figure}

\subsection{The NLP  corrections from the electromagnetic dipole operator}
\label{section£ºNLP O7 effects}

The first type of the next-to-leading power (NLP) correction originates from the subleading terms
of the hard-collinear quark propagator due to the (anti)-collinear photon emission off
the constituent light-flavour quark of the $B_q$-meson
as displayed in Figure \ref{fig: Feynman diagrams}(a).
Such NLP contribution from the electromagnetic penguin operator $P_7$ can be determined from
the QCD correlation function (\ref{QCD correlator: 7})
\begin{eqnarray}
T_{7, \, \alpha \beta}^{\rm hc, \, NLP} &=&
2 \, e_q \, \overline{m}_b(\nu)  \,  m_{B_q} \, \int d^4 x \,
\int {d^4 {\ell} \over (2 \pi)^4 } \, {\rm exp} \left [ {i \, (q - \ell) \cdot x} \right ] \,
{(q - \ell)^{\mu} \over \ell^2 + i 0} \nonumber \\
&&   \times  \, \langle  0 |  \bar q(x) \,\, \gamma_{\beta}^{\perp} \,\, \gamma_{\mu}  \,\,
\slashed {\bar n} \,\, \gamma_{\alpha}^{\perp} \,\, P_R  \,\, h_v(0)|  \bar B_q   \rangle  \,,
\label{T7 NLP: hc}
\end{eqnarray}
where the symmetry factor ``2" arises from the exchanges of two identical particles in the final states.
Applying the light-cone decomposition of the Dirac matrix $\gamma_{\mu}$ \cite{Becher:2014oda}
\begin{eqnarray}
\gamma_{\mu}  = {\slashed n \over 2} \, \bar n_{\mu}
+ {\slashed {\bar n} \over 2} \,  n_{\mu} + \gamma_{\mu}^{\perp} \,,
\label{gamma matrix decomposition}
\end{eqnarray}
and employing the general parametrization of the hadronic matrix element
(\ref{T7 NLP: hc}), one can immediately demonstrate that the yielding contributions
from the second and third terms in  (\ref{gamma matrix decomposition})
vanish in the four-dimensional space.
We are therefore led to the reduced formula
\begin{eqnarray}
T_{7, \, \alpha \beta}^{\rm hc, \, NLP} &=&
\left [ i \, e_q \, \overline{m}_b(\nu)  \,  m_{B_q}  \right ] \, \int d^4 x \,
\int {d^4 {\ell} \over (2 \pi)^4 } \, {\rm exp} \left [ {i \, (q - \ell) \cdot x} \right ] \,
{\bar n _{\mu} \over \ell^2 + i 0} \nonumber \\
&&   \times  \, {\partial \over \partial x_{\mu}} \, \langle  0 |  \bar q(x) \,\, \gamma_{\beta}^{\perp} \,\, \slashed n  \,\,
\slashed {\bar n} \,\, \gamma_{\alpha}^{\perp} \,\, P_R  \,\, h_v(0)|  \bar B_q   \rangle
\nonumber \\
&=&  \left [ - {e_q \, \overline{m}_b(\nu)  \,  m_{B_q} \over 4 \, \pi^2} \right ] \,
\int d^4 x \,\,    {e^{i \, q \cdot x} \over x^2} \, \left (2 \, v_{\mu} - n_{\mu} \right )
\nonumber \\
&& \times \,  {\partial \over \partial x_{\mu}} \,
\, \langle  0 |  \bar q(x) \,\, \gamma_{\beta}^{\perp} \,\, \slashed n  \,\,
\slashed {\bar n} \,\, \gamma_{\alpha}^{\perp} \,\, P_R  \,\, h_v(0)|  \bar B_q   \rangle \,,
\label{T7 NLP hc: before EOM}
\end{eqnarray}
with the aid of the standard method of  the integration by parts (IBP)
and the precise relation  of the four-vectors  $\bar n _{\mu} = 2 \, v_{\mu} - n_{\mu}$.
Taking advantage of the well-known  operator identity
due to the classical equations of motion \cite{Kawamura:2001jm,Kawamura:2001bp}
\begin{eqnarray}
v_{\mu} \,  {\partial \over \partial x_{\mu}} \,
\left [  \bar q(x) \,\Gamma \,  h_v(0) \right ]
= i \, \int_0^1 d u \, \bar u \, \bar q(x) \,g_s \, G_{\alpha \beta}(u x) \,
x^{\alpha} \, v^{\beta} \, \Gamma \,  h_v(0)
+ (v \cdot \partial) \,  \left [  \bar q(x) \,\Gamma \,  h_v(0) \right ]  \,,
\end{eqnarray}
where the derivative with respect to the total translation is defined by
\begin{eqnarray}
\partial_{\mu} \left [  \bar q(x) \,\Gamma \,  h_v(0) \right ]
= {\partial \over \partial y^{\mu}} \,
\left \{  \bar q(x+y) \, [x+y, y] \,\Gamma \,  h_v(y) \right \} \bigg |_{y=0} \,,
\qquad
\left [ x_1, x_2 \right ] \equiv Y_s(x_1)  \, Y_s^{\dag}(x_2) \,,
\hspace{0.5 cm}
\end{eqnarray}
the subleading power HQET matrix element (\ref{T7 NLP hc: before EOM})
can be further computed as follows
\begin{eqnarray}
T_{7, \, \alpha \beta}^{\rm hc, \, NLP} &=&
\left [ - {e_q \, \overline{m}_b(\nu)  \, f_{B_q} \,  m_{B_q}^2 \over 4 \, \pi^2} \right ] \,
\left [g_{\alpha \beta}^{\perp}  - i \, \varepsilon_{\alpha \beta}^{\perp} \right ] \,
\int d^4 x \,\,    {e^{i \, q \cdot x} \over x^2} \nonumber \\
&& \, \times \left \{ i \, \bar n \cdot x \, \int_0^1 du \, \bar u \,
\hat{\Psi}_4(v \cdot x, \, u \, v \cdot x, \, \mu)
- 2 \,  \left [ \bar{\Lambda}  -  i \, {\partial \over \partial \bar n \cdot x} \right ]
\, \hat{\phi}_B^{+}(v \cdot x, \mu)  \right \}  \,,
\label{T7 NLP hc: coordinate space formula}
\end{eqnarray}
where the hadronic parameter $\bar{\Lambda}$  characterizes the mass splitting
between the heavy quark and the physical heavy-hadron state under discussion
\cite{Manohar:2000dt}
\begin{eqnarray}
m_{B_q}  =  m_b + \bar{\Lambda} - {\lambda_1 + 3 \, \lambda_2(m_b) \over m_b}
+ {\cal O} \left ( {1 \over m_b^2} \right )  \,.
\end{eqnarray}
Moreover, the coordinate-space factorization formula (\ref{T7 NLP hc: coordinate space formula})
at LO in the strong coupling has been established by implementing  the  systematic parametrization
for the light-cone hadronic matrix element of the three-body effective operator
at the twist-six accuracy \cite{Braun:2017liq}
\begin{eqnarray}
&& \langle 0 | \bar q_{\alpha}(\tau_1 \, n) \, g_s \, G_{\mu \nu}(\tau_2 \, n) \,
h_{v \, \beta}(0) | \bar B_q \rangle \nonumber \\
&& = {\tilde{f}_{B_q}(\mu) \, m_{B_q} \over 4} \,
\bigg [ (1 + \slashed v) \, \bigg \{ (v_{\mu} \gamma_{\nu} - v_{\nu} \gamma_{\mu})  \,
\left [\hat{\Psi}_A(\tau_1, \tau_2, \mu) - \hat{\Psi}_V(\tau_1, \tau_2, \mu) \right ]
- i \, \sigma_{\mu \nu} \, \hat{\Psi}_V(\tau_1, \tau_2, \mu) \nonumber  \\
&& \hspace{0.4 cm}
- (n_{\mu} \, v_{\nu} - n_{\nu} \, v_{\mu} ) \, \hat{X}_A(\tau_1, \tau_2, \mu)
+ (n_{\mu} \, \gamma_{\nu} - n_{\nu} \, \gamma_{\mu} ) \,
\left [ \hat{W}(\tau_1, \tau_2, \mu)  + \hat{Y}_A(\tau_1, \tau_2, \mu)   \right ] \nonumber \\
&& \hspace{0.4 cm} + \, i \, \epsilon_{\mu \nu \alpha \beta} \,
n^{\alpha} \, v^{\beta}  \, \gamma_5 \, \hat{\tilde{X}}_A(\tau_1, \tau_2, \mu)
- \, i \, \epsilon_{\mu \nu \alpha \beta} \,
n^{\alpha} \, \gamma^{\beta}  \, \gamma_5 \, \hat{\tilde{Y}}_A(\tau_1, \tau_2, \mu)  \nonumber \\
&&  \hspace{0.4 cm}  - \, (n_{\mu} \, v_{\nu} - n_{\nu} \, v_{\mu} ) \,
\slashed  n \, \hat{W}(\tau_1, \tau_2, \mu)
+ \, (n_{\mu} \, \gamma_{\nu} -  n_{\nu} \, \gamma_{\mu} ) \,
\slashed  n \, \hat{Z}(\tau_1, \tau_2, \mu)   \bigg \}  \,
\gamma_5 \bigg ]_{\beta \, \alpha}  \,.
\label{definition: 3P B-meson LCDA}
\end{eqnarray}
The emerging eight independent invariant functions can be expressed in terms of the
HQET distribution amplitudes with the definite collinear twist
\begin{eqnarray}
\hat{\Phi}_3 &=& \hat{\Psi}_A - \hat{\Psi}_V \,,
\qquad  \hspace{3.8 cm}
\hat{\Phi}_4 = \hat{\Psi}_A + \hat{\Psi}_V \,,
\nonumber \\
\hat{\Psi}_4  &=& \hat{\Psi}_A + \hat{X}_A \,,
\qquad  \hspace{3.8 cm}
\hat{\tilde{\Psi}}_4  = \hat{\Psi}_V - \hat{\tilde{X}}_A \,,
\nonumber \\
\hat{\tilde{\Phi}}_5 &=&  \hat{\Psi}_A + \hat{\Psi}_V + 2 \, \hat{Y}_A - 2 \, \hat{\tilde{Y}}_A + 2 \, W \,,
\qquad
\hat{\Psi}_5 =  - \hat{\Psi}_A + \hat{X}_A - 2 \, \hat{Y}_A  \,,
\nonumber \\
\hat{\tilde{\Psi}}_5 &=&  - \hat{\Psi}_V -  \hat{\tilde{X}}_A + 2 \, \hat{\tilde{Y}}_A  \,,
\qquad  \hspace{2.2 cm}
\hat{\Phi}_6 =  \hat{\Psi}_A - \hat{\Psi}_V + 2 \, \hat{Y}_A  + 2 \, \hat{W}
+ 2 \, \hat{\tilde{Y}}_A - 4 \, \hat{Z} \,.
\hspace{1.0 cm}
\end{eqnarray}
In the light of the momentum-space representations of the two-particle
and three-particle $B_q$-meson distribution amplitudes
\begin{eqnarray}
&& \hat{\phi}_B^{+}(v \cdot x, \mu)  = \int_0^{\infty} \, d \omega \,
e^{-i \, \omega \, v \cdot x} \,  \phi_B^{+}(\omega, \mu) \,,
\nonumber \\
&& \hat{\Psi}_4(v \cdot x, \, u \,\, v \cdot x, \, \mu)
=  \int_0^{\infty} \, d \omega_1  \, \int_0^{\infty} \, d \omega_2  \,
e^{-i \,(\omega_1 + u \, \omega_2) \, v \cdot x} \,
\Psi_4(\omega_1, \omega_2,  \mu)  \,,
\end{eqnarray}
the resulting factorized expression for the NLP amplitude (\ref{T7 NLP hc: before EOM})
can be written as
\begin{eqnarray}
T_{7, \, \alpha \beta}^{\rm hc, \, NLP} &=&
\left [ - 2 \, i \, e_q \, \overline{m}_b(\nu)  \, f_{B_q} \,  m_{B_q}  \right ] \,
\left [g_{\alpha \beta}^{\perp}  - i \, \varepsilon_{\alpha \beta}^{\perp} \right ] \,
\bigg  \{  {1 \over 2} - \left ( {\bar \Lambda \over \lambda_{B_q}} \right )
\nonumber \\
&& + \int_0^{\infty} \, d \omega_1  \, \int_0^{\infty} \, d \omega_2  \,
{1 \over \omega_2} \, \left [ {1 \over \omega_2}  \,\,
\ln {\omega_1 \over \omega_1 + \omega_2} + {1 \over \omega_1} \right ] \,
\Psi_4(\omega_1, \omega_2,  \mu)  \bigg \} \,,
\label{T7 NLP hc: momentum space formula}
\end{eqnarray}
where the first inverse moment of the leading-twist $B_q$-meson LCDA
\begin{eqnarray}
\lambda_{B_q}^{-1}(\mu) =   \int_0^{\infty} \, d \omega \,\,
{\phi_B^{+}(\omega, \mu)  \over \omega}
\label{definition£ºinverse moment}
\end{eqnarray}
fulfills  the established relation under the renormalization-scale evolution
in the LO approximation (see \cite{Galda:2020epp} for the NLL improvement)
\begin{eqnarray}
\lambda_{B_q}(\mu) = \lambda_{B_q}(\mu_0) \,
\bigg \{ 1 + {\alpha_s(\mu_0) \, C_F \over 4 \, \pi} \,
\ln {\mu \over \mu_0} \,
\left [ 4 \, \widehat{\sigma}_{B_q}^{(1)}(\mu_0)
+ 4 \, \ln{\sqrt{\mu_0 \, \mu}  \,\, e^{\gamma_E} \over \lambda_{B_q}(\mu_0)}  - 2 \right ]
+ {\cal O}(\alpha_s^2) \bigg \} \,.
\end{eqnarray}
We have introduced the following definition of the inverse-logarithmic moments
$\widehat{\sigma}_{B_q}^{(n)}$ \cite{Beneke:2018wjp}
\begin{eqnarray}
\widehat{\sigma}_{B_q}^{(n)} (\mu) = \lambda_{B_q}(\mu) \,
\int_{0}^{\infty} \, {d \omega \over \omega} \,
\left [ \ln \left ( {\lambda_{B_q}(\mu) \over \omega} \right )
- \gamma_E \right ]^{n} \, \phi_{B}^{+}(\omega, \mu)  \,.
\label{definition£ºinverse-logarithmic moment}
\end{eqnarray}
Accordingly, the obtained subleading power corrections to the two helicity form factors reads
\begin{eqnarray}
F_{7, L}^{\rm hc, \, NLP} &=&
 - {2 \, e_q \, \overline{m}_b(\nu)  \, f_{B_q} \over   m_{B_q}^2}  \,
\bigg  \{  \int_0^{\infty} \, d \omega_1  \, \int_0^{\infty} \, d \omega_2  \,
{1 \over \omega_2} \, \left [ {1 \over \omega_2}  \,\,
\ln {\omega_1 \over \omega_1 + \omega_2} + {1 \over \omega_1} \right ] \,
\Psi_4(\omega_1, \omega_2,  \mu)  \nonumber \\
&& \hspace{3.0 cm} + \,  {1 \over 2} - \left ( {\bar \Lambda \over \lambda_{B_q}} \right )  \bigg \} \,,
\nonumber \\
F_{7, R}^{\rm hc, \, NLP} &=&  0\,.
\label{NLP factorization formula of F7hc}
\end{eqnarray}
Apparently, the NLP hard-collinear contribution defined by (\ref{T7 NLP: hc})
preserves the large-recoil symmetry  $F_{i, \, \|}^{(p)} = F_{i, \, \perp}^{(p)}$
of the transversality  amplitudes, in accordance with the analogous observation
for the power suppressed effect in $B_q \to \gamma \ell \bar \ell$
due to the $B$-type insertion of $P_7$ \cite{Beneke:2020}.

In addition, it would be interesting to investigate the subleading power
correction to the hadronic matrix element $T_{7, \alpha \beta}$
from the non-vanishing light-quark mass.
A straightforward calculation at tree level leads to the following expression
\begin{eqnarray}
T_{7, \, \alpha \beta}^{m_q, \, {\rm NLP}} &=&
\left [ - i \, e_q \, \overline{m}_b(\nu)  \, m_q \,  f_{B_q} \,  m_{B_q}  \right ] \,
\left [g_{\alpha \beta}^{\perp}  - i \, \varepsilon_{\alpha \beta}^{\perp} \right ] \,
\int_0^{\infty} \, d \omega \, {\phi_B^{-}(\omega, \mu)  \over \omega} \,,
\label{T7 NLP mq}
\end{eqnarray}
where the convolution integral over the variable $\omega$ is unfortunately divergent
based upon the asymptotic behaviour of $\phi_B^{-}(\omega, \mu) $.
Employing  the prescription described in \cite{Beneke:2003zv},
we parameterize this contribution  by introducing
an unknown complex quantity $X_{\rm NLP}$
\begin{eqnarray}
{\cal I}^{m_q}_{\rm NLP} &=& \int_0^{\infty} \, d \omega \, {\phi_B^{-}(\omega, \mu)  \over \omega}
= \left [  \int_0^{\Lambda_{\rm UV}}
+ \int_{\Lambda_{\rm UV}}^{\infty} \right ] \,\,
d \omega \,\, {\phi_B^{-}(\omega, \mu)  \over \omega}
\nonumber \\
&=& \left [  \phi_B^{-}(0, \mu) \, X_{\rm NLP}
+ \int_0^{\Lambda_{\rm UV}} d \omega  \,
{\phi_B^{-}(\omega, \mu) - \phi_B^{-}(0, \mu) \over \omega}  \right ]
+ \int_{\Lambda_{\rm UV}}^{\infty}  \,
d \omega \,\, {\phi_B^{-}(\omega, \mu)  \over \omega}  \,,
\end{eqnarray}
where an arbitrary perturbative scale $\Lambda_{\rm UV}$
is introduced to guarantee  the ultraviolet finiteness of
the newly defined phenomenological parameter.
For concreteness, the nonperturbative model  for
the coefficient $X_{\rm NLP}$ adopted in this work reads
(see \cite{Beneke:2007zz} for further discussions)
\begin{eqnarray}
X_{\rm NLP}  = \left [ 1 + \varrho_{S} \,\,  {\exp} (i \, \varphi_S) \right ] \,
\ln  \left ( {\Lambda_{\rm UV} \over \Lambda_{\rm h}} \right )  \,,
\qquad  \varrho_{S} \in [0, 1] \,,
\qquad \Lambda_{\rm h} = 0.5 \, {\rm GeV} \,.
\end{eqnarray}
The achieved results of the  helicity form factors
are given by
\begin{eqnarray}
F_{7, L}^{m_q, \, {\rm NLP}} &=&
 - {e_q \, \overline{m}_b(\nu)  \,f_{B_q} \,  \over  m_{B_q}^2}  \,
\left (  m_q \,  {\cal I}^{m_q}_{\rm NLP} \right )   \,,
\nonumber \\
F_{7, R}^{m_q, \, {\rm NLP}} &=&  0 \,,
\label{NLP formula of the light-quark mass effect}
\end{eqnarray}
which implies the anticipated scaling  rule in the heavy quark expansion
\begin{eqnarray}
F_{7, L}^{m_q, \, {\rm NLP}} : F_{7, L}^{\rm hc, \, NLP}
= m_q :  \Lambda_{\rm QCD} \,.
\end{eqnarray}

We now turn to evaluate the NLP correction from the power suppressed term
in the ${\rm SCET_{I}}$  representation of the heavy-to-light
transition current appearing in (\ref{QCD correlator: 7})
\begin{eqnarray}
J^{\rm (A2)} \supset
(\bar \xi_{\rm \overline{hc}} \, W_{\rm \overline{hc}}) \,
\gamma_{\alpha}^{\perp} \,\,  P_L \,\,
\left ( { i \, \overrightarrow{\slashed D}_{\top} \over 2 \, m_b} \right )
\, \,   h_v   + ... \,,
\label{JA2 current}
\end{eqnarray}
which arises from the HQET expansion of the heavy quark field in QCD \cite{Mannel:2020fts}
\begin{eqnarray}
b(x) &=&  e^{- i \, m_b \, v \cdot x } \,
\bigg [ 1 + { i \, \overrightarrow{{\slashed D}}_{\top}  \over 2 \, m_b}
+  { (v\cdot \overrightarrow{D}) \, \overrightarrow{{\slashed D}}_{\top}  \over 4 \, m_b^2}
- { \overrightarrow{{\slashed D}}_{\top} \, \overrightarrow{{\slashed D}}_{\top} \over 8 \, m_b^2}
+ {\cal O} \left ({1 \over m_b^3} \right ) \bigg ] \, h_v(x) \,,
\nonumber \\
D_{\top}^{\mu} &\equiv&  D^{\mu}
- (v\cdot D)  \, v^{\mu} \,.
\end{eqnarray}
As we will not go beyond the accuracy  of $\Lambda_{\rm QCD}/m_b$
in the derivative expansion, the lowest-order equation of motion in HQET
(see \cite{Falk:1993dh} for the generalization at NLO)
\begin{eqnarray}
i \, v \cdot \overrightarrow{D} \, h_v = 0
\end{eqnarray}
evidently permits the replacement
$i\, \overrightarrow{\slashed D}_{\top} \, h_v \to i \, \overrightarrow{\slashed D} \, h_v$
in the effective weak  current (\ref{JA2 current}).
The corresponding NLP correction to the correlation function
(\ref{QCD correlator: 7}) can be derived as follows
\begin{eqnarray}
T_{7, \, \alpha \beta}^{A2, \, {\rm NLP}} &=&
\left [ - {i \, e_q \, m_{B_q}^2  \over 2} \right ] \,
\, \int d^4 x \, \int {d^4 {\ell} \over (2 \pi)^4 } \,
{\rm exp} \left [ {i \, (q - \ell) \cdot  x} \right ] \,
{1 \over \ell^2 + i 0} \nonumber \\
&&   \times  \, \langle  0 |  \bar q(x) \,\, \gamma_{\beta}^{\perp} \,\,
\slashed {n} \,\, \slashed {\bar n} \,\,  \gamma_{\alpha}^{\perp}  \,\,
\overrightarrow{\slashed{D}}  \,\, P_L  \,\, h_v(0)|  \bar B_q   \rangle   \nonumber \\
&=&   \left [ - i \, e_q \, m_{B_q}^2 \right ] \,
\, \int d^4 x \, \int {d^4 {\ell} \over (2 \pi)^4 } \,
{\rm exp} \left [ {i \, (q - \ell) \cdot x} \right ] \,
{1 \over \ell^2 + i 0} \nonumber \\
&&   \times  \, \langle  0 |  \bar q(x) \,\,
\left [ \gamma_{\beta}^{\perp} \,\, \gamma_{\alpha}^{\perp} \,\,
\bar n_{\rho} \,\, P_R
+ \gamma_{\rho}^{\perp} \,\, \gamma_{\alpha}^{\perp} \,\,
\gamma_{\beta}^{\perp} \,\, P_L  \right ] \,\,
\slashed {n}  \,\, {\overrightarrow{D}}^{\rho} \,\, h_v(0)|  \bar B_q   \rangle \,.
\label{NLP JA2 amplitude}
\end{eqnarray}
Implementing further the classical HQET operator identities \cite{Kawamura:2001bp,Kawamura:2001jm}
\begin{eqnarray}
\bar q(x) \, \Gamma \, {\overrightarrow{D}}_{\rho} \,  h_v(0)
&=& \partial_{\rho} \left [ \bar q(x) \, \Gamma \, h_v(0) \right ] \,
+ \, i \, \int_0^1 d u \, \bar u \,\, \bar q(x) \,g_s \, G_{\lambda \rho}(u x) \,
x^{\lambda} \, \Gamma \,  h_v(0) \nonumber \\
&& - \, {\partial \over \partial x^{\rho}} \bar q(x) \, \Gamma \, h_v(0)  \,,
\nonumber \\
{\partial \over \partial x_{\rho}} \bar q(x) \, \gamma_{\rho} \, \Gamma \,   h_v(0)
&=& - i \, \int_0^1 d u \, u \,\, \bar q(x) \,g_s \, G^{\lambda \rho}(u x) \,
x_{\lambda} \, \gamma_{\rho} \, \Gamma \,  h_v(0)  \,,
\end{eqnarray}
the subleading power  amplitude  $T_{7, \, \alpha \beta}^{A2, \, {\rm NLP}}$
can be cast in the following form
\begin{eqnarray}
T_{7, \, \alpha \beta}^{A2, \, {\rm NLP}} &=&
{e_q \, m_{B_q}^2  \over 4 \, \pi^2}  \,
\int d^4 x \,\,    {e^{i \, q \cdot x} \over x^2} \,
\bigg \{   \left \langle 0 \left | \partial^{\rho}
\left [\bar q(x)  \, \Gamma_{{\rm A}, \, \alpha \beta \rho} \, h_v(0) \right ]
\right | \bar{B}_q \right \rangle  \nonumber \\
&& \hspace{1.5 cm} +  \, \left  \langle 0  \left | n_{\rho} \, {\partial \over \partial x_{\rho}} \,
\bar q(x)  \, \Gamma_{{\rm B}, \, \alpha \beta} \, h_v(0)  \right | \bar{B}_q \right \rangle
\nonumber \\
&& \hspace{1.5 cm}  + \,  i \, \int_0^1 d u \,\,
\left  \langle 0  \left |  \bar q(x) \,g_s \, G^{\lambda \rho}(u x) \,
x_{\lambda} \,\, \Gamma_{{\rm C}, \, \alpha \beta \rho} \,  h_v(0)
\right | \bar{B}_q \right \rangle\bigg \}   \,,
\end{eqnarray}
with
\begin{eqnarray}
\Gamma_{{\rm A}, \, \alpha \beta \rho} &=&
\left ( \gamma_{\rho}^{\perp} \,\, \gamma_{\alpha}^{\perp} \,\,
\gamma_{\beta}^{\perp}  \,\,  P_L
\, -   \, \gamma_{\beta}^{\perp} \,\, \gamma_{\alpha}^{\perp} \,\,
n_{\rho}  \,\, P_R  \right ) \,\, \slashed {n} \,,
\nonumber \\
\Gamma_{{\rm B}, \, \alpha \beta}   &=&
\left ( \gamma_{\beta}^{\perp} \,\, \gamma_{\alpha}^{\perp} \,\, \slashed {n}
\, +  \,  \gamma_{\alpha}^{\perp} \,\, \gamma_{\beta}^{\perp}  \,\, \slashed {\bar n} \right ) \, P_L  \,,
\nonumber \\
\Gamma_{{\rm C}, \, \alpha \beta \rho} &=&
\bar u \,\, \Gamma_{{\rm A}, \, \alpha \beta \rho}
\, +  \, u \,\, \gamma_{\rho} \,\, \gamma_{\alpha}^{\perp}
\,\, \gamma_{\beta}^{\perp}  \,\, \slashed {n}  \,\,  P_R  \,.
\end{eqnarray}
Applying the definitions of the two-particle and three-particle
$B_q$-meson distribution amplitudes (\ref{definition: 2P B-meson LCDA})
and (\ref{definition: 3P B-meson LCDA})
and carrying out the essential integrations immediately gives rise to
\begin{eqnarray}
T_{7, \, \alpha \beta}^{A2, \, {\rm NLP}} &=&
\left [  i \, e_q \, \, f_{B_q} \,  m_{B_q}^2  \right ] \,
\left [g_{\alpha \beta}^{\perp}  - i \, \varepsilon_{\alpha \beta}^{\perp} \right ] \,
\bigg  \{  {1 \over 2} \left ( {\bar \Lambda \over \lambda_{B_q}} \right ) - 1
\nonumber \\
&& + \int_0^{\infty} \, d \omega_1  \, \int_0^{\infty} \, d \omega_2  \,\,
{1 \over \omega_1 \, (\omega_1 + \omega_2) }  \,\,
\Phi_3(\omega_1, \omega_2,  \mu)  \bigg \}  \,.
\end{eqnarray}
We explicitly verify that substituting the  covariant derivative
$\overrightarrow{\slashed{D}}$ by the transverse derivative
$\overrightarrow{\slashed D}_{\top}$  in the tree-level
expression (\ref{NLP JA2 amplitude}) indeed yields the identical
factorization formula.
It is  then straightforward to identify the resulting NLP corrections
to the helicity form factors
\begin{eqnarray}
F_{7, L} ^{A2, \, {\rm NLP}} &=&
{e_q \, \, f_{B_q}  \over   m_{B_q}} \,
\bigg  \{  \int_0^{\infty} \, d \omega_1  \, \int_0^{\infty} \, d \omega_2  \,\,
{1 \over \omega_1 \, (\omega_1 + \omega_2) }  \,\,
\Phi_3(\omega_1, \omega_2,  \mu)
\, + \, {1 \over 2} \left ( {\bar \Lambda \over \lambda_{B_q}} \right )
\, - \, 1 \bigg \} \,,
\nonumber  \\
F_{7, R} ^{A2, \, {\rm NLP}} &=&  0 \,.
\label{NLP factorization formula of F7A2}
\end{eqnarray}

We are now in a position to compute the subleading power contributions
to the QCD matrix element (\ref{QCD correlator: 7}) from the higher-twist
$B_q$-meson distribution amplitudes with the perturbative factorization technique.
As emphasized in \cite{Ball:1998sk,Ball:1998ff}, the consistent QCD calculations of the higher twist
corrections  will necessitate  taking into account the non-minimal partonic configurations
with additional quark-gluon fields  and the non-vanishing partonic transverse momenta
in the leading Fock state simultaneously.
Including the off-light-cone corrections for the non-local two-body HQET operators
up to the ${\cal O}(x^2)$ accuracy \cite{Braun:2017liq}
\begin{eqnarray}
&& \langle 0 | (\bar q_s Y_s)_{\beta}(x) \,\,\,  (Y_s^{\dag} h_v)_{\alpha}(0) |   \bar B_q \rangle
\nonumber \\
&& = -  {i \, \tilde{f}_{B_q}(\mu) \,  m_{B_q} \over 4} \,
\int_0^{\infty} d \omega \, e^{-i \omega v \cdot x} \, \bigg [  {1+ \slashed v \over 2}   \,
\bigg \{  2 \, \left [ \phi_B^{+}(\omega, \mu)  + x^2 \, g_B^{+}(\omega, \mu)  \right ]
\nonumber \\
&& \hspace{0.5 cm} \, -  {1 \over v \cdot x} \,
\left [ (\phi_B^{+}(\omega, \mu)-\phi_B^{-}(\omega, \mu))
+ x^2 \, ( g_B^{+}(\omega, \mu)-g_B^{-}(\omega, \mu))  \right ]\,
\slashed {x}  \bigg  \} \, \gamma_5  \bigg  ]_{\alpha \beta} \,,
\label{definition: 2P B-meson LCDA with x2 correction}
\end{eqnarray}
we can readily derive the factorized expression of the yielding
higher-twist contribution
\begin{eqnarray}
T_{7, \, \alpha \beta}^{\rm 2PHT, \, NLP}
= \left [ 4 \, i \, e_q \, \overline{m}_b(\nu) \, f_{B_q} \, m_{B_q} \right ] \,
\left [g_{\alpha \beta}^{\perp}  - i \, \varepsilon_{\alpha \beta}^{\perp} \right ] \,
\int_0^{\infty} \, d \omega \, {g_B^{+}(\omega, \mu) \over \omega^2}  \,.
\label{2PHT factorization formula: original}
\end{eqnarray}
Employing the non-trivial relation of the coordinate-space  distribution amplitudes
\begin{eqnarray}
\hat{g}_B^{+}(\tau, \mu) &=& {1 \over 4 \, \tau^2} \, \left [ \hat{\phi}_B^{+}(\tau, \mu)
- \hat{\phi}_B^{-}(\tau, \mu) \right ]
- {1 \over 2 \, \tau} \, \left ( {d \over d \tau} + i \, \bar{\Lambda} \right ) \,
\hat{\phi}_B^{+}(\tau, \mu)  \nonumber \\
&& - \, {1 \over 2} \, \int_0^1 d u \, \bar u \, \hat{\Psi}_4(\tau, u \, \tau, \mu)  \,,
\end{eqnarray}
we can express the twist-four LCDA $g_B^{+}(\omega, \mu)$ in terms of the ``genuine"
three-particle distribution amplitude of the same twist
and the Wandzura-Wilczek  contribution \cite{Wandzura:1977qf}
related to the lower twist two-particle distribution amplitudes
\begin{eqnarray}
g_B^{+}(\omega, \mu) = \bar{g}_B^{+}(\omega, \mu)
- {1 \over 2} \, \int_0^{\omega} \, d \omega_1 \, \int_0^1 d u \,
{\bar u \over u} \,\, \Psi_4 \left (\omega, {\omega - \omega_1 \over u}, \mu \right )\,.
\label{relation for gBplus}
\end{eqnarray}
The manifest expression of $\bar{g}_B^{+}(\omega, \mu)$
is in analogy to $\hat{g}_B^{-}(\omega, \mu)$ introduced in \cite{Lu:2018cfc}
\begin{eqnarray}
\bar{g}_B^{+}(\omega, \mu) = {1 \over 4} \,
\int_{\omega}^{\infty} \, d \rho \,
\bigg \{ (\rho - \omega)  \,
\left [\phi_B^{-}(\rho, \mu) - \phi_B^{+}(\rho, \mu)  \right ]
- 2 \, (\bar \Lambda - \rho) \, \phi_B^{+}(\rho, \mu) \bigg \}   \,.
\end{eqnarray}
Substituting the established relation (\ref{relation for gBplus})
into the tree-level factorization formula (\ref{2PHT factorization formula: original})
leads to the equivalent form in the following
\begin{eqnarray}
T_{7, \, \alpha \beta}^{\rm 2PHT, \, NLP}
&=& \left [ 4 \, i \, e_q \, \overline{m}_b(\nu) \, f_{B_q} \, m_{B_q} \right ] \,
\left [g_{\alpha \beta}^{\perp}  - i \, \varepsilon_{\alpha \beta}^{\perp} \right ] \,
\bigg [ \int_0^{\infty} \, d \omega \,\,  {\bar{g}_B^{+}(\omega, \mu) \over \omega^2}
\nonumber \\
&& \,\, - \,  {1 \over 2} \, \int_0^{\infty} \, d \omega_1 \,\,
\int_0^{\infty} \, d \omega_2 \, \int_0^1 d u \,
{\bar u \over (\omega_1 + u\, \omega_2)^2} \,\,
\Psi_4 \left (\omega_1, \omega_2, \mu \right ) \bigg ] \,.
\label{2PHT factorization formula}
\end{eqnarray}
On the other hand, the subleading twist three-particle corrections
to the vacuum-to-$B_q$-meson matrix element (\ref{QCD correlator: 7})
can be computed from the light-cone expansion of the quark propagator
in the background soft-gluon filed \cite{Balitsky:1987bk}
\begin{eqnarray}
\langle 0 | {\rm T} \, \{\bar q (x), q(0) \} | 0\rangle
\supset   i \, g_s \, \int_0^{\infty} \,\, {d^4 \ell \over (2 \pi)^4} \,
{e^{- i \, \ell \cdot x} \over \ell^2 + i 0} \,
\int_0^1 \, d u \, \left  [ u \, x_{\mu} \, \gamma_{\nu}
 - \frac{\slashed \ell \,\, \sigma_{\mu \nu}}{2 \, \left (\ell^2 + i 0 \right )}  \right ]
\, G^{\mu \nu}(u \, x) \,. \hspace{0.6 cm}
\end{eqnarray}
Applying the general parametrization of the light-cone HQET matrix element
(\ref{definition: 3P B-meson LCDA}) allows us to write down the factorized
expression of the higher-Fock-state contribution
\begin{eqnarray}
T_{7, \, \alpha \beta}^{\rm 3PHT, \, NLP}
&=&  \left [ - \, i \, e_q \, \overline{m}_b(\nu) \, f_{B_q} \, m_{B_q}  \right ] \,
\left [g_{\alpha \beta}^{\perp}  - i \, \varepsilon_{\alpha \beta}^{\perp} \right ] \,
\int_0^{\infty} \, d \omega_1 \,\,
\int_0^{\infty} \, d \omega_2 \, \int_0^1 d u \,
{1  \over (\omega_1 + u\, \omega_2)^2} \,\,
\nonumber \\
&& \,\, \times \, \left [ (2 u -1) \, \Psi_4 \left (\omega_1, \omega_2, \mu \right )
- \tilde{\Psi}_4 \left (\omega_1, \omega_2, \mu \right )   \right ] \,.
\label{3PHT factorization formula}
\end{eqnarray}
Putting together the achieved two-particle and three-particle higher twist contributions,
we can readily derive the factorization formula for the resulting NLP correction
\begin{eqnarray}
T_{7, \, \alpha \beta}^{\rm HT, \, NLP}
&=&  \left [ - \, i \, e_q \, \overline{m}_b(\nu) \, f_{B_q} \, m_{B_q} \right ] \,
\left [g_{\alpha \beta}^{\perp}  - i \, \varepsilon_{\alpha \beta}^{\perp} \right ] \,
\bigg \{  -4 \,   \int_0^{\infty} \, d \omega \,\,  {\bar{g}_B^{+}(\omega, \mu) \over \omega^2}
\nonumber \\
&& \,\, + \,  \int_0^{\infty} \, d \omega_1 \,\,
\int_0^{\infty} \, d \omega_2 \, \int_0^1 d u \,
{1  \over (\omega_1 + u\, \omega_2)^2}  \,
\left [  \Psi_4 \left (\omega_1, \omega_2, \mu \right )
- \tilde{\Psi}_4 \left (\omega_1, \omega_2, \mu \right )   \right ] \bigg \}   \,.
\hspace{0.8 cm}
\label{HT factorization formula: original}
\end{eqnarray}
According to the momentum-space relation for the HQET
distribution amplitudes \cite{Bell:2013tfa}
\begin{eqnarray}
\omega \, \phi_B^{-}(\omega, \mu)
+ \int_0^{\omega} d \eta \,
\left [ \phi_B^{+}(\eta, \mu) - \phi_B^{-}(\eta, \mu)  \right ]
= 2 \, \int_0^{\omega} d \eta \,  \int_{\omega - \eta}^{\infty} \,
{d \xi \over \xi} \,\, \left[  {\partial \over \partial \xi}  \Phi_3(\eta, \xi) \right ]   \,,
\end{eqnarray}
we are then able to rewrite the first convolution integral appearing in
(\ref{HT factorization formula: original}) as follows
\begin{eqnarray}
\int_0^{\infty} \, d \omega \,\,  {\bar{g}_B^{+}(\omega, \mu) \over \omega^2}
= {1 \over4}  \, \bigg \{ 2 \,  {\bar{\Lambda} \over \lambda_{B_q}} - 1
- \int_0^{\infty} d \omega \, \ln \omega  \, \Delta \phi_B^{-}(\omega, \mu) \bigg \} \,,
\label{EOM relation: gBplusbar}
\end{eqnarray}
with
\begin{eqnarray}
\Delta \phi_B^{-}(\omega, \mu)
&=&  - \omega {d \over d \omega} \, \phi_B^{-}(\omega, \mu)
- \phi_B^{+}(\omega, \mu)
\nonumber \\
&=& 2 \, \bigg[ - \int_0^{\infty} \, {d \omega_2 \over \omega_2^2} \, \Phi_3(\omega, \omega_2, \mu)
+ \int_0^{\omega} \, {d \omega_2 \over \omega_2^2} \, \Phi_3(\omega - \omega_2, \omega_2, \mu)
\nonumber \\
&& \hspace{0.5 cm} +  \int_0^{\omega} \, {d \omega_2 \over \omega_2} \,
{d \over d \omega} \Phi_3(\omega - \omega_2, \omega_2, \mu)  \bigg ] \,.
\end{eqnarray}
The newly defined distribution amplitude $\Delta \phi_B^{-}(\omega, \mu)$
can be further constructed from the ``genuine" twist-three
nonperturbative function  $\phi_{B}^{- \, \rm tw3}(\omega, \mu)$
\begin{eqnarray}
\Delta \phi_B^{-}(\omega, \mu) = - \omega {d \over d \omega} \,
\phi_{B}^{- \, \rm tw3}(\omega, \mu)
=  - \omega {d \over d \omega} \,
\left [ \phi_{B}^{-}(\omega, \mu)
- \phi_{B}^{- \, \rm WW}(\omega, \mu)  \right ] \,,
\end{eqnarray}
where the Wandzura-Wilczek term $\phi_{B}^{- \, \rm WW}(\omega, \mu)$
can be expressed in terms of the leading twist distribution amplitude
$\phi_{B}^{+}(\omega, \mu)$ \cite{Beneke:2000wa}
\begin{eqnarray}
\phi_{B}^{- \, \rm WW}(\omega, \mu) =
\int_{\omega}^{\infty} \, \, {d \rho  \over \rho}  \, \phi_{B}^{+}(\rho, \mu)
= \int_0^1 \, \, {d \eta  \over\eta} \, \phi_{B}^{+} \left ({\omega \over \eta}, \mu \right) \,.
\end{eqnarray}
The integral representation of the higher twist distribution amplitude $(\Psi_4 - \tilde{\Psi}_4)$
motivated by the corresponding renormalization group equation
suggests the following decomposition
\begin{eqnarray}
\left [ \Psi_4 - \tilde{\Psi}_4 \right ] (\omega_1, \omega_2, \mu)
= \left [ \Psi_4 - \tilde{\Psi}_4 \right ]^{\rm tw3}(\omega_1, \omega_2, \mu)
+  \left [ \Psi_4 - \tilde{\Psi}_4 \right ]^{\rm tw4}(\omega_1, \omega_2, \mu)  \,,
\end{eqnarray}
where the twist-three contribution can be deduced from $\Phi_3$ \cite{Beneke:2018wjp}
\begin{eqnarray}
\int_0^1 d u \, \left [ \hat{\Psi}_4 - \hat{\tilde{\Psi}}_4 \right ]^{\rm tw3}(z, u \, z, \mu)
={1 \over z^2} \,  \hat{\phi}_{B}^{- \, \rm tw3}(z, \mu)\,,
\end{eqnarray}
We can then derive an equivalent form of  the second convolution integral
in (\ref{HT factorization formula: original})
\begin{eqnarray}
&& \int_0^{\infty} \, d \omega_1 \,\,
\int_0^{\infty} \, d \omega_2 \, \int_0^1 d u \,
{1  \over (\omega_1 + u\, \omega_2)^2}  \,
\left [  \Psi_4 \left (\omega_1, \omega_2, \mu \right )
- \tilde{\Psi}_4 \left (\omega_1, \omega_2, \mu \right )   \right ]
\nonumber \\
&& = \int_0^{\infty} d \omega \, \ln \omega \, \Delta \phi_B^{-}(\omega, \mu)
+ \int_0^{\infty} \, d \omega_1 \,\,
\int_0^{\infty} \, d \omega_2 \,
{1  \over \omega_1 \, (\omega_1 + \omega_2)}  \,
\left [ \Psi_4 - \tilde{\Psi}_4 \right ]^{\rm tw4}(\omega_1, \omega_2, \mu)  \,,
\hspace{0.8 cm}
\end{eqnarray}
which together with the obtained relation (\ref{EOM relation: gBplusbar}) leads to
\begin{eqnarray}
T_{7, \, \alpha \beta}^{\rm HT, \, NLP}
&=&  \left [ - \, i \, e_q \, \overline{m}_b(\nu) \, f_{B_q} \, m_{B_q} \right ] \,
\left [g_{\alpha \beta}^{\perp}  - i \, \varepsilon_{\alpha \beta}^{\perp} \right ] \,
\bigg \{ 1 - 2 {\bar \Lambda \over \lambda_{B_q}}
+  2 \, \int_0^{\infty} d \omega \, \ln \omega \, \Delta \phi_B^{-}(\omega, \mu)
\nonumber \\
&& \,\, + \,  \int_0^{\infty} \, d \omega_1 \,\,
\int_0^{\infty} \, d \omega_2 \,
{1  \over \omega_1 \, (\omega_1 + \omega_2)}  \,
\left [ \Psi_4 - \tilde{\Psi}_4 \right ]^{\rm tw4}(\omega_1, \omega_2, \mu)   \bigg \}   \,.
\label{HT factorization formula: Version II}
\end{eqnarray}
Implementing the exact identity for the coordinate-space distribution amplitudes
\begin{eqnarray}
\int_0^{1} \, d  u \, \left [ \Psi_4 - \tilde{\Psi}_4 \right ]^{\rm tw4} (z, \, u z, \,  \mu)
= - {1 \over z} \, \left \{ \int_0^1 d u \,\, u \, \hat{\phi}_B^{+ \prime}(u z, \mu)
+  \hat{\phi}_B^{+ \prime}(z, \mu)
+ 2 \, i \, \bar{\Lambda} \,  \hat{\phi}_B^{+}(z, \mu)   \right \}
&& \nonumber \\
-   \int_0^1 d u \,\, \left \{  \left [ \Psi_4 + \tilde{\Psi}_4 \right ](z, \, u z, \,  \mu)
+ \left [ \Psi_4 + \tilde{\Psi}_4 \right ](u z, \,z, \,   \mu)   \right \},
&& \hspace{0.8 cm}
\end{eqnarray}
we are ready to establish the momentum-space relation accordingly
\begin{eqnarray}
&& \int_0^{\infty} \, d \omega_1 \,\, \int_0^{\infty} \, d \omega_2 \,
{1  \over \omega_1 \, (\omega_1 + \omega_2)}  \,
\left [ \Psi_4 - \tilde{\Psi}_4 \right ]^{\rm tw4}(\omega_1, \omega_2, \mu)
\nonumber \\
&& = 2 \, {\bar{\Lambda} \over \lambda_{B_q}} - 2
- \int_0^{\infty} \, d \omega_1 \,\, \int_0^{\infty} \, d \omega_2 \,
{1  \over \omega_1 \, \omega_2}  \,
\left [ \Psi_4 + \tilde{\Psi}_4 \right ] (\omega_1, \omega_2, \mu)  \,,
\end{eqnarray}
which implies another useful representation of the higher twist factorization formula
\begin{eqnarray}
T_{7, \, \alpha \beta}^{\rm HT, \, NLP}
&=&  \left [ - \, i \, e_q \, \overline{m}_b(\nu) \, f_{B_q} \, m_{B_q}  \right ] \,
\left [g_{\alpha \beta}^{\perp}  - i \, \varepsilon_{\alpha \beta}^{\perp} \right ] \,
\bigg \{ -1 +  2 \, \int_0^{\infty} d \omega \, \ln \omega \, \Delta \phi_B^{-}(\omega, \mu)
\nonumber \\
&& \,\, - \,  \int_0^{\infty} \, d \omega_1 \,\,
\int_0^{\infty} \, d \omega_2 \,
{1  \over \omega_1 \, \omega_2}  \,
\left [ \Psi_4 + \tilde{\Psi}_4 \right ] (\omega_1, \omega_2, \mu)   \bigg \}   \,.
\label{HT factorization formula: Version III}
\end{eqnarray}
Neglecting the nonleading  partonic configuration corrections
from the four-body light-cone HQET operators of the types
$\bar q \, G \, G \, h_v$ and $\bar q \, q \, \bar q \, h_v$
facilitates the construction of the approximate expression for
the twist-four three-particle distribution amplitudes
\begin{eqnarray}
\left (1 - \omega_2 \, {\partial \over \partial \omega_2} \right)  \,
\left [ \Psi_4 + \tilde{\Psi}_4 \right ] (\omega_1, \omega_2, \mu)
\simeq  -2 \,\omega_1 \, {\partial \over \partial \omega_1} \,
\Phi_4(\omega_1, \omega_2, \mu) \,.
\end{eqnarray}
We are then led to an even more compact factorization formula at tree level
\begin{eqnarray}
T_{7, \, \alpha \beta}^{\rm HT, \, NLP}
& \simeq &  \left [ - \, i \, e_q \, \overline{m}_b(\nu) \, f_{B_q} \, m_{B_q}  \right ] \,
\left [g_{\alpha \beta}^{\perp}  - i \, \varepsilon_{\alpha \beta}^{\perp} \right ] \,
\bigg \{ -1 +  2 \, \int_0^{\infty} d \omega \, \ln \omega \, \Delta \phi_B^{-}(\omega, \mu)
\nonumber \\
&& \,\, - \,  2 \, \int_0^{\infty} \, d \omega_2 \,
{1  \over \omega_2}  \,
\Phi_4 (\omega_1=0, \omega_2, \mu)   \bigg \}   \,,
\label{HT factorization formula: Version IV}
\end{eqnarray}
in analogy to the factorized expression for the subleading twist corrections
to the radiative leptonic $B \to \gamma \ell \bar \nu_{\ell}$ decays
as displayed in \cite{Beneke:2018wjp}.
 As a consequence, the resulting higher twist corrections to the helicity
 form factors can be written as
\begin{eqnarray}
F_{7, L} ^{{\rm HT}, \, {\rm NLP}} & \simeq &
\left [ - \, {e_q \, \overline{m}_b(\nu) \, f_{B_q} \,  \over  m_{B_q}^2} \right ] \,
\bigg \{ -1  +  \, 2 \, \int_0^{\infty} d \omega \, \ln \omega \,
\Delta \phi_B^{-}(\omega, \mu)
- 2 \, \int_0^{\infty} \, d \omega_2 \, {\Phi_4 (0, \omega_2, \mu)   \over \omega_2}  \,
\bigg \}  \,,
\nonumber  \\
F_{7, R} ^{{\rm HT}, \, {\rm NLP}} &=&  0 \,.
\label{NLP facorization formula of higher twist effects}
\end{eqnarray}

Furthermore, the subleading power contribution to the QCD correlation function
(\ref{QCD correlator: 7}) due to the electromagnetic current of the bottom quark
can be computed as follows
\begin{eqnarray}
T_{7, \, \alpha \beta}^{e_b, \, {\rm NLP}}
= \left [ - \, i \, e_q \, f_{B_q} \, m_{B_q}^2  \right ] \,
\left [g_{\alpha \beta}^{\perp}  - i \, \varepsilon_{\alpha \beta}^{\perp} \right ] \,,
\end{eqnarray}
which evidently generates the local NLP corrections to the two helicity form factors
\begin{eqnarray}
F_{7, L} ^{e_b, \, {\rm NLP}} & = &
- \, {e_q \,  f_{B_q} \,  \over  m_{B_q}} \,,
\nonumber  \\
F_{7, R} ^{e_b, \, {\rm NLP}} &=&  0 \,.
\end{eqnarray}
It is perhaps worth mentioning that the (anti-)collinear photon emission from the heavy quark
gives rise to the power suppressed correction preserving the large-recoil symmetry relation
of the transversality amplitudes, in contrast to the counterpart contribution to
the radiative decay $B \to \gamma \ell \bar \nu_{\ell}$ form factors  \cite{Beneke:2011nf}.

\subsection{The NLP corrections from the four-quark operators}
\label{subsection: weak annihilation effect}

Bearing in mind that we only aim at evaluating the subleading power contributions
to the double radiative $B_q$-meson decay form factors at ${\cal O}(\alpha_s^0)$ accuracy,
it is therefore unnecessary to take into account the NLP correction to the QCD correlation function
(\ref{QCD correlator: 1-6 and 8}) defined by the chromomagnetic dipole operator $P_8$.
By contrast, there exist  two different types of the power suppressed contributions
to the hadronic matrix elements of the four-quark operators already at tree level.
On the one hand, the factorizable quark-loop diagrams with an insertion of the QCD
penguin operators in Figure \ref{fig: Feynman diagrams}(b) yield the NLP contributions
analogous to the discussions presented in Section \ref{section£ºNLP O7 effects}
\footnote{Here we do not include the additional power suppressed contribution
due to the soft gluon radiation off the quark loop as explored in detail in
\cite{Khodjamirian:2010vf,Khodjamirian:2012rm} with the dispersion technique.},
which can be readily achieved by replacing the Wilson coefficient
$C_7(\mu)$ multiplying each individual subleading power correction
with the effective coefficient $C_7^{\rm eff}(\mu)$.
On the other hand, the QCD matrix elements of the four-quark operators further
generate the weak-annihilation-type of the subleading power corrections
as displayed in Figure \ref{fig: Feynman diagrams}(c), with no energetic photons
coupling to the partonic constituents of the initial $B_q$-meson state directly.
The resulting factorization formulae of the weak-annihilation contributions
to the two helicity form factors of $\bar B_q \to \gamma \gamma$ can be written as
\begin{eqnarray}
\sum_{i=1}^{6} \, C_i \, F_{i, \, L}^{(p), \,{\rm WA, \, NLP}} &=&
{f_{B_q} \,  \over  m_{B_q}} \,
\left [ {\cal F}^{(p), \,{\rm WA}}_{V} - {\cal F}^{(p), \,{\rm WA}}_{A} \right ] \,,
\nonumber  \\
\sum_{i=1}^{6}  \, C_i \, F_{i, R}^{(p), \,{\rm WA, \, NLP}} &=&
{f_{B_q} \,  \over  m_{B_q}} \,
\left [ {\cal F}^{(p), \,{\rm WA}}_{V} + {\cal F}^{(p), \,{\rm WA}}_{A} \right ]  \,.
\end{eqnarray}
The explicit expressions of the one-loop hard matching coefficient functions
with contributions from a complete set of the hadronic operators are given by
\begin{eqnarray}
{\cal F}^{(p), \,{\rm WA}}_{V} &=&
e_u^2 \, \left [C_F \, C_1 + C_2 \right ] \, H_{LL}(m_p)
 + \,  e_u^2 \, \left [ 12 \, N_c \, C_5 \right ] \,
 \left [ H_{LL}(m_c) + H_{LL}(0) \right ]
\nonumber \\
&& + \,  e_d^2 \, \left [ C_F \, (C_4 + 16 \, C_6)
+  C_3 + 8 \, C_5 \, ( 3\, N_c +2 )  \right ] \, H_{LL}(0) \,
\nonumber \\
&& + \,  e_d^2 \, \left [C_F \, (C_4 + 16 \, C_6)
+  C_3 + 4 \, C_5 (3 \, N_c + 4) \right ] \, H_{LL}(m_b)
\nonumber \\
&& + \, \left ( {m_b \over m_{B_q}} \right ) \, e_d^2 \,
\left [C_F \, (C_4 + 4 \, C_6) + (C_3 + 4 C_5) \right ]
\, H_{LR}^{(1)}(m_b) \,,
\nonumber \\
{\cal F}^{(p), \,{\rm WA}}_{A} &=&
\left ( {m_b \over m_{B_q}} \right ) \, e_d^2 \,
\left [C_F \, (C_4 + 4 \, C_6) + (C_3 + 4 C_5) \right ]
\, H_{LR}^{(2)}(m_b) \,,
\label{1-loop WA functions}
\end{eqnarray}
where we have introduced the following conventions for perturbative loop functions
\begin{eqnarray}
H_{LL}(m_q) &=& - \left [ y_q \,\,  C_0 \left ({1 \over y_q} \right) + {1 \over 2}  \right ] \,,
\qquad    \hspace{2.3 cm}
H_{LR}^{(1)}(m_q) = C_0 \left ({1 \over y_q} \right) \,,
\nonumber \\
H_{LR}^{(2)}(m_q) &=& - \, \left [ \left (4 \, y_q - 1 \right ) \, C_0 \left ({1 \over y_q} \right) + 2 \right ] \,,
\qquad \hspace{2.3 cm}
y_q = \left ( {m_q \over m_{B_q}} \right )^2   \,,
\end{eqnarray}
with
\begin{eqnarray}
C_0(x) = -2 \,\arctan^2 \left (\frac{1}{\sqrt{{4 \over x} -1}} \right ) \,.
\end{eqnarray}
It is apparent that the subleading power weak annihilation contributions from
the QCD penguin operators will spoil the large-recoil symmetry relation
between the two transversity amplitudes perturbatively.
In particular, the non-trivial strong phase of the perturbative kernel $H_{LL}(m_c)$
arises from the discontinuities in the variable $(p+q)^2$,
which can be  understood from the final-state rescattering mechanism
$\bar B_q \to H_c \, H_{\bar c}^{\prime} \to \gamma \, \gamma$ at hadronic level
with $H_c$ and $H_{\bar c}^{\prime}$ standing for the appropriate
(anti-)charm-hadron states \cite{Choudhury:1998rb,Liu:1999qz}
(see also \cite{Khodjamirian:2012rm} for further discussions
in the context of $B \to K \ell \ell$).
We also mention in passing that the weak annihilation diagrams involving
the light-quark loop will not generate the perturbative strong phase
at ${\cal O}(\alpha_s^0)$ in the limit $m_q \to 0$, due to the helicity
and angular momentum conservations.

\subsection{Final factorized expressions for the NLP corrections}

We now summarize the tree-level factorization formulae of the various subleading power
corrections to the helicity form factors considered so far
\begin{eqnarray}
\sum_{i=1}^{8} \, C_i \, F_{i, \, L}^{(p), \,{\rm fac, \, NLP}}
&=& C_7^{\rm eff} \, \left [   F_{7, \, L}^{\rm hc, \, NLP}
+  F_{7, \, L}^{m_q, \, {\rm NLP}}
+  F_{7, \, L}^{A2, \, {\rm NLP}}
+ F_{7, \, L}^{\rm HT, \, NLP}
+   F_{7, \, L}^{e_b, \, {\rm NLP}}  \right ]
\nonumber \\
&& \, + \, {f_{B_q} \,  \over  m_{B_q}} \,
\left [ {\cal F}^{(p), \,{\rm WA}}_{V} - {\cal F}^{(p), \,{\rm WA}}_{A} \right ]  \,,
\nonumber \\
\sum_{i=1}^{8} \, C_i \, F_{i, \, R}^{(p), \,{\rm fac, \, NLP}}
&=& {f_{B_q} \,  \over  m_{B_q}} \,
\left [ {\cal F}^{(p), \,{\rm WA}}_{V} + {\cal F}^{(p), \,{\rm WA}}_{A} \right ]   \,,
\label{final form of the factorized NLP effect}
\end{eqnarray}
which together with the NLL resummation improved factorization formulae
(\ref{NLL factorization formula of the helicity FFs})
at leading power in the heavy quark expansion
constitute the main technical results of this paper.

\section{The resolved photon corrections from the dispersion approach}
\label{section:soft NLP}

The major objective of this section is to compute the power suppressed
soft contributions to the double radiative $B_q$-meson decay amplitudes
with the OPE-controlled dispersion technique.
We will implement the nonperturbative prescription originally constructed for
accessing the long-distance photon correction to the pion-photon form factor
\cite{Khodjamirian:1997tk} by investigating the QCD correlation function
responsible for the more general transition  $B_q \to \gamma^{\ast} \gamma$
with an off-shell and transversely polarized photon
\begin{eqnarray}
\tilde{T}_{7, \alpha \beta} &=&
2 \, \overline{m}_b(\nu) \,
\int d^4 x \, e^{i q \cdot x} \, \langle 0 |
{\rm T} \left  \{ j^{\rm em}_{\beta}(x),
\bar q_L(0) \sigma_{\mu \alpha} \, p^{\mu}  b_R(0) \right \}| \bar B_q(p+q) \rangle \big |_{q^2 < 0}
\nonumber \\
&& +  \left [ p \leftrightarrow q, \alpha \leftrightarrow \beta \right ] \,
\nonumber \\
&=&  - {i \, e_q \, \overline{m}_b(\nu) \, m_{B_q}^2  \over 2} \,
\left \{   \left ( g_{\alpha \beta}^{\perp} - i \, \varepsilon_{\alpha \beta}^{\perp} \right )  \,
\tilde{F}_{7, L}(\bar n \cdot q, n \cdot q)
+  \left [ p \leftrightarrow q, \alpha \leftrightarrow \beta \right ]  \right \} \,.
\label{definition: generalized hadronic tensor}
\end{eqnarray}
It is then straightforward to derive the leading power factorization formula
for the generalized form factor $\tilde{F}_{7, L}(\bar n \cdot q, n \cdot q)$
in the $\Lambda_{\rm QCD}/m_b$ expansion for the hard-collinear $q^2$ regime
\begin{eqnarray}
\tilde{F}_{7, L}(\bar n \cdot q, n \cdot q) &=&  \hat{U}_1(m_b, \mu_{\rm h}, \mu) \,\hat{U}_2(m_b, \mu_{\rm h}, \mu) \,
f_{B_q} \, K^{-1}( m_b, \mu_{\rm h}) \, C_{T_1}^{(\rm A0)} \,
\nonumber \\
&& \, \times \, \int_0^{\infty} \, d \omega \, { \phi_B^{+}(\omega, \mu) \over \omega - n \cdot q - i 0} \,
\tilde{J}(\bar n \cdot q, n \cdot q, \omega, \mu)   \,,
\label{factorization formula for the generalized FF}
\end{eqnarray}
where the short-distance matching coefficient $\tilde{J}$ at the one-loop accuracy is given by \cite{Wang:2016qii}
\begin{eqnarray}
\tilde{J} &=& 1 + {\alpha_s(\mu) \, C_F \over 4 \pi} \,
\bigg \{  \left [ \ln^2 \left ( {\mu^2 \over \bar n \cdot q \,\, (\omega - n \cdot q)} \right )
- {\pi^2 \over 6} - 1 \right ]
- \, {n \cdot q \over \omega} \,  \ln \left (  {n \cdot q - \omega \over n \cdot q} \right ) \,
\nonumber \\
&& \hspace{3.0 cm} \, \times
\left [ \ln \left ( {\mu^2 \over -q^2} \right )
+ \ln^2 \left ( {\mu^2 \over \bar n \cdot q \,\, (\omega - n \cdot q)} \right )
+ 3\right ]  \bigg \} + {\cal O}(\alpha_s^2) \,.
\end{eqnarray}
Isolating the contributions of the lowest-lying hadronic states
and applying the standard definitions of the vector-meson decay constant
and the heavy-to-light $B_q$-meson decay form factors
\begin{eqnarray}
\langle 0 | \bar q \, \gamma_{\mu} \, q | V(k, \epsilon) \rangle
&=& i \, a_V^{(q)} \, f_V \, m_V \, \epsilon_{\mu}(k) \,,
\nonumber \\
\langle  V(k, \, \epsilon^{\ast})  | \bar q \, \sigma_{\mu \nu} \, q^{\nu} \, b | \bar B(k+q) \rangle
&=& 2 \, a_V^{(q)} \, T_1(q^2) \,  \varepsilon_{\mu \nu \rho \sigma}  \,
\epsilon^{\ast \nu}(k) \, k^{\rho} \, q^{\sigma} \,,
\nonumber \\
\langle  V(k, \, \epsilon^{\ast})  | \bar q \, i \, \sigma_{\mu \nu} \, \gamma_5 \, q^{\nu} \, b | \bar B(k+q) \rangle
&=&  a_V^{(q)} \, T_2(q^2) \, \left [ (m_B^2 - m_V^2) \, \epsilon_{\mu}^{\ast}(k)
- (\epsilon^{\ast} \cdot q) \, (2 \, k +q)_{\mu}  \right ]
\nonumber  \\
&& \, + \,  a_V^{(q)} \,  T_3(q^2) \, (\epsilon^{\ast} \cdot q) \,
\left [q_{\mu} - {q^2 \over m_B^2 - m_V^2} \, (2 \, k +q)_{\mu} \right ] \,,
\hspace{0.8 cm}
\end{eqnarray}
we can further derive the hadronic dispersion relation for the hadronic tensor
(\ref{definition: generalized hadronic tensor})
\begin{eqnarray}
\tilde{T}_{7, \alpha \beta}  &=&
- i \, \overline{m}_b(\nu) \, m_{B_q}^2  \,
\left ( g_{\alpha \beta}^{\perp} - i \, \varepsilon_{\alpha \beta}^{\perp} \right ) \,
\bigg \{ \sum_{V} \, \frac{c_V \, f_V \, m_V \, T_1^{B_q \to V}(0)}
{\bar n \cdot q \left (m_V^2/\bar n \cdot q - n \cdot q - i \,0 \right ) }
\nonumber \\
&& \,  + \int_{\omega_s}^{\infty} d \omega^{\prime} \,
{\rho^{\rm had}(\bar n \cdot q, \omega^{\prime}) \over \omega^{\prime} - n \cdot q - i 0} \bigg \}
+  \,   \left [ p \leftrightarrow q, \alpha \leftrightarrow \beta \right ] \,,
\label{dispersion relation for the generalized hadronic tensor}
\end{eqnarray}
with the aid of an exact QCD identity $T_1^{B_q \to V}(0)=T_2^{B_q \to V}(0)$.
The constant $c_V$ is determined by the flavour factor $a_V^{(f)}$
and the electric charge $e_f$ entering the QED quark-current (\ref{QED current})
\begin{eqnarray}
c_V =  a_V^{(q)} \, \sum_{f=u, \, d, \, s} \,  a_V^{(f)} \, e_f
=  \left\{
\begin{array}{l}
- {1 \over 3}  \,, \qquad \hspace{1.5 cm}
({\rm for} \,\, V=\phi) \vspace{0.5 cm} \\
- {1 \over 2}   \,,
 \qquad  \hspace{1.5 cm}
({\rm for} \,\, V=\rho)  \vspace{0.5 cm} \\
{1 \over 6}   \,,
\qquad  \hspace{1.8 cm}
({\rm for} \,\, V=\omega)
\end{array}
 \hspace{0.5 cm} \right.
\end{eqnarray}
where the non-vanishing coefficients  $a_V^{(f)}$ relevant to
the radiative  transitions $b \to (s, \, d) \, \gamma$ read
\begin{eqnarray}
a_{\phi}^{(s)} = 1 \,,
\qquad
a_{\rho}^{(u)} = - a_{\rho}^{(d)} = {1 \over \sqrt{2}} \,,
\qquad
a_{\omega}^{(u)} = a_{\omega}^{(d)} = {1 \over \sqrt{2}} \,.
\end{eqnarray}
Matching the two different representations of the correlation function
(\ref{definition: generalized hadronic tensor}) with the parton-hadron
duality ansatz for the physical spectral density and performing the Borel transformation with respect to
the variable $n \cdot q$ leads to the desired sum rules for the tensor $B_q \to V$ decay form factor in QCD
\begin{eqnarray}
 \sum_{V} \, {c_V \, f_V \, m_V \over \bar n \cdot q} \,
{\rm exp} \left [ - {m_V^2  \over \bar n \cdot q \, \omega_M}  \right ] \,  \, T_1^{B_q \to V}(0)
= {e_q \over 2} \, {1 \over \pi} \, \int_{0}^{\omega_s} \, d \omega^{\prime} \,
e^{- \omega^{\prime} / \omega_M} \,\, {\rm Im}_{\omega^{\prime}}  \,
\tilde{F}_{7, L}(\bar n \cdot q, \omega^{\prime}) \,.
\hspace{0.8 cm}
\label{LCSR of B to V FFs}
\end{eqnarray}
The factorized expression (\ref{factorization formula for the generalized FF})
allows us to construct the QCD spectral function immediately with the master formulae
collected in Appendix A of \cite{Wang:2016qii} (see also \cite{Beneke:2018wjp})
\begin{eqnarray}
{\rm Im}_{\omega^{\prime}}  \, \tilde{F}_{7, L}(\bar n \cdot q, \omega^{\prime})
= \hat{U}_1(m_b, \mu_{\rm h}, \mu) \,\hat{U}_2(m_b, \mu_{\rm h}, \mu) \,
f_{B_q} \, K^{-1}(m_b,  \mu_{\rm h}) \, C_{T_1}^{(\rm A0)} \,\,
\phi_{B, \, {\rm eff}}^{+}(\bar n \cdot q, \omega^{\prime}, \mu) \,,
\hspace{0.8 cm}
\end{eqnarray}
with the effective heavy-meson ``distribution amplitude" encoding both the soft and
hard-collinear strong interaction dynamics
\begin{eqnarray}
\phi_{B, \, {\rm eff}}^{+}(\bar n \cdot q, \omega^{\prime}, \mu)
&=& \phi_{B}^{+}(\omega^{\prime}, \mu)
+  {\alpha_s(\mu) \, C_F \over 4 \, \pi} \,
\bigg \{  \left [ \ln^2 {\mu^2 \over \bar n \cdot q \,\omega^{\prime} }
+ {\pi^2 \over 6} - 1  \right ] \,   \phi_{B}^{+}(\omega^{\prime}, \mu)
\nonumber \\
&& \, + \, \left [  2 \, \ln {\mu^2 \over \bar n \cdot q \,\omega^{\prime} } + 3 \right ] \,
\omega^{\prime} \, \int_{\omega^{\prime}}^{\infty} \, d \omega \,
\ln {\omega - \omega^{\prime} \over \omega^{\prime}} \,
{d \over d \omega} \, {\phi_{B}^{+}(\omega, \mu) \over \omega}
\nonumber \\
&& \, - \, 2 \,\ln {\mu^2 \over \bar n \cdot q \,\omega^{\prime} }  \,
\int_{0}^{\omega^{\prime}} \, d \omega \,
\ln {\omega^{\prime} - \omega \over \omega^{\prime}} \,
{d \over d \omega} \, \phi_{B}^{+}(\omega, \mu)
\nonumber \\
&& \, + \, \int_{0}^{\omega^{\prime}} \, d \omega \,
\ln^2 {\omega^{\prime} - \omega \over \omega^{\prime}} \,
{d \over d \omega} \, \left [\left ({\omega^{\prime} \over \omega} + 1 \right ) \,
\phi_{B}^{+}(\omega, \mu)  \right ]   \bigg \} \,.
\label{1-loop effective B-meson DA}
\end{eqnarray}
The resulting predictions from the SCET sum rules (\ref{LCSR of B to V FFs}) of
the  tensor transition form factors\footnote{The isospin-symmetry relations $m_{\rho}=m_{\omega}$,
$f_{\rho}=f_{\omega}$ and $T_1^{\bar B_d \to \rho}=T_1^{\bar B_d \to \omega}$
have been implemented in our light-cone sum rule (LCSR)
calculations of $b \to d$ transition form factors. }
$T_1^{\bar B_s \to \phi}$ and $T_1^{\bar B_d \to \rho}$
with the intervals of the Borel mass and the threshold parameter determined in \cite{Gao:2019lta}
are explicitly verified to be compliable with the numerical results obtained from
an alternative LCSR approach \cite{Straub:2015ica}.
Substituting the obtained sum rules  (\ref{LCSR of B to V FFs}) into the hadronic dispersion relation
(\ref{dispersion relation for the generalized hadronic tensor})
and taking the light-cone limit for the four-momentum  $q_{\mu} \to \bar n \cdot q \, n_{\mu} /2$
gives rise to the nonperturbatively refined spectral representation
\begin{eqnarray}
T_{7, \alpha \beta}  &=&
- { i \, e_q \, \overline{m}_b(\nu) \, m_{B_q}^2 \over 2}\,
\left ( g_{\alpha \beta}^{\perp} - i \, \varepsilon_{\alpha \beta}^{\perp} \right ) \,
\bigg \{ {1 \over \pi} \, \, \int_{0}^{\infty} \, { d \omega^{\prime} \over \omega^{\prime}} \,
{\rm Im}_{\omega^{\prime}}  \, \tilde{F}_{7, L}(\bar n \cdot q, \omega^{\prime})
\nonumber \\
&&  \, + \,  {1 \over \pi} \, \int_{0}^{\omega_s} \, d \omega^{\prime}  \,
\left [ {\bar n \cdot  q \over m_V^2}  \,
{\rm exp} \left ( {m_V^2 - \bar n \cdot  q  \, \omega^{\prime} \over \bar n \cdot  q \, \omega_M}  \right )
- {1 \over \omega^{\prime}} \right ]  \,
{\rm Im}_{\omega^{\prime}}  \, \tilde{F}_{7, L}(\bar n \cdot q, \omega^{\prime})  \bigg \}
\nonumber \\
&& +  \left [ p \leftrightarrow q, \alpha \leftrightarrow \beta \right ]  \,,
\end{eqnarray}
where the first and third lines precisely correspond to the SCET factorization formula
of the leading power contribution for the correlation function (\ref{QCD correlator: 7}).
Consequently, we can readily identify the resulting NLP corrections to the helicity
form factors of $B_q \to \gamma \gamma$ as follows
\begin{eqnarray}
F_{7, \, L}^{\rm soft, \, NLP} &=& - {e_q \, \overline{m}_b(\nu)  \over m_{B_q}} \,
\,  {1 \over \pi} \, \int_{0}^{\omega_s} \, d \omega^{\prime}  \,
\left [ {\bar n \cdot  q \over m_V^2}  \,
{\rm exp} \left ( {m_V^2 - \bar n \cdot  q  \, \omega^{\prime} \over \bar n \cdot  q \, \omega_M}  \right )
- {1 \over \omega^{\prime}} \right ]  \,
{\rm Im}_{\omega^{\prime}}  \, \tilde{F}_{7, L}(\bar n \cdot q, \omega^{\prime})  \,,
\nonumber \\
F_{7, \, R}^{\rm soft, \, NLP} &=& 0 \,.
\label{final formula of soft correction}
\end{eqnarray}
According to the power counting scheme for the effective threshold
and the Borel parameter
\begin{eqnarray}
\omega_s = {s_0 \over \bar n \cdot q} \sim
{\cal O} \left ( {\Lambda_{\rm QCD}^2 \over  m_{B_q}} \right )  \,,
\qquad
\omega_M = {M^2 \over \bar n \cdot q} \sim
{\cal O} \left ( {\Lambda_{\rm QCD}^2 \over  m_{B_q}} \right )
\end{eqnarray}
the  peculiar NLP mechanism responsible for (\ref{final formula of soft correction})
indeed arises from the long-distance correlation of the electromagnetic current $j_{\beta}^{\rm em}(x)$
and the dipole operator $P_7(0)$ with $x^2 \sim 1/ \Lambda_{\rm QCD}^2$,
which cannot be probed with the standard light-cone OPE technique without introducing the additional
nonperturbative distribution amplitudes.
Inspecting the factorization property of the subleading power correction in question
with the heavy-meson and (anti)-collinear-photon LCDAs simultaneously indicates that
the soft-collinear convolution integrals appearing in the factorized expression for the effective matrix element
of the yielding ${\rm A0}$-type $\rm {SCET}_{I}$ current develop rapidity divergences,
in analogy to the counterpart for the semileptonic $B_q \to V$ form factors with a transversely polarized
vector meson  \cite{Beneke:2000wa,Beneke:2003pa}.
Under this circumstance,  evaluating the NLP ``resolved" photon corrections in QCD
can be addressed independently  by employing the LCSR method with
the photon distribution amplitudes \cite{Ball:2002ps},
as already proposed in \cite{Wang:2018wfj} in the context of $B \to \gamma \, \ell \, \bar \nu_{\ell}$.
We are now ready to summarize the power suppressed soft contributions from
both the magnetic penguin operators and the four-quark operators
\begin{eqnarray}
\sum_{i=1}^{8} \, C_i \, F_{i, \, L}^{(p), \,{\rm soft, \, NLP}}
&=& \left ( {V_{7, \, \rm eff}^{(p)}  \over C_{T_1}^{(\rm A0)}} \right )
\, F_{7, \, L}^{\rm soft, \, NLP} \,,
\nonumber \\
\sum_{i=1}^{8} \, C_i \, F_{i, \, R}^{(p), \,{\rm soft, \, NLP}}
&=&  0  \,,
\end{eqnarray}
where the effective hard function $V_{7, \, \rm eff}^{(p)}$ at the accuracy
of ${\cal O}(\alpha_s)$ has been presented in (\ref{combined hard function at NLO}).

Collecting the different pieces together, the yielding $\bar B_q \to \gamma \gamma$
amplitude can be eventually expressed in terms of two independent helicity amplitudes
\begin{eqnarray}
\bar {\cal A}(\bar B_q \to \gamma \gamma)
= \left [ - i \, {4 \, G_F \over \sqrt{2}}  \right ]
\, {\alpha_{\rm em} \over 4 \pi}  \, m_{B_q}^3 \,
\epsilon_1^{\ast \alpha}(p)  \, \epsilon_2^{\ast \beta}(q) \,
\left [ ( g_{\alpha \beta}^{\perp} - i \, \varepsilon_{\alpha \beta}^{\perp} )  \,  \bar {\cal A}_{L}
- ( g_{\alpha \beta}^{\perp} + i \, \varepsilon_{\alpha \beta}^{\perp} )  \,  \bar {\cal A}_{R}  \right ]  \,,
\hspace{0.8 cm}
\label{final amplitude of anti-B-meson decay}
\end{eqnarray}
where the manifest expressions of $\bar {\cal A}_{L}$ and $\bar {\cal A}_{R}$ can be derived
in the following \footnote{For later convenience we further introduce the shorthand notations
for the separate amplitudes
\begin{eqnarray}
\bar {\cal A}_{L}^{\rm X} = \sum_{p=u,  c} \, V_{p b} \, V_{p q}^{\ast} \,
\sum_{i=1}^8 \, C_i \, F_{i, \,  L}^{(p), \, \rm X} \,,
\qquad
\bar {\cal A}_{R}^{\rm X} = \sum_{p=u,  c} \, V_{p b} \, V_{p q}^{\ast} \,
\sum_{i=1}^8 \, C_i \, F_{i, \, R}^{(p), \, \rm X}  \,.
\label{definition: helicity amplitudes}
\end{eqnarray}}
\begin{eqnarray}
\bar {\cal A}_{L} &=& \sum_{p=u,  c} \, V_{p b} \, V_{p q}^{\ast} \,
\sum_{i=1}^8 \, C_i \, \left [ F_{i, \,  L}^{(p), \, \rm LP}
+  F_{i, \,  L}^{(p), \,  \rm  {fac, \, NLP}}
+  F_{i, \,  L}^{(p), \,  \rm  {soft, \, NLP}}   \right ] \,,
\nonumber \\
\bar {\cal A}_{R} &=& \sum_{p=u,  c} \, V_{p b} \, V_{p q}^{\ast} \,
\sum_{i=1}^8 \, C_i \, \left [ F_{i, \,  R}^{(p), \, \rm LP}
+  F_{i, \,  R}^{(p), \,  \rm  {fac, \, NLP}}
+  F_{i, \,  R}^{(p), \,  \rm  {soft, \, NLP}}   \right ] \,.
\label{definition: helicity amplitudes}
\end{eqnarray}
In order to facilitate the phenomenological explorations of
the time-dependent decay observables  in the presence of
the neutral meson-antimeson mixing, we further introduce
the transversity amplitudes with definite CP transformation properties
\begin{eqnarray}
\bar {\cal A}_{\|} =  {1 \over \sqrt{2}} \,
\left ( \bar {\cal A}_{L} - \bar {\cal A}_{R} \right )  \,,
\qquad
\bar {\cal A}_{\perp} =  {1 \over \sqrt{2}} \,
\left ( \bar {\cal A}_{L} + \bar {\cal A}_{R} \right ) \,,
\end{eqnarray}
which allows for writing the exclusive radiative decay amplitude
(\ref{final amplitude of anti-B-meson decay}) in an alternative form
\begin{eqnarray}
\bar {\cal A}(\bar B_q \to \gamma \gamma)
= \left [ - i \, {4 \, G_F \over \sqrt{2}}  \right ]
\, {\alpha_{\rm em} \over 4 \pi}  \, m_{B_q}^3 \,
\epsilon_1^{\ast \alpha}(p)  \, \epsilon_2^{\ast \beta}(q) \,
\left \{ \sqrt{2} \, \left [ g_{\alpha \beta}^{\perp}  \,  \bar {\cal A}_{\|}
- i \,  \varepsilon_{\alpha \beta}^{\perp}   \,  \bar {\cal A}_{\perp}  \right ]  \right \} \,,
\label{final transversity amplitude of anti-B-meson decay}
\end{eqnarray}

\section{Numerical analysis}
\label{section:numerics}

Having at our disposal the improved  results of the helicity form factors
including both the NLL resummation of the leading power
contributions and the newly obtained NLP corrections,
we are now ready to investigate their numerical implications
on a number of observables for the exclusive radiative $B_q \to \gamma \gamma$
decay processes of experimental interest.
To this end, we will proceed by specifying the different types of theory inputs
(Wilson coefficients, quark masses, $B_q$-meson LCDAs, CKM parameters),
which will be essential to determine the two independent helicity amplitudes
$\bar {\cal A}_{L}$ and $\bar {\cal A}_{R}$ numerically.

\subsection{Theory inputs}
\label{section: theory inputs}

\begin{table}
\centering
\renewcommand{\arraystretch}{2.0}
\resizebox{\columnwidth}{!}{
\begin{tabular}{|l|ll||l|ll|}
\hline
\hline
  Parameter
& Value
& Ref.
&  Parameter
& Value
& Ref.
\\
\hline
\hline
  $G_F$                                         & $1.166379 \times 10^{-5} \,\, {\rm GeV}^{-2} $
                                                                                        & \cite{Tanabashi:2018oca} 
& $m_W$                                         & $80.379 \pm 0.012$ GeV                & \cite{Tanabashi:2018oca} 
\\
  $m_Z$                                         & $91.1876 \pm 0.0021$  GeV             & \cite{Tanabashi:2018oca} 
& $\overline{m}_t(\overline{m}_t)$              & $163.51 \pm 0.55$  GeV                & \cite{Tanabashi:2018oca} 
\\
  $\alpha_s^{(5)}(m_Z)$                         & $0.1188 \pm 0.0017$                   & \cite{Tanabashi:2018oca} 
& $\alpha_{\rm em}^{(5)}(m_Z)^{-1}$             & $127.952 \pm 0.009$                   & \cite{Tanabashi:2018oca} 
\\
  $\sin^2 \theta_W$                             & $0.23121 \pm 0.00004$                 & \cite{Tanabashi:2018oca} 
&              &                    &  
\\
\hline
\hline
  $\overline{m}_b(\overline{m}_b)$              & $4.198 \pm 0.012$  GeV                & \cite{Tanabashi:2018oca} 
& $m_{b}^{\rm PS}(2 \rm GeV)$                   & $4.532^{+0.013}_{-0.039}$  GeV        & \cite{Beneke:2014pta} 
\\
  $\overline{m}_c(3 \, \rm GeV)$                &  $0.988 \pm 0.007$  GeV               &  \cite{Aoki:2019cca}
& $m_{c}^{\rm PS}(1 \, \rm GeV)$                &  $1.39 \pm 0.05$  GeV                 &  \cite{Beneke:2020}
\\
  $\overline{m}_d(2 \, {\rm GeV})$              & $4.71 \pm 0.09$  MeV                  & \cite{Tanabashi:2018oca} 
& $\overline{m}_s(2 \, {\rm GeV})$              & $92.9 \pm 0.7$   MeV                  & \cite{Tanabashi:2018oca}
\\
\hline
\hline
   $m_{B_d}$                                    & $5279.64 \pm 0.13$ MeV                & \cite{Tanabashi:2018oca} 
&  $m_{B_s}$                                    & $5366.88 \pm 0.17$ MeV                & \cite{Tanabashi:2018oca} 
\\
  $f_{B_d}|_{N_f = 2+1+1}$                          & $190.0 \pm 1.3$ MeV                   & \cite{Aoki:2019cca} 
& $f_{B_s}|_{N_f = 2+1+1}$                      & $230.3 \pm 1.3$ MeV                   & \cite{Aoki:2019cca} 
\\
  $\tau_{B_d}$                                  & $(1.519 \pm 0.004)$ ps                & \cite{Tanabashi:2018oca} 
& $\tau_{B_s}$                                  & $(1.527 \pm 0.011)$ ps                & \cite{Tanabashi:2018oca} 
\\
  $x_{d}$                                       & $0.769 \pm 0.004$                     & \cite{Tanabashi:2018oca} 
& $x_{s}$                                       & $26.89 \pm 0.07$                      & \cite{Tanabashi:2018oca} 
\\
  $y_{d}$                                       & $0.0023^{+0.0005}_{-0.0008}$          & \cite{King:2019lal,Aoki:2019cca} 
& $y_{s}$                                       &  $0.0645\pm 0.003$                     & \cite{Tanabashi:2018oca} 
\\
\hline
\hline
  $\lambda_{B_d}(\mu_0)$                        & $(350 \pm 150)$ MeV                  & \cite{Beneke:2020}  
& $\lambda_{B_s}(\mu_0)$                        & $(400 \pm 150)$ MeV                  & \cite{Beneke:2020}
\\
  $\widehat{\sigma}_{B_d}^{(1)}(\mu_0)$                & $0.0 \pm 0.7$                 & \cite{Beneke:2020}
& $\widehat{\sigma}_{B_s}^{(1)}(\mu_0)$                & $0.0 \pm 0.7$                 & \cite{Beneke:2020}
\\
  $\widehat{\sigma}_{B_d}^{(2)}(\mu_0)$                & $0.0 \pm 6.0$                 & \cite{Beneke:2020}
& $\widehat{\sigma}_{B_s}^{(2)}(\mu_0)$                & $0.0 \pm 6.0$                 & \cite{Beneke:2020}
\\
  $2 \, \lambda_E^2(\mu_0) + \lambda_H^2(\mu_0)$       & $0.25 \pm 0.15$               & \cite{Beneke:2018wjp}
& $\lambda_E^2(\mu_0)/\lambda_H^2(\mu_0)$              & $0.50 \pm 0.10$               & \cite{Beneke:2018wjp}
\\
\hline
\hline
  $\lambda$                                     & $0.22650 \pm 0.00048$                 & \cite{Tanabashi:2018oca} 
& $A$                                           & $0.790^{+0.017}_{-0.012}$             & \cite{Tanabashi:2018oca} 
\\
  $\bar \rho$                                   & $0.141^{+0.016}_{-0.017}$             & \cite{Tanabashi:2018oca} 
& $\bar \eta$                                   & $0.357 \pm  0.011$                    & \cite{Tanabashi:2018oca} 
\\
\hline
\hline
  $m_{\rho}$                                           & $(775.26 \pm 0.25)$ MeV       & \cite{Tanabashi:2018oca}
& $m_{\phi}$                                           & $(1019.461 \pm 0.016)$ MeV    & \cite{Tanabashi:2018oca}
\\
  $\{M_{\rho}^2,  s_{\rho}^0 \}$                      & $\left \{ 1.5 \pm 0.5, \, 1.2 \pm 0.1 \right \} \, {\rm GeV}^2$
                                                                                        & \cite{Gao:2019lta}  
& $\{M_{\phi}^2,  s_{\phi}^0 \}$                      & $\left \{ 1.9 \pm 0.5, \, 1.6 \pm 0.1 \right \} \, {\rm GeV}^2$
                                                                                        & \cite{Gao:2019lta}  
\\
\hline
\hline
\end{tabular}
}
\renewcommand{\arraystretch}{1.0}
\caption{Summary of the numerical values for the theory input  parameters
implemented in the phenomenological analysis of the double radiative
decay observables. }
\label{table: input parameters}
\end{table}

For definiteness we will employ the three-loop running of
the strong coupling constant in the $\overline{\rm MS}$ scheme
by adopting the interval of $\alpha_s^{(5)}(m_Z)$ displayed in
Table \ref{table: input parameters} (equivalently,
$\Lambda_{\rm QCD}^{n_f=5} = 218^{+21}_{-19} \, {\rm MeV}$)
and taking  the renormalization-scale dependent quark flavour number
with the threshold values $\mu_{4}=4.8 \, {\rm GeV}$
and $\mu_{3}=1.2 \, {\rm GeV}$ for crossing $n_f=4$ and $n_f=3$, respectively.
The Wilson coefficients $C_i(\nu)$ in the weak effective Hamiltonian
(\ref{weak effective  Hamiltonian}) for the hard scale $\nu \simeq m_b$
at the NLL accuracy  will be evaluated from the  NLO matching expressions
at the initial scale  $\mu_W=m_W$ \cite{Chetyrkin:1996vx,Bobeth:1999mk}
with the renormalization-group formalism constructed in
\cite{Beneke:2001at,Beneke:2004dp,Bobeth:2003at,Huber:2005ig}.
The yielding results of $C_i(\nu)$ at the default scale $\nu=4.8 \, {\rm GeV}$
in the NLL approximation then read
\begin{eqnarray}
C_{1,...,8, \, \rm NLL}^{\rm eff} =
\left  \{-0.292, 1.007, -0.00448, -0.0809, 0.000330, 0.000852, -0.304, -0.167  \right \} \,,
\hspace{0.8 cm}
\end{eqnarray}
which are in excellent agreement with the numerical values displayed in \cite{Beneke:2004dp}
(see also \cite{Beneke:2020}) with slightly different choices of the SM input parameters.
The $\overline {\rm MS}$ mass of the bottom quark in the magnetic dipole operators
$P_{7, \, 8}$ collected in Table \ref{table: input parameters}
is determined from the lattice QCD simulations with $N_f=2+1+1$ flavours of sea quarks.
By contrast, the bottom-quark mass appearing in the perturbative hard and hard-collinear
functions from ${\rm QCD \to SCET_{I} \to SCET_{II}}$ matching is usually understood to be
the pole mass due to on-shell kinematics \cite{Beneke:2001at,Beneke:2004dp,Beneke:2004rc}.
However, converting the precisely known $\overline {\rm MS}$-scheme heavy quark mass to the counterpart
pole-scheme mass will lead to the numerical results sensitive to the loop order
of the matching relation in the corresponding perturbative expansion.
We will therefore use the potential-subtracted (PS) renormalization scheme \cite{Beneke:1998rk}
(see \cite{Hoang:2008yj} for an alternative short-distance quark-mass scheme) for both the bottom- and charm-quark masses
consistently in the short-distance functions displayed in (\ref{1-loop hard function}),
(\ref{combined hard function at NLO}),  (\ref{1-loop jet function}), (\ref{1-loop K function}),
(\ref{1-loop WA functions}) and (\ref{1-loop effective B-meson DA})
following the prescription of \cite{Beneke:2001at,Beneke:2020}.

We now turn to discuss the nonperturbative hadronic inputs entering the factorized expressions
of the $B_q \to \gamma \gamma$ helicity amplitudes (\ref{definition: helicity amplitudes}).
The leptonic decay constants of the pseudoscalar $B_d$- and $B_s$-mesons in QCD
are taken from the lattice averages of $N_f=2+1+1$ results in the isospin-symmetry limit \cite{Aoki:2019cca}
(see \cite{Bazavov:2017lyh} for further discussions on the strong-isospin breaking effect due to $m_u \neq m_d$
and \cite{Carrasco:2015xwa,Beneke:2019slt} for the systematic strategies
to take into account the  electromagnetic corrections).
In addition, the two-particle and three-particle $B_q$-meson distribution amplitudes in HQET
apparently serve as the fundamental ingredients of  the derived factorization formulae,
encoding the strong interaction dynamics from the soft-scale fluctuation,
for both the LO and NLO terms in the heavy quark expansion.
Following \cite{Beneke:2018wjp} we will introduce the general three-parameter ansatz for
the leading-twist LCDA  $\phi_B^{+}(\omega, \mu_0)$
\begin{eqnarray}
\phi_B^{+}(\omega, \mu_0) &=& \int_0^{\infty} d s \, \sqrt{w \, s} \,\,
J_{1}(2 \, \sqrt{w \, s}) \, \eta_{+}(s, \mu_0) \,,
\nonumber \\
\eta_{+}(s, \mu_0) &=&  {}_{1}F_{1}(\alpha; \beta; -s \, \omega_0) \,,
\label{model of B-meson DA}
\end{eqnarray}
which can be analytically evolved into the hard-collinear scale $\mu_{\rm hc}$
under the renormalization-group flows at the LL accuracy in terms of
the generalized hypergeometric ${}_{2}F_{2}(a, b; c, d; x)$ functions.
The shape parameter $\lambda_{B_q}$ defined in (\ref{definition£ºinverse moment})
and the associated inverse-logarithmic moments $\widehat{\sigma}_{B_q}^{(n)}$
(\ref{definition£ºinverse-logarithmic moment})
can be determined straightforwardly with the following identities
\begin{eqnarray}
\lambda_{B_q}(\mu_0) &=& \left ( {\alpha-1 \over \beta-1} \right ) \, \omega_0 \,,
\qquad
\widehat{\sigma}_{B_q}^{(1)} (\mu_0) =  \psi (\beta-1) - \psi (\alpha-1)
+ \ln  \left ( {\alpha-1 \over \beta-1} \right ) \,,
\nonumber \\
\widehat{\sigma}_{B_q}^{(2)} (\mu_0) &=&
\left [  \widehat{\sigma}_{B_q}^{(1)} (\mu_0)  \right ]^2 +
\psi^{(1)}(\alpha-1) - \psi^{(1)}(\beta-1)  + {\pi^2 \over 6} \,.
\end{eqnarray}
In spite of the intensive investigations of understanding the dynamical aspects of
the twist-two $B_q$-meson distribution amplitude (as well as the counterpart
transverse-momentum-dependent wavefunction) with distinct techniques and strategies
\cite{Braun:2003wx,Wang:2015vgv,Wang:2017jow,Li:2012nk,Li:2012md,Beneke:2011nf,Wang:2016qii,Wang:2018wfj,Beneke:2018wjp},
the current theory constraints on $\lambda_{B_q}(\mu_0)$ still turn out to be  far from satisfactory
due to the uncontrollable systematic uncertainties.
Consequently, we will vary the input value of $\lambda_{B_d}(\mu_0)$ in the conservative interval
as presented in Table \ref{table: input parameters} and further assign approximately ${\cal O} (15 \, \%)$
SU(3)-flavour symmetry breaking effect\footnote{This pattern can be intuitively understood in the weak
binding approximation (i.e., the static limit for both the bottom quark and the light anti-quark
in the $\bar B_q$-system)  from the typical values of the constituent down- and strange- quark masses
as discussed in \cite{Lin:1989vj,Lin:1990kw,Reina:1997my,Herrlich:1991bq,Chang:1997fs}
before the advent of the QCD factorization approach for exclusive heavy-hadron decays.
Furthermore, the existing QCD sum rule calculation of the ratio
$\lambda_{B_s}(\mu_0)/ \lambda_{B_d}(\mu_0)=1.19 \pm 0.14$ \cite{Khodjamirian:2020hob}
also supports the estimated flavour-symmetry violating effect in Table  \ref{table: input parameters}.}
for the ratio  $\lambda_{B_s}(\mu_0)/ \lambda_{B_d}(\mu_0)$ as recently advocated in \cite{Beneke:2020}.

Moreover, we will employ the phenomenological models for
the subleading-twist heavy-meson distribution amplitudes
satisfying the asymptotic behaviours at small quark and gluon momenta
and the classical equations of motion constraints \cite{Beneke:2018wjp}
(see \cite{Braun:2017liq,Lu:2018cfc} for the particular two-parameter models
of the following ansatz)
\begin{eqnarray}
\Phi_3(\omega_1, \omega_2, \mu_0) &=&
- {1 \over 2} \, \kappa(\mu_0) \, \left [ \lambda_E^2 (\mu_0) - \lambda_H^2 (\mu_0) \right ] \,
\omega_1 \, \omega_2^2 \, f^{\prime}(\omega_1 + \omega_2) \,,
\nonumber \\
\Phi_4(\omega_1, \omega_2, \mu_0) &=&
{1 \over 2} \, \kappa(\mu_0) \, \left [ \lambda_E^2 (\mu_0) + \lambda_H^2 (\mu_0) \right ] \,
\omega_2^2 \, f(\omega_1 + \omega_2) \,,
\nonumber \\
\Psi_4(\omega_1, \omega_2, \mu_0) &=&
\kappa(\mu_0) \,  \lambda_E^2 (\mu_0) \, \omega_1 \,  \omega_2 \, f(\omega_1 + \omega_2) \,,
\nonumber \\
\tilde{\Psi}_4(\omega_1, \omega_2, \mu_0) &=&
\kappa(\mu_0) \,  \lambda_H^2 (\mu_0) \, \omega_1 \,  \omega_2 \, f(\omega_1 + \omega_2) \,,
\nonumber \\
\phi_B^{- \, \rm WW}(\omega, \mu_0) &=&
\int_{\omega}^{\infty} d \rho \, f(\rho) \,,
\nonumber \\
\phi_B^{-\, \rm tw3}(\omega, \mu_0)
&=&  {1 \over 6} \, \kappa(\mu_0)  \,
\left [ \lambda_E^2 (\mu_0) - \lambda_H^2 (\mu_0) \right ] \,
\left [ \omega^2  \, f^{\prime}(\omega) + 4 \, \omega \, f(\omega)
- 2 \, \int_{\omega}^{\infty} d \rho \, f(\rho)  \right ] \,,
\hspace{1.0 cm}
\label{ansatz for higher-twist DAs}
\end{eqnarray}
where the normalization condition and the first three moments of
the function $f(\omega)$ read
\begin{eqnarray}
\int_{0}^{\infty} d \omega \,  f(\omega) &=& \lambda_{B_q}^{-1}(\mu_0)  \,,
\qquad
\int_{0}^{\infty} d \omega \, \omega\, f(\omega) = 1 \,,
\qquad
\int_{0}^{\infty} d \omega \, \omega^2 \, f(\omega) = {4 \over 3} \, \bar \Lambda \,,
\nonumber \\
\qquad
\kappa^{-1}(\mu_0) &=& {1 \over 2} \, \int_{0}^{\infty} d \omega \, \omega^3 \, f(\omega)
= \bar \Lambda^2 +  {1 \over 6} \, \left [ 2 \, \lambda_E^2 (\mu_0) + \lambda_H^2 (\mu_0) \right ] \,.
\end{eqnarray}
The HQET parameters $\lambda_E^2$ and $\lambda_H^2$ are defined by the hadronic matrix element
of the three-body effective local operator \cite{Grozin:1996pq,Braun:2017liq}
\begin{eqnarray}
&& \langle 0 | \bar q (0) \, g_s \, G_{\mu \nu} \, \Gamma \, h_v(0)| \bar B_q(v) \rangle
\nonumber \\
&& = - {\tilde{f}_{B_q} \, m_{B_q} \over 6} \,
{\rm Tr} \left \{ \gamma_5 \, \Gamma \,\,
\left (  {1 + \slashed v \over 2} \right ) \,
\left [ \lambda_H^2  \, \left (  i \, \sigma_{\mu \nu} \right )
+ (\lambda_H^2 - \lambda_E^2) \,
\left ( v_{\mu} \, \gamma_{\nu} - v_{\nu} \, \gamma_{\mu} \right ) \right ]  \right \}   \,.
\end{eqnarray}
Applying their renormalization-group equations at the one-loop order \cite{Grozin:1996hk,Nishikawa:2011qk}
\begin{eqnarray}
{d \over d \ln \mu} \,
\left(
\begin{array}{c}
\lambda_E^2(\mu) \\
\lambda_H^2(\mu) \\
\end{array}
\right)
+ {\alpha_s(\mu) \over 4 \, \pi}  \, \gamma_{\rm EH}^{(0)}  \,
\left(
\begin{array}{c}
\lambda_E^2(\mu) \\
\lambda_H^2(\mu) \\
\end{array}
\right) = 0
\,,
\label{1-loop RGE of lambdaE and lamdbaH}
\end{eqnarray}
we can readily write down the corresponding solutions in the LL approximation
\begin{eqnarray}
\left(
\begin{array}{c}
\lambda_E^2(\mu) \\
\lambda_H^2(\mu) \\
\end{array}
\right)
= \hat{V} \,
\left [  \left ( {\alpha_s(\mu) \over \alpha_s(\mu_0)} \right )^{\gamma_i^{(0)}/(2 \, \beta_0)} \right ]_{\rm diag}
\,  \hat{V}^{-1} \,\,\,
\left(
 \begin{array}{c}
\lambda_E^2(\mu_0) \\
\lambda_H^2(\mu_0) \\
\end{array}
\right)\,,
\label{1-loop solution to lambda-E and lambda-H}
\end{eqnarray}
where the matrix $\hat{V}$ diagonalize the evolution kernel  $\gamma_{\rm EH}^{(0)}$
with the yielding eigenvalues as irrational numbers remarkably
\begin{eqnarray}
\gamma_{\pm}^{(0)}= \left ( {8 \over 3} \, C_F + 2 \, N_c  \right ) \pm
{1 \over 6} \, \sqrt{64 \, C_F^2 - 144 \, N_C \, C_F + 90 \, N_C^2}
= {1 \over 9} \, \left ( 86  \pm \sqrt{{1565 \over 2}} \right )\,.
\end{eqnarray}
The available QCD sum rule predictions for these two dimensionful quantities
\begin{eqnarray}
\left \{  \lambda_E^2(\mu_0),  \,\, \lambda_H^2(\mu_0) \right \}
= \left\{
\begin{array}{l}
\left \{  (0.11 \pm 0.06) \, {\rm GeV^2}, \,\, (0.18 \pm 0.07) \, {\rm GeV^2}  \right \}  \,,
\qquad \hspace{1.5 cm} \text{\cite{Grozin:1996pq}}
\vspace{0.5 cm} \\
\left \{  (0.03 \pm 0.02) \, {\rm GeV^2}, \,\, (0.06 \pm 0.03) \, {\rm GeV^2}  \right \}    \,,
 \qquad  \hspace{1.5 cm}
\text{\cite{Nishikawa:2011qk}}
\end{array}
\hspace{0.8 cm} \right. \,
\label{QCDSR for lambdaE and lambdaH}
\end{eqnarray}
are observed to differ significantly  in magnitude due to the sizeable radiative correction
to the quark-gluon condensate contribution and the yet higher power correction from
the dimension-six operator $\langle \bar q q \rangle^2$ included in
the improved analysis \cite{Nishikawa:2011qk}.
Taking advantage of the default ansatz (\ref{ansatz for higher-twist DAs})
for the subleading twist $B_q$-meson distribution amplitudes allows for the enormous
simplification of the factorized expressions for the power suppressed contributions
at tree level derived in Section \ref{section£ºNLP O7 effects}
with the interesting relations
\begin{eqnarray}
&& \int_0^{\infty} \, d \omega_1  \, \int_0^{\infty} \,
{d \omega_2  \over \omega_2} \, \left [ {1 \over \omega_2}  \,\,
\ln {\omega_1 \over \omega_1 + \omega_2} + {1 \over \omega_1} \right ] \,
\Psi_4(\omega_1, \omega_2,  \mu)
= \left ( {12 - \pi^2 \over 6} \right ) \,\kappa(\mu) \, \lambda_E^2(\mu) \,,
\hspace{1.0 cm}
\label{NLP integral relation I}
\\
&& {\cal I}^{m_q}_{\rm NLP} = \lambda_{B_q}^{-1}(\mu) \, \bigg \{
\left [ 1 - {1 \over 3} \, \kappa(\mu) \,
\left [ \lambda_E^2(\mu) - \lambda_H^2(\mu) \right ] \right ] \,
\left [ X_{\rm NLP} - \widehat{\sigma}_{B_q}^{(1)}(\mu)
- \ln {\Lambda_{\rm UV} \over \lambda_{B_q}(\mu)} - \gamma_E \right ]
\label{NLP integral relation II}
\nonumber \\
&& \hspace{1.5 cm} + \,  {1 \over 2} \, \kappa(\mu) \,
\left [ \lambda_E^2(\mu) - \lambda_H^2(\mu) \right ]  \bigg  \} \,,
\\
&& \int_0^{\infty} \, d \omega_1  \, \int_0^{\infty} \, d \omega_2  \,
{1 \over \omega_1 (\omega_1 + \omega_2)}  \,
\Phi_3(\omega_1, \omega_2,  \mu)
= {1 \over 3} \, \kappa(\mu) \, \left [ \lambda_E^2(\mu) - \lambda_H^2(\mu) \right ]    \,,
\label{NLP integral relation III}
\\
&& \int_0^{\infty} \, d \omega \, \ln \omega \, \Delta \phi_{B}^{-}(\omega, \mu)
= {1 \over 6} \, \kappa(\mu) \, \left [ \lambda_E^2(\mu) - \lambda_H^2(\mu) \right ] \,,
\label{NLP integral relation IV}
\\
&& \int_0^{\infty} \, d \omega_2  \,{\Phi_4(0, \omega_2,  \mu) \over \omega_2}
= {1 \over 2} \, \kappa(\mu) \, \left [ \lambda_E^2(\mu) + \lambda_H^2(\mu) \right ]  \,,
\label{NLP integral relation V}
\end{eqnarray}
which are apparently independent of the precise shape of the profile function $f(\omega)$.
Keeping in mind that the  predicted ratio $\lambda_E^2(\mu_0)/\lambda_H^2(\mu_0)$
shown in  (\ref{QCDSR for lambdaE and lambdaH}) is insensitive to  the higher-order
corrections in the sum rule calculations and the normalization constant $\kappa(\mu_0)$
is actually determined by the particular combination $2 \, \lambda_E^2(\mu_0)+ \lambda_H^2(\mu_0)$,
it proves to be most advantageous to take the aforementioned two quantities as the independent
hadronic inputs parametrizing the chromoelectric and chromomagnetic matrix elements rather than
$\lambda_E^2(\mu_0)$ and $\lambda_H^2(\mu_0)$ themselves as already discussed in \cite{Beneke:2018wjp}.
The renormalization-scale dependence of the first logarithmic moment $\widehat{\sigma}_{B_q}^{(1)}(\mu)$
entering the expression of  ${\cal I}^{m_q}_{\rm NLP}$ can be readily determined from
the evolution equation of the HQET distribution amplitude $\phi_{B}^{+}(\omega, \mu)$ \cite{Lange:2003ff}
\begin{eqnarray}
\widehat{\sigma}_{B_q}^{(1)}(\mu) =
\widehat{\sigma}_{B_q}^{(1)}(\mu_0)
+ {\alpha_s(\mu_0) \, C_F \over \pi} \, \ln { \mu \over \mu_0} \,
\left ( \left [ \widehat{\sigma}_{B_q}^{(1)}(\mu_0)  \right ]^2 -  \widehat{\sigma}_{B_q}^{(2)}(\mu_0) \right )
+ {\cal O}(\alpha_s^2) \,,
\end{eqnarray}
which can be also derived from the one-loop evolution equation of the moment
$\widehat{\sigma}_{B_q}^{(1)}(\mu)$ presented in \cite{Bell:2013tfa,Wang:2015vgv}
alternatively.
Following \cite{Beneke:2019slt}, we will  set the ``effective mass" of the $B_q$-meson state
appearing in the subleading power factorization formulae
(\ref{NLP factorization formula of F7hc}) and (\ref{NLP factorization formula of F7A2})
as $\bar \Lambda=m_{B_q}-m_b$ with $m_b= (4.8 \pm 0.1) \, {\rm GeV}$ numerically
(see \cite{Lee:2005gza,Feldmann:2014ika} for further discussions
on the scheme dependence of this  HQET  quantity).

In addition, we will vary the hard scale $\mu_{\rm h}$ of the short-distance  matching coefficient
$V_{7, \rm eff}^{(p)}$ and the QCD renormalization scale $\nu$ in the interval $[m_b/2, 2 \, m_b]$
around the default value $m_b$.
In the same vein, the numerical value of the hard-collinear scale will be varied
in the range $\mu = (1.5 \pm 0.5) \, {\rm GeV}$ as widely employed in
the exclusive $B$-meson decay phenomenology \cite{Beneke:2005gs,Beneke:2011nf,Gao:2019lta,Beneke:2020}.
The CKM matrix elements in the effective weak Hamiltonian (\ref{weak effective  Hamiltonian})
are further evaluated from the four Wolfenstein parameters determined in \cite{Tanabashi:2018oca}
with the expanded matching relations up to the accuracy of ${\cal O}(\lambda^9)$ \cite{Charles:2004jd}.
Following \cite{Gao:2019lta} the adopted input values of the Borel parameter $M^2$ and the effective threshold $s_0$
appearing in the subleading power soft contributions to  the  helicity form factors
of $\bar B_q \to \gamma \gamma$ (\ref{final formula of soft correction}) are consistent with
the intervals implemented in the two-point QCD sum rules for the Gegenbauer moments  \cite{Ball:1998sk}
and the leptonic decay constant \cite{Ball:1996tb} of  the corresponding transversely polarized vector meson.

\subsection{Theory predictions for the helicity amplitudes}
\label{section: results of decay amplitudes}

\begin{figure}
\begin{center}
\includegraphics[width=1.0 \columnwidth]{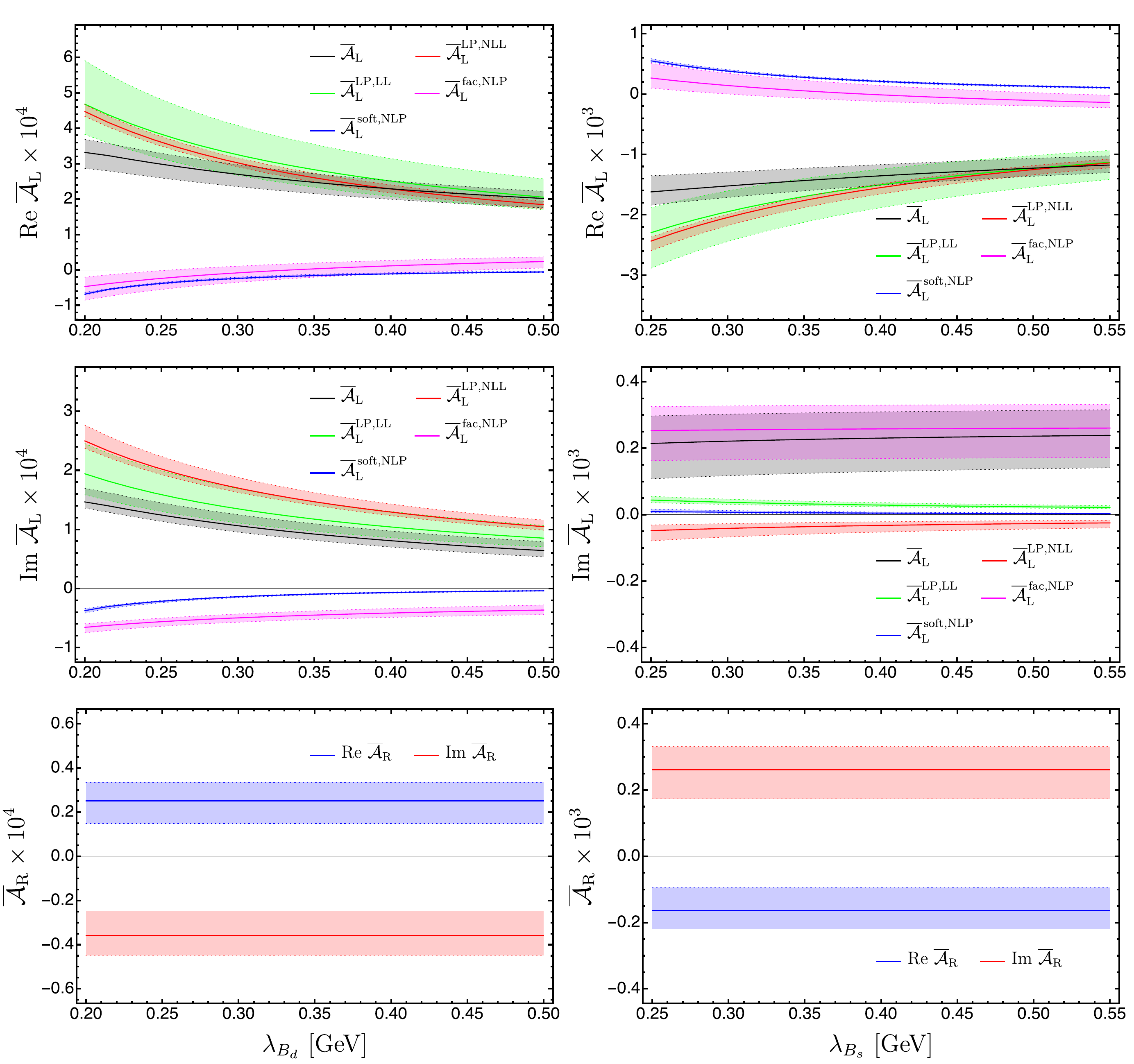}
\vspace*{0.1cm}
\caption{Breakdown of the distinct QCD mechanisms governing the two helicity amplitudes
for $\bar B_d \to \gamma \gamma$ [left panel] and for  $\bar B_s \to \gamma \gamma$ [right  panel]
with the theory uncertainties from the variations of the hard and hard-collinear scales
in the default intervals.}
\label{fig: Breakdown of the full helicity amplitudes}
\end{center}
\end{figure}

\begin{figure}
\begin{center}
\includegraphics[width=1.0 \columnwidth]{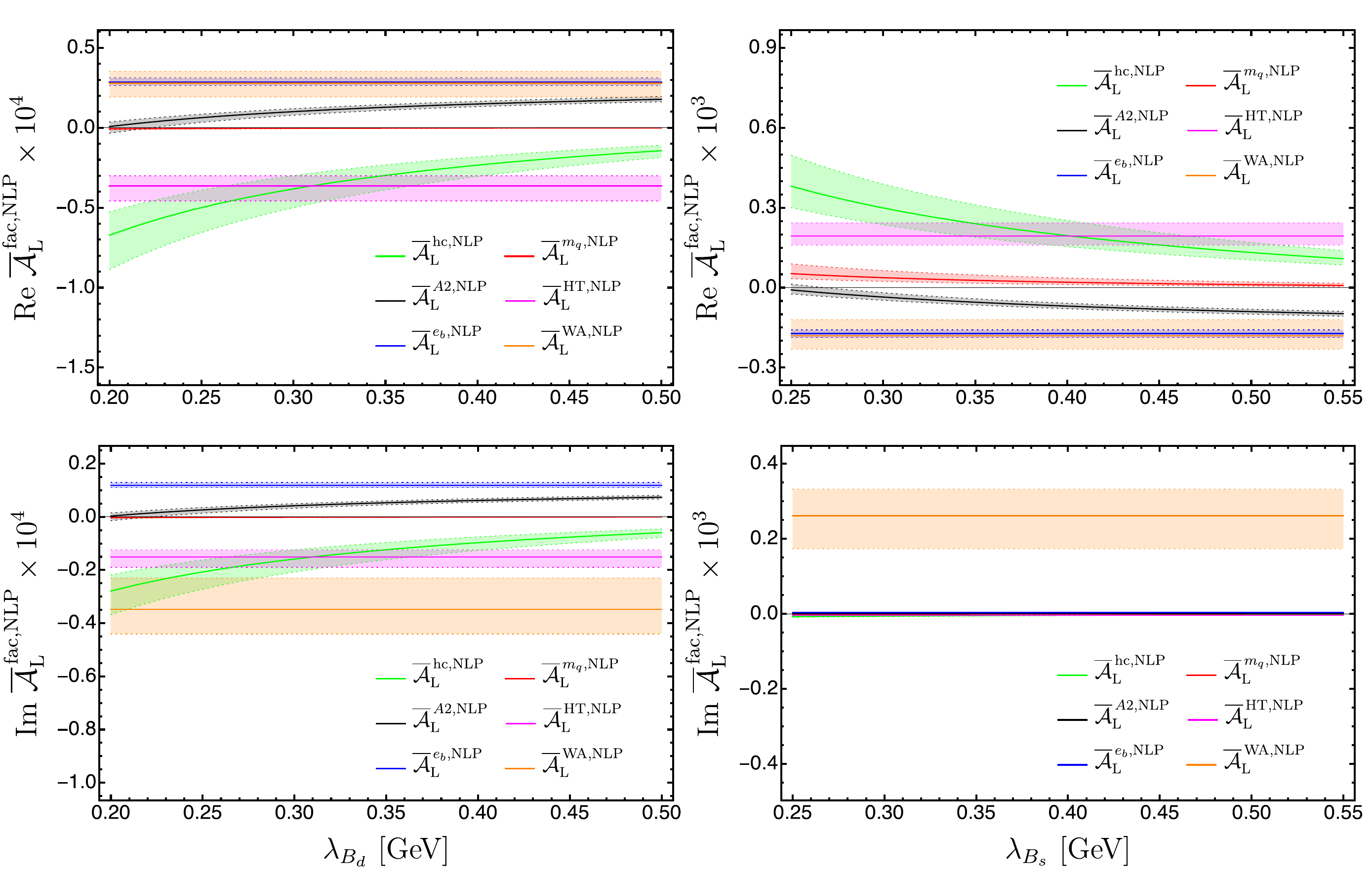}
\vspace*{0.1cm}
\caption{Theory predictions of the individual pieces yielding  the subleading power
corrections to the left-handed helicity amplitude in the heavy quark expansion
for $\bar B_d \to \gamma \gamma$ [left panel] and for  $\bar B_s \to \gamma \gamma$ [right  panel]
with the  uncertainties from the variations of the hard and hard-collinear scales
in the default intervals. }
\label{fig: Breakdown of the NLP helicity amplitudes}
\end{center}
\end{figure}

We now proceed to explore the phenomenological impacts of the newly derived
NLP corrections at ${\cal O}(\alpha_s^0)$ and the NLL resummation improved
leading power contributions on the exclusive radiative $B_q \to \gamma \gamma$
helicity amplitudes.  To develop a transparent understanding of the higher order
corrections from distinct dynamic mechanisms, we display in Figure
\ref{fig: Breakdown of the full helicity amplitudes} the $\lambda_{B_q}$ dependence
of the leading power contributions at the LL and NLL accuracy,
the combined local and non-local NLP corrections (\ref{final form of the factorized NLP effect})
from the QCD factorization approach and the subleading power soft contributions
estimated with the dispersion technique for both $b \to d \, \gamma$ and $b \to s \, \gamma$ transitions,
including the theory uncertainties from the variations
of the hard scales $\mu_{\rm h}$ and $\nu$ as well as the hard-collinear scale $\mu$.
Evidently, the radiative corrections with renormalization-group improvement can significantly
reduce the perturbative scale uncertainties for the LL predictions of the helicity amplitude
$\bar {\cal A}_{L}(\bar B_q \to \gamma \gamma)$  at leading power in the $\Lambda_{\rm QCD}/m_b$ expansion.
In general, the perturbative QCD correction to ${\rm Re} \, \bar {\cal A}_{L}(\bar B_q \to \gamma \gamma)$
(with $q=d,\, s$) at ${\cal O} (\alpha_s)$ can shift the corresponding LL prediction
by an amount of approximately $(2 \sim 10) \%$ for the allowed interval of $\lambda_{B_q}$
\footnote{It is interesting to remark that the NLL QCD corrections will bring about
the destructive (constructive) impacts on the helicity amplitude
${\rm Re} \, \bar {\cal A}_{L}(\bar B_d \to \gamma \gamma)$
(${\rm Re} \, \bar {\cal A}_{L}(\bar B_s \to \gamma \gamma)$)
due to the subtle interplay of the perturbative effects entering the
two  short-distance functions $C_{7}^{\rm eff} \, C_{T_1}^{(\rm A0)}$
and $\sum \limits_{i=1, \, 2, \, 8} \, C_{i}^{\rm eff} \, F_{i, 7}^{(p)}$,
which are responsible for the dominant contributions of the effective hard
vertex functions $V_{7, \rm eff}^{(p)}$ (with $p=u, \, c$)
in (\ref{combined hard function at NLO}).}.
Moreover, the yielding LL and NLL scale uncertainty bands only overlap marginally
for ${\rm Im} \, \bar {\cal A}_{L}(\bar B_d \to \gamma \gamma)$
and even turn out to be well separated for  ${\rm Im} \, \bar {\cal A}_{L}(\bar B_s \to \gamma \gamma)$,
while adopting the central values for all the remaining input parameters.
The latter can be attributed to the fact that the negligible LL contribution to
${\rm Im} \, \bar {\cal A}_{L}(\bar B_s \to \gamma \gamma)$
only arises from the heavily suppressed  CKM factors
${\rm Im} \left [ V_{ub} \, V_{us}^{\ast} \right ]$
and ${\rm Im} \left [ V_{cb} \, V_{cs}^{\ast} \right ]$
in comparison with the counterpart amplitude ${\rm Re} \, \bar {\cal A}_{L}(\bar B_s \to \gamma \gamma)$,
while the dominant NLL correction to  ${\rm Im} \, \bar {\cal A}_{L}(\bar B_s \to \gamma \gamma)$
originating  from the emerged strong phase of the two-loop contribution
of the current-current operator $P_{2}^{c}$ is apparently free of such CKM suppression.
In addition, the power suppressed soft corrections to both ${\rm Re} \, \bar {\cal A}_{L}(\bar B_q \to \gamma \gamma)$
and ${\rm Im} \bar {\cal A}_{L}(\bar B_q \to \gamma \gamma)$  will  gives rise to the destructive interferences
with the appropriate leading power contributions for the central input parameters, in analogy to the previous observation
for $\bar B \to \gamma  \ell  \bar \nu_{\ell}$ \cite{Wang:2016qii,Beneke:2018wjp}.
By contrast, the factorizable NLP correction to the amplitude ${\rm Re} \, \bar {\cal A}_{L}(\bar B_q \to \gamma \gamma)$
can flip the sign numerically with the evolution of the inverse moment $\lambda_{B_q}$£¬
due to the nontrivial competing mechanisms between the two distinct sectors of
$\{ \bar {\cal A}_{L}^{A2, \, {\rm NLP}}, \,\,  \bar{\cal A}_{L}^{e_b,  \, {\rm NLP}}, \,\, \bar{\cal A}_{L}^{\rm WA, \, NLP} \}$
and $\{ \bar {\cal A}_{L}^{\rm hc, \, NLP}, \,\,  \bar{\cal A}_{L}^{m_q,  \, {\rm NLP}}, \,\, \bar{\cal A}_{L}^{\rm HT, \, NLP} \}$
as displayed in Figure \ref{fig: Breakdown of the NLP helicity amplitudes}.
However, the imaginary part of the factorizable NLP amplitude
${\rm Im} \, \bar {\cal A}_{L}^{\rm  fac, \, NLP}(\bar B_d \to \gamma \gamma)$
will constantly lead to $(26 \sim 35) \%$  reduction of the leading power contribution,
because the overwhelming power corrections from
$\{ \bar {\cal A}_{L}^{\rm hc, \, NLP}, \,\,  \bar{\cal A}_{L}^{\rm HT, \, NLP}, \,\, \bar{\cal A}_{L}^{\rm WA, \, NLP} \}$
remain as the negative quantities in the entire region of $\lambda_{B_d} \in [200, 500] \, {\rm MeV}$.
Furthermore, the major contribution to the subleading power amplitude
${\rm Im} \, \bar {\cal A}_{L}^{\rm  fac, \, NLP}(\bar B_s \to \gamma \gamma)$ stems from
the weak-annihilation type of the charm-loop diagrams generated by the four-quark operators $P_{1, 2}^{c}$,
as a result of the strong suppression of the remaining factorized NLP corrections from
the weak phases of the CKM matrix elements entering the effective Hamiltonian of $b \to s \gamma$.

In addition, it has already become evident from the factorization formula
(\ref{final form of the factorized NLP effect})
that the right-handed helicity amplitude $\bar {\cal A}_{R}(\bar B_q \to \gamma \gamma)$
results from the local subleading power contribution
$\sum \limits_{i=1}^{6}  \, C_i \, F_{i, R}^{(p), \,{\rm WA}}$
completely, in contrast to the large-recoil symmetry breaking effects
for $\bar B_q \to \gamma \ell \bar \ell$ \cite{Beneke:2020}.
The sizeable uncertainties of the yielding  predictions for
${\rm Re} \, \bar {\cal A}_{R}(\bar B_q \to \gamma \gamma)$
and ${\rm Im} \, \bar {\cal A}_{R}(\bar B_q \to \gamma \gamma)$
reflect the uncancelled renormalization-scale dependence of
the Wilson coefficients $C_i(\nu)$ for the tree-level calculation
described in Section \ref{subsection: weak annihilation effect},
which is expected to be compensated by the unknown $\nu$-dependent hard functions
at ${\cal O}(\alpha_s)$ requiring the explicit evaluations of the two-loop
one-particle irreducible diagrams depicted in Figure 4 of \cite{Bosch:2002bv}.
It needs to be pointed out further that the obtained results
of ${\rm Re} \, \bar {\cal A}_{R}(\bar B_d \to \gamma \gamma)$
and ${\rm Im} \, \bar {\cal A}_{R}(\bar B_d \to \gamma \gamma)$
appear to own different signs from the counterpart predictions
for $\bar B_s \to \gamma \gamma$ due to the contributing CKM matrix elements
${\rm Re} \,\left [  V_{cb} \, V_{cd}^{\ast} \right ] < 0$ and
${\rm Re} \, \left [ V_{cb} \, V_{cs}^{\ast} \right ]  > 0$.

Inspecting the various NLP corrections to the factorizable $\bar B_q \to \gamma \gamma$
amplitudes displayed in Figure \ref{fig: Breakdown of the NLP helicity amplitudes} numerically
reveals the notable sensitivity of  the real parts for
$\bar{\cal A}_{L}^{\rm hc, \, NLP}, \,\,   \bar {\cal A}_{L}^{m_q, \, {\rm NLP}}$
and $\bar {\cal A}_{L}^{A2, \, {\rm NLP}}$  on the actual value of $\lambda_{B_q}$,
which together with the remaining NLP corrections independent of the inverse moment
characterizes the diverse facets of the strong interaction dynamics encoded in
the subleading power heavy-hadron decay amplitudes.
It is also worth mentioning that the non-local strange-quark mass correction to
${\rm Re} \, \bar {\cal A}_{L}(\bar B_s \to \gamma \gamma)$ is of minor numerical
importance when compared with the local subleading power effect from the (anti-)collinear
photon radiation off the bottom quark in virtue of the smallness of
the nonperturbative quantity $m_s / \lambda_{B_s}$ in reality.
Thanks to the established CKM hierarchy \cite{Tanabashi:2018oca}
\begin{eqnarray}
{\rm Im} \left [ V_{tb} V_{td}^{\ast} \right ] : {\rm Re} \left [ V_{tb} V_{td}^{\ast} \right ]
= 0.415 : 1 \,, \nonumber
\end{eqnarray}
adding the power suppressed helicity amplitudes ${\rm Im} \, \bar {\cal A}_{L}^{A2, \, \rm{NLP}}$
and ${\rm Im} \, \bar {\cal A}_{L}^{e_b, \, \rm{NLP}}$ for the radiative
$\bar B_d \to \gamma \gamma$ decay process together
will fall short of the magnitude   of the  counterpart weak-annihilation contribution
${\rm Im} \, \bar {\cal A}_{L}^{\rm WA, \,NLP}$ consistently.

\begin{figure}
\begin{center}
\includegraphics[width=1.0 \columnwidth]{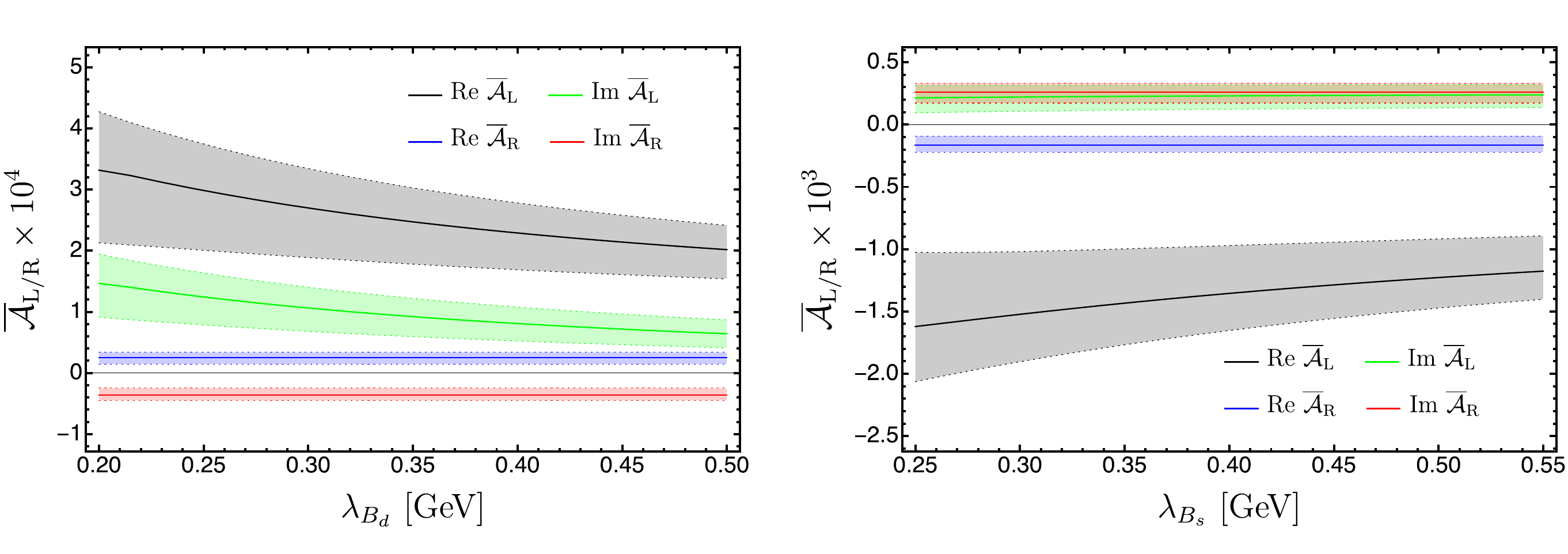}
\vspace*{0.1cm}
\caption{Theory predictions of the two helicity amplitudes for $\bar B_d \to \gamma \gamma$
[left panel] and for  $\bar B_s \to \gamma \gamma$ [right  panel]
with the combined  uncertainties by adding the separate errors
due to the variations of all the input parameters collected in
Section \ref{section: theory inputs} in quadrature. }
\label{fig: final helicity amplitudes with errors}
\end{center}
\end{figure}

We further present in Figure \ref{fig: final helicity amplitudes with errors}
the final theory predictions for the helicity amplitudes of $\bar B_q \to \gamma \gamma$,
as the analytical functions of the inverse moment $\lambda_{B_q}$,
with the combined uncertainties  by adding in quadrature
all the separate uncertainties from varying the input parameters
discussed in Section \ref{section: theory inputs}.
It is straightforward to identify the dominant theory uncertainties
for the left-handed helicity amplitudes arising from the hard and hard-collinear
scale fluctuations and the inverse-logarithmic moments $\widehat{\sigma}_{B_q}^{(1)}$
and $\widehat{\sigma}_{B_q}^{(2)}$, in addition to the particular sensitivity to $\lambda_{B_q}$
\footnote{Varying the HQET parameters $\lambda_E^2$
and $\lambda_H^2$ entering the factorized expressions of the NLP amplitudes
$\bar {\cal A}_{L}^{\rm hc, \,NLP}$, $\bar {\cal A}_{L}^{A2, \, \rm{NLP}}$
and $\bar {\cal A}_{L}^{\rm HT, \,NLP}$ will only bring about the subdominant uncertainties,
since the numerically important contributions stem from the constant factors
and the ``twist-two" terms proportional to $\bar \Lambda / \lambda_{B_q}$
in the curly brackets of (\ref{NLP factorization formula of F7hc}),
(\ref{NLP factorization formula of F7A2}) and
(\ref{NLP facorization formula of higher twist effects}). }.
In contrast, the yielding theory uncertainties for the right-handed helicity amplitudes
are dominated by the hard-scale variations of the Wilson coefficients $C_i(\nu)$
entering the tree-level factorization formulae of the weak-annihilation corrections.
Moreover, our limited knowledge of the nonperturbative coefficient $X_{\rm NLP}$
parameterizing the rapidity divergence of the NLP correction
(\ref{NLP formula of the light-quark mass effect})
will not generate the sizeable uncertainty due to the smallness
of the ratio $m_q/\lambda_{B_q}$ on top of the $\lambda_{B_q}/m_{B_q}$ suppression
in comparison with the counterpart charmless two-body
$B_q$-meson decays in QCD factorization \cite{Beneke:2003zv}.

\subsection{Phenomenological observables for $B_q \to \gamma \gamma$}

Prior to addressing the phenomenological implications of the improved
calculations of the double radiative decay amplitudes on the experimental
observables, we will construct the systematic expressions for investigating
the CP-averaged branching fractions, the polarization fractions
and  the time-dependent CP asymmetries
taking into account the neutral-meson mixing based upon the standard formalism
presented in \cite{Tanabashi:2018oca,Nierste:2009wg}.
In doing so, we are required  to derive the decay amplitude for the counterpart
channel $B_q \to \gamma \gamma$ by applying the CP-transformation for the  amplitude
(\ref{final transversity amplitude of anti-B-meson decay})
\begin{eqnarray}
{\cal A}(B_q \to \gamma \gamma)
= \left [ - i \, {4 \, G_F \over \sqrt{2}}  \right ]
\, {\alpha_{\rm em} \over 4 \pi}  \, m_{B_q}^3 \,
\epsilon_1^{\ast \alpha}(p)  \, \epsilon_2^{\ast \beta}(q) \,\,
\left \{ \sqrt{2} \, \left [ - \, g_{\alpha \beta}^{\perp} \,  {\cal A}_{\|}
- i \, \varepsilon_{\alpha \beta}^{\perp}   \,  {\cal A}_{\perp}  \right ] \right \} \,,
\hspace{0.8 cm}
\label{final amplitude of B-meson decay}
\end{eqnarray}
where we have adopted the phase convention
${\rm CP} \, | \bar B_q \rangle = -  | B_q \rangle$
and the two transversity amplitudes ${\cal A}_{\|}$
and ${\cal A}_{\perp}$ can be determined from
the corresponding ``barred" amplitudes with
all the weak phases conjugated.
Solving the Schr\"{o}dinger equation governing
the time evolution of the $B_q-\bar B_q$ system
immediately leads to the time-dependent decay rates
\cite{Dunietz:1986vi,Dunietz:2000cr}
\begin{eqnarray}
\Gamma^{\chi}  (B_q(t) \to \gamma \, \gamma) &=&
{\cal N}_{f} \, \left |{\cal A}^{\chi} (B_q \to \gamma \gamma) \right |^2 \,\,
{1 + |\lambda_{f}^{\chi}|^2 \over 2} \, {\rm exp} \left [ - \Gamma_{q} \, t \right ] \,
\bigg \{ \cosh { \Delta \Gamma_{q} \, t \over 2}
+  {\cal A}_{\Delta \Gamma}^{\chi} \, \sinh { \Gamma_{q} \, t \over 2}
\nonumber \\
&& \, + \, {\cal A}_{\rm CP}^{{\rm dir}, \, \chi} \, \cos (\Delta m_q \, t)
+ {\cal A}_{\rm CP}^{{\rm mix}, \, \chi} \, \sin (\Delta m_q \, t)
\bigg \}  \,,
\nonumber \\
\bar \Gamma^{\chi}  (\bar B_q(t) \to \gamma \, \gamma) &=&
{\cal N}_{f} \, \left |{\cal A}^{\chi} (B_q \to \gamma \gamma) \right |^2 \,\,
{1 + |\lambda_{f}^{\chi}|^2 \over 2} \,
{{\rm exp} \left [ - \Gamma_{q} \, t \right ] \over 1 - a_q} \, \bigg \{ \cosh { \Delta \Gamma_{q} \, t \over 2}
+  {\cal A}_{\Delta \Gamma}^{\chi} \, \sinh { \Gamma_{q} \, t \over 2}
\nonumber \\
&& \, - \, {\cal A}_{\rm CP}^{{\rm dir}, \, \chi} \, \cos (\Delta m_q \, t)
- {\cal A}_{\rm CP}^{{\rm mix}, \, \chi} \, \sin (\Delta m_q \, t)
\bigg \}  \,,
\label{time-dependent decay rate}
\end{eqnarray}
where ${\cal N}_{f}$ stands for the time-independent normalization factor
arising from the phase-space integration and the index $\chi$ is introduced to characterize
the parallel and perpendicular polarization configurations of the two-photon states.
Apparently, the two decay amplitudes ${\cal A}^{\|} (B_q \to \gamma \gamma)$
and ${\cal A}^{\perp} (B_q \to \gamma \gamma)$ for the final states possessing the definite
CP-parities can be obtained from (\ref{final amplitude of B-meson decay})
by  keeping the individual terms proportional to
${\cal A}_{\|}$ and ${\cal A}_{\perp}$, respectively.
The average width $\Gamma_q$ of the two bottom-meson mass eigenstates depends upon the diagonal matrix elements
of the decay matrix with $\Gamma_{11}^{q}=\Gamma_{22}^{q}$ thanks to the CPT invariance of the effective
Hamiltonian dictating the neutral-meson mixing pattern.
The mass and width splittings between the light and heavy mass eigenstates
can be expressed in terms of the dispersive and absorptive matrix elements
$M_{12}^{q}$ and $\Gamma_{12}^{q}$.
The $B_q$-mixing observables $\Delta m_q$ and $\Delta \Gamma_q$ further facilitate
the construction of the two important quantities
\begin{eqnarray}
x_q= {\Delta m_q \over \Gamma_q} \,, \qquad
y_q= {\Delta \Gamma_q \over 2 \, \Gamma_q} \,,
\end{eqnarray}
whose numerical results with uncertainties from the world average values of the experimental measurements
\cite{Tanabashi:2018oca} (with the combination procedures described
in \cite{Amhis:2019ckw})\footnote{As an exception, we prefer to adopt the SM prediction
for the decay rate difference $\Delta \Gamma_d$ \cite{King:2019lal}
with the nonperturbive  inputs from the lattice simulations \cite{Aoki:2019cca} to
derive the interval of $y_d$ satisfying the approximation relation
$y_q : x_q \approx y_s : x_s$ \cite{Tanabashi:2018oca,Anikeev:2001rk},
which holds to the first order in $|\Gamma_{12}^{q}/M_{11}^{q}|$
by stimulatingly switching off the SU(3)-flavour symmetry violating effects
of the ``bag" parameters and the subleading  power corrections to the
relevant $\Delta B=2$ matrix elements in the heavy quark expansion
(see \cite{Artuso:2015swg} for an overview of the current theory status).}
have been already summarized in Table \ref{table: input parameters}.
The small quantity $a_q$ characterizing the CP violation in mixing
can be defined by the nondiagonal matrix elements in the following form
\begin{eqnarray}
a_q = 1 - \left |{q \over p} \right |^2 \,, \qquad
{q \over p}  =  - \frac{\Delta m_q + i \, \Delta \Gamma_q /2}
{2 \,M_{12}^{q} - i \,\Gamma_{12}^{q} }
= - \frac{2 \,M_{12}^{q \, \ast} - i \,\Gamma_{12}^{q \, \ast}}
{\Delta m_q + i \, \Delta \Gamma_q /2} \,.
\label{a_q and q over p}
\end{eqnarray}
The SM predictions for such flavour specific asymmetries  \cite{Jubb:2016mvq,Artuso:2015swg}
\begin{eqnarray}
a_d = (-4.7 \pm 0.6) \times 10^{-4} \,, \qquad
a_s = (2.22 \pm 0.27) \times 10^{-5} \,,
\end{eqnarray}
imply that we can safely drop out the tiny corrections of ${\cal O}(a_q^n)$
(with $n\geq 1$) for the radiative $B_q \to \gamma \gamma$ decay phenomenology.
The phase-convention independent combination $\lambda_f$
and the appearing three CP violation observables ${\cal A}_{\rm CP}^{{\rm dir}, \, \chi}$,
${\cal A}_{\rm CP}^{{\rm mix}, \, \chi}$ and ${\cal A}_{\Delta \Gamma}^{\chi}$
are further defined as
\begin{eqnarray}
\lambda_f^{\chi} = {q \over p} \, \frac{\bar {\cal A}^{\chi} (\bar B_q \to \gamma \gamma)}
{{\cal A}^{\chi} (B_q \to \gamma \gamma)}\,,
& \qquad &
{\cal A}_{\rm CP}^{{\rm dir}, \, \chi} =
\frac{1 - |\lambda_f^{\chi}|^2} {1 + |\lambda_f^{\chi}|^2}  \,,
\nonumber \\
{\cal A}_{\rm CP}^{{\rm mix}, \, \chi} = -\frac{2  \, {\rm Im} \, \lambda_f^{\chi}}
{1 + |\lambda_f^{\chi}|^2} \,,
& \qquad &
 {\cal A}_{\Delta \Gamma}^{\chi} = -\frac{2  \, {\rm Re} \, \lambda_f^{\chi}}
{1 + |\lambda_f^{\chi}|^2}   \,,
\label{definition: three CP asymmetries}
\end{eqnarray}
which evidently reveals an exact relation
\begin{eqnarray}
|{\cal A}_{\rm CP}^{{\rm dir}, \, \chi}|^2 +
|{\cal A}_{\rm CP}^{{\rm mix}, \, \chi}|^2 +
|{\cal A}_{\Delta \Gamma}^{\chi}|^2 = 1 \,.
\end{eqnarray}
Expanding the obtained expression (\ref{a_q and q over p}) for
the phase-convention dependent quantity $q/p$
in terms of the small parameter $|\Gamma_{12}^{q} / M_{12}^{q}|$
leads to an approximate solution \cite{Anikeev:2001rk}
\begin{eqnarray}
{q \over p} = - {M_{12}^{q \, \ast} \over |M_{12}^{q}| }  \,
\left [ 1 - {a_q \over 2} \right ]
+ {\cal O} \left ( \left | {\Gamma_{12}^{q} \over  M_{12}^{q}} \right  |^2\right ) \,.
\end{eqnarray}
The SM contributions of the off-diagonal matrix element $M_{12}^{q}$
turn out to be dominated by the $\Delta B=2$ box diagrams
with internal top and anti-top quarks due to the very strong GIM cancellation \cite{Glashow:1970gm}
for the yielding two terms proportional to  $\lambda_c^2$ and $\lambda_c \, \lambda_t$
(with $\lambda_p= V_{p b} \, V_{p q}^{\ast}$),
which can be traced back to the peculiar analytical behaviours of
the associated perturbative loop function
$F \left ( m_{p_1}^2/m_W^2, m_{p_2}^2/m_W^2 \right )$ \cite{Inami:1980fz, Artuso:2015swg}.
We are then led to conclude that  the imaginary part of $M_{12}^{q}$ originates to a rather good
approximation from the CKM factor $V_{t b} \, V_{t q}^{\ast}$ uniquely with the standard OPE technique,
which further enables us to write down an even illuminating relation for the $B_q$-meson mixing
\begin{eqnarray}
{q \over p} \simeq  -  \frac{ V_{t b}^{\ast} \, V_{t q}} { V_{t b} \, V_{t q}^{\ast}}
= - {\rm exp} \left [ i \, {\rm arg} (V_{t b}^{\ast} \, V_{t q})^2  \right ] \,.
\end{eqnarray}

According to (\ref{time-dependent decay rate}),
we can immediately derive the CP-averaged  branching fraction of $B_q \to \gamma  \gamma$
for the flavour-tagged measurement at an $e^{+} \, e^{-}$ collider
with equal numbers of the produced bottom and anti-bottom mesons \cite{Harrison:1998yr,DescotesGenon:2011pb}
\begin{eqnarray}
{\cal BR}^{\chi} (B_q \to \gamma  \gamma) &=&
{\cal N}_f  \,  \left |{\cal A}^{\chi} (B_q \to \gamma \gamma) \right |^2 \,
{1 + |\lambda_{f}^{\chi}|^2 \over 2 } \,
\int_{-\infty}^{+\infty} d t \,  \,
{\rm exp} \left [ - \Gamma_{q} \, |t| \right ] \,
\nonumber \\
&& \, \times \, \left \{  \cosh { \Delta \Gamma_{q} \, t \over 2}
+  {\cal A}_{\Delta \Gamma}^{\chi} \, \sinh { \Gamma_{q} \, t \over 2}  \right \}
+ {\cal O} (a_q) \nonumber \\
&=&  \left ( {{\cal N}_f  \over \Gamma_q} \right ) \,
\left |{\cal A}^{\chi} (B_q \to \gamma \gamma) \right |^2 \,
 \left ( 1 + |\lambda_{f}^{\chi}|^2  \right )  \,
\left ( {1 \over 1 - y_q^2} \right ) + {\cal O} (a_q)   \,
\nonumber \\
&=&  \langle {\cal BR}^{\chi}(B_q \to \gamma  \gamma) \rangle_{\rm th} \,
\left ( {1 \over 1 - y_q^2} \right )  + {\cal O} (a_q)    \,.
\end{eqnarray}
Interestingly, the resulting correction factor of the time-integrated branching ratio
due to the neutral-meson mixing is independent of the polarization  index  $\chi$
for the coherent bottom-meson pair production at the SuperKEKB accelerator.
The CP-averaged branching fraction in the limit $\Delta \Gamma_q \to 0$
can be readily computed from the corresponding transversity amplitudes
\begin{eqnarray}
\langle {\cal BR}^{\chi}(B_q \to \gamma  \gamma) \rangle_{\rm th}
= {\tau_{B_q} \over 2} \, \frac{\alpha_{\rm em}^2 \, G_F^2 \, m_{B_q}^5} {16 \, \pi^3} \,
\left ( |{\cal A}_{\chi}|^2 +  |\bar {\cal A}_{\chi}|^2 \right )  \,.
\end{eqnarray}
Consequently, we can proceed to construct a typical set of phenomenological observables
in analogy to the two-body hadronic $B \to V V$ decays \cite{Beneke:2006hg}
\begin{eqnarray}
{\cal BR} (B_q \to \gamma  \gamma) =
\sum_{\chi=\|, \, \perp} \, {\cal BR}^{\chi} (B_q \to \gamma  \gamma) \,,
\qquad
f_{\chi} =  \frac{{\cal BR}^{\chi} (B_q \to \gamma  \gamma)}
{{\cal BR} (B_q \to \gamma  \gamma) }  \,,
\end{eqnarray}
where the two polarization fractions are apparently not linearly independent
due to the normalization condition $f_{\|} \, + \,  f_{\perp}  = 1$.
Applying the definition for the time-dependent CP asymmetry
for the neutral $B_q$-meson decaying into CP eigenstates yields \cite{Nierste:2009wg}
\begin{eqnarray}
A_{\rm CP}^{\chi} (t) &=& \frac{\bar \Gamma^{\chi}  (\bar B_q(t) \to \gamma \gamma)
- \Gamma^{\chi}  (B_q(t) \to \gamma \gamma)}
{\bar \Gamma^{\chi}  (\bar B_q(t) \to \gamma \gamma)
+ \Gamma^{\chi}  (B_q(t) \to\gamma \gamma)}
\nonumber \\
&=& - \frac{{\cal A}_{\rm CP}^{{\rm dir}, \, \chi} \, \cos(\Delta m_q \, t)
+ {\cal A}_{\rm CP}^{{\rm mix}, \, \chi} \, \sin(\Delta m_q \, t)}
{\cosh (\Delta \Gamma_q \, t /2) + {\cal A}_{\Delta \Gamma}^{\chi} \, \sinh (\Delta \Gamma_q \, t /2)}  \,.
\end{eqnarray}
The quantity ${\cal A}_{\rm CP}^{{\rm dir}, \, \chi}$ describes the direct CP-violating effect
as a consequence of
$\left |\bar {\cal A}^{\chi} : {\cal A}^{\chi} \right| \neq 1 $,
which requires the appearance of at least two individual terms
with distinct strong and weak phases simultaneously  contributing to
the  transversity  amplitude ${\cal A}^{\chi}$.
By contrast, ${\cal A}_{\rm CP}^{{\rm mix}, \, \chi}$ encodes
the CP-violating  information due to interference between  decays
with and without mixing for the final states common to
both the bottom and anti-bottom meson decays (i.e., by virtue of
the two different decay mechanisms of $B_q \to \gamma \gamma$
and $B_q \to \bar B_q \to \gamma \gamma$).
The observable  ${\cal A}_{\Delta \Gamma}^{\chi}$
\footnote{As pointed out in \cite{DeBruyn:2012wj}, the mass-eigenstate rate asymmetry
${\cal A}_{\Delta \Gamma}^{\chi}$ is also essential to determine the effective
lifetime for the untagged  $B_q \to f$ decay.} takes the extreme values $\pm 1$ for
the so-called ``golden modes" with CP-eigenstate final states,
whose decay amplitudes contain only  one CKM structure such that $|\lambda_f^{\chi}| = 1$.

\begin{figure}
\begin{center}
\includegraphics[width=1.0 \columnwidth]{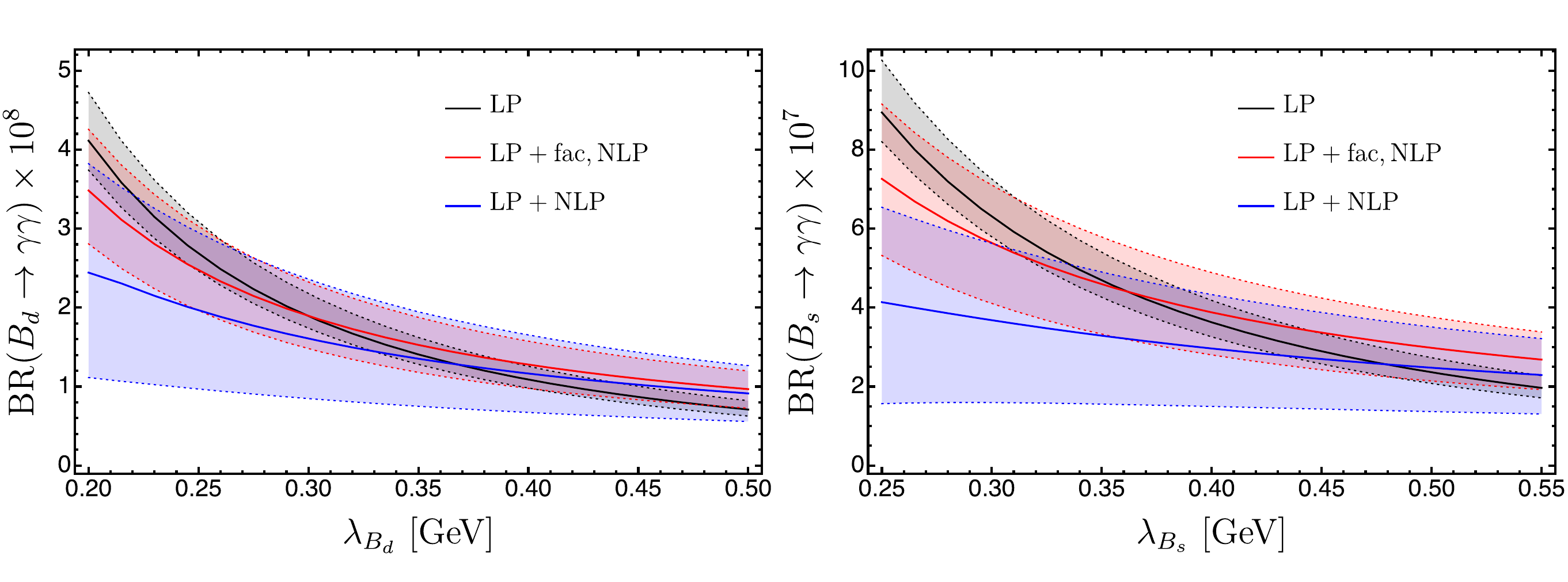}
\vspace*{0.1cm}
\caption{Theory predictions for the $\lambda_{B_q}$-dependence of the
CP-averaged branching fractions  of  $\bar B_d \to \gamma \gamma$ [left panel]
and $\bar B_s \to \gamma \gamma$ [right  panel],
where the numerical results obtained with the NLL resummation
improved leading power approximations [grey bands]
and with adding further solely  the  factorizable NLP corrections
[red bands] are also shown for a comparison.}
\label{fig: BR results with errors}
\end{center}
\end{figure}

We now present the improved theory predictions for the CP-averaged branching fractions
of $B_q \to \gamma \gamma$ in the presence of the neutral-meson mixing
in Figure \ref{fig: BR results with errors}, which summarizes our main phenomenological results
for confronting the forthcoming precision Belle II measurements
with an ultimate integrated luminosity of $50 \, {\rm ab}^{-1}$ \cite{Kou:2018nap}.
The factorizable NLP correction to ${\cal BR} (B_q \to \gamma  \gamma)$ appears to reduce the counterpart
leading power contribution by approximately an amount of $(15-20) \, \%$
at $\lambda_{B_d}= 200 \, {\rm MeV}$ ($\lambda_{B_s}= 250 \, {\rm MeV}$).
However, such destructive interference mechanism will gradually evolve into
the constructive intervention pattern with the increase of $\lambda_{B_q}$
(as large as $35\, \%$ enhancement at the maximal value of the inverse moment)
due to the emergence of the zero-crossing regime for the factorizable NLP correction
to the amplitude ${\rm Re} \, \bar {\cal A}_L(\bar B_q \to \gamma \gamma)$
as discussed in Section \ref{section: results of decay amplitudes} together with
the hierarchy structure
\begin{eqnarray}
&& |{\rm Re} \, \bar {\cal A}_L(\bar B_q \to \gamma \gamma)|:
|{\rm Im} \, \bar {\cal A}_L(\bar B_q \to \gamma \gamma)|
\in [2.3, 3.3] \,,
\nonumber \\
&& |{\rm Re} \, \bar {\cal A}_L(\bar B_s \to \gamma \gamma)|:
|{\rm Im} \, \bar {\cal A}_L(\bar B_s \to \gamma \gamma)|
\in [5.5, 7.6] \,,
\end{eqnarray}
where the quoted uncertainties arise from the variations of $\lambda_{B_q}$ merely.
The NLP resolved photon contribution to the branching  fraction
of  $B_q \to \gamma \gamma$ will result in the sizeable reduction of
the corresponding ``${\rm LP + fac, \, NLP}$" prediction
with the default ansatz of the $B_q$-meson distribution amplitude
(\ref{model of B-meson DA}) at $\alpha = \beta$
(namely, the exponential model \cite{Grozin:1996pq,Braun:2017liq,Lu:2018cfc}):
numerically $(6-28) \, \%$ and $(15-43) \, \%$ corrections for
$b \to d \, \gamma$ and $b \to s \, \gamma$ transitions, respectively.
Inspecting the distinct sources of the yielding theory uncertainties
as collected  in Tables \ref{Results of Bd to mu mu gamma}
and \ref{Results of Bs to mu mu gamma}, the currently poor understanding
towards the two-particle and three-particle $B_q$-meson distribution amplitudes
in HQET remains the most significant ingredient of  preventing us from
achieving the phenomenologically encouraging precision up to date,
which can be competitive with the experimental prospects of the anticipated
Belle II measurements (with the estimated precision of $9.6 \, \%$ and $23 \, \%$
for  ${\cal BR} (B_d \to \gamma  \gamma)$  and ${\cal BR} (B_s \to \gamma  \gamma)$
approximately).

\begin{table}[t]
\centering
\renewcommand{\arraystretch}{2.0}
\resizebox{\columnwidth}{!}{
\begin{tabular}{|c||c||c||c|c|c|c|c|c|c|}
\hline
\hline
& Central Value & Total Error & $\lambda_{B_d}$
& $\{ \widehat{\sigma}_{B_d}^{(1)}, \, \widehat{\sigma}_{B_d}^{(2)} \}$ & $\mu$
& $\nu $ & $\mu_{\rm h}$   &  $\bar \Lambda $ & $m^{\rm PS}_c$ \\
\hline
\hline
$10^8 \times {\cal BR}$
   & 1.352 & $_{-0.745}^{+1.242}$ &
   $_{-0.439}^{+1.091}$ & $_{-0.516}^{+0.503}$ &
   $_{-0.127}^{+0.032}$ & $_{-0.245}^{+0.235}$ &
   $_{-0.051}^{+0.167}$ & $_{-0.110}^{+0.098}$ &
   $_{-0.030}^{+0.030}$ \\
\hline
\hline
$f_{\|}$
   & 0.386 & $_{-0.058}^{+0.060}$ &
   $_{-0.030}^{+0.033}$ & $_{-0.038}^{+0.019}$ &
   $_{-0.007}^{+0.001}$ & $_{-0.030}^{+0.044}$ &
   $_{-0.003}^{+0.008}$ & $_{-0.006}^{+0.005}$ &
   $_{-0.002}^{+0.002}$ \\
\hline
$f_\perp$
   & 0.614 & $_{-0.060}^{+0.058}$ &
   $_{-0.033}^{+0.030}$ & $_{-0.019}^{+0.038}$ &
   $_{-0.001}^{+0.007}$ & $_{-0.044}^{+0.030}$ &
   $_{-0.008}^{+0.003}$ & $_{-0.005}^{+0.006}$ &
   $_{-0.002}^{+0.002}$ \\
\hline
\hline
${\cal A}^{\rm dir, \, \|}_{\rm CP}$
   & 0.130 & $_{-0.027}^{+0.044}$ & $_{-0.007}^{+0.012}$ &
   $_{-0.010}^{+0.016}$ & $_{-0.006}^{+0.008}$ &
   $_{-0.019}^{+0.029}$ & $_{-0.012}^{+0.024}$ &
   $_{-0.005}^{+0.006}$ & $_{-0.004}^{+0.004}$ \\
\hline
${\cal A}^{\rm mix, \, \|}_{\rm CP}$
   & $-0.155$ & $_{-0.060}^{+0.036}$ & $_{-0.016}^{+0.009}$ &
   $_{-0.022}^{+0.013}$ & $_{-0.011}^{+0.008}$ &
   $_{-0.039}^{+0.025}$ & $_{-0.032}^{+0.016}$ &
   $_{-0.008}^{+0.006}$ & $_{-0.008}^{+0.008}$ \\
\hline
${\cal A}^{\|}_{\Delta \Gamma}$
   & $-0.979$ & $_{-0.009}^{+0.017}$ & $_{-0.002}^{+0.004}$ &
   $_{-0.003}^{+0.006}$ & $_{-0.002}^{+0.003}$ &
   $_{-0.006}^{+0.011}$ & $_{-0.004}^{+0.009}$ &
   $_{-0.002}^{+0.002}$ & $_{-0.002}^{+0.002}$ \\
\hline
${\cal A}^{\rm dir, \, \perp}_{\rm CP}$
   & 0.356 & $_{-0.063}^{+0.076}$ & $_{-0.037}^{+0.030}$ &
   $_{-0.038}^{+0.061}$ & $_{-0.008}^{+0.017}$ &
   $_{-0.023}^{+0.010}$ & $_{-0.009}^{+0.015}$ &
   $_{-0.011}^{+0.014}$ & $_{-0.019}^{+0.019}$ \\
\hline
${\cal A}^{\rm mix, \, \perp}_{\rm CP}$
   & 0.083 & $_{-0.058}^{+0.079}$ & $_{-0.034}^{+0.041}$ &
   $_{-0.026}^{+0.008}$ & $_{-0.003}^{+0.000}$ &
   $_{-0.037}^{+0.061}$ & $_{-0.012}^{+0.028}$ &
   $_{-0.001}^{+0.000}$ & $_{-0.007}^{+0.007}$ \\
\hline
${\cal A}^{\perp}_{\Delta \Gamma}$
   & 0.931 & $_{-0.030}^{+0.019}$ & $_{-0.009}^{+0.009}$ &
   $_{-0.024}^{+0.013}$ & $_{-0.007}^{+0.003}$ &
   $_{-0.001}^{+0.001}$ & $_{-0.009}^{+0.004}$ &
   $_{-0.005}^{+0.004}$ & $_{-0.008}^{+0.008}$ \\
\hline
\hline
\end{tabular}
}
\caption{Theory predictions of the CP-averaged branching fraction,
the two polarization fractions and the six CP-violating observables
for $B_d \to \gamma \gamma$ in the presence of the neutral-meson mixing
with the total uncertainties obtained by adding all separate uncertainties
in quadrature, where the numerically important individual uncertainties
for each physical quantity are further displayed for completeness.}
\label{Results of Bd to mu mu gamma}
\end{table}

\begin{table}[t]
\centering
\renewcommand{\arraystretch}{2.0}
\resizebox{\columnwidth}{!}{
\begin{tabular}{|c||c||c||c|c|c|c|c|c|c|}
\hline
\hline
& Central Value & Total Error & $\lambda_{B_s}$ & $\{\widehat{\sigma}_{B_s}^{(1)}, \, \widehat{\sigma}_{B_s}^{(2)} \}$
& $\mu$  & $\nu $ & $\mu_{\rm h}$ &$\bar \Lambda $ & $m^{\rm PS}_c$ \\
\hline
\hline
$10^7 \times {\cal BR}$
   & 2.964 & $_{-1.614}^{+1.800}$ &
   $_{-0.671}^{+1.173}$ & $_{-1.163}^{+1.071}$ &
   $_{-0.421}^{+0.164}$ & $_{-0.709}^{+0.672}$ &
   $_{-0.114}^{+0.358}$ & $_{-0.272}^{+0.258}$ &
   $_{-0.077}^{+0.076}$ \\
\hline
\hline
$f_{\|}$
   & 0.361 & $_{-0.065}^{+0.064}$ &
   $_{-0.024}^{+0.026}$ & $_{-0.047}^{+0.023}$ &
   $_{-0.013}^{+0.004}$ & $_{-0.033}^{+0.052}$ &
   $_{-0.004}^{+0.010}$ & $_{-0.008}^{+0.006}$ &
   $_{-0.003}^{+0.003}$ \\
\hline
$f_{\perp}$
   & 0.639 & $_{-0.064}^{+0.065}$ &
   $_{-0.026}^{+0.024}$ & $_{-0.023}^{+0.047}$ &
   $_{-0.004}^{+0.013}$ & $_{-0.052}^{+0.033}$ &
   $_{-0.010}^{+0.004}$ & $_{-0.006}^{+0.008}$ &
   $_{-0.003}^{+0.003}$ \\
\hline
\hline
$10^2 \times {\cal A}^{\rm dir, \, \|}_{\rm CP}$
   & $-0.635$ & $_{-0.249}^{+0.151}$ &
   $_{-0.084}^{+0.048}$ & $_{-0.098}^{+0.055}$ &
   $_{-0.048}^{+0.034}$ & $_{-0.162}^{+0.102}$ &
   $_{-0.119}^{+0.060}$ & $_{-0.036}^{+0.030}$ &
   $_{-0.018}^{+0.019}$ \\
\hline
$10^2 \times {\cal A}^{\rm mix, \, \|}_{\rm CP}$
   & 0.763 & $_{-0.198}^{+0.338}$ &
   $_{-0.063}^{+0.114}$ & $_{-0.072}^{+0.132}$ &
   $_{-0.045}^{+0.064}$ & $_{-0.133}^{+0.221}$ &
   $_{-0.078}^{+0.161}$ & $_{-0.039}^{+0.048}$ &
   $_{-0.040}^{+0.040}$ \\
\hline
$10^4 \times \left ( {\cal A}^{\|}_{\Delta \Gamma} + 1 \right ) $
   & 0.493 & $_{-0.230}^{+0.459}$ &
   $_{-0.076}^{+0.150}$ & $_{-0.085}^{+0.176}$ &
   $_{-0.055}^{+0.082}$ & $_{-0.152}^{+0.310}$ &
   $_{-0.093}^{+0.218}$ & $_{-0.047}^{+0.062}$ &
   $_{-0.042}^{+0.043}$ \\
\hline
$10^2 \times {\cal A}^{\rm dir, \, \perp}_{\rm CP}$
   & $-1.888$ & $_{-0.409}^{+0.273}$ &
   $_{-0.072}^{+0.059}$ & $_{-0.342}^{+0.201}$ &
   $_{-0.124}^{+0.062}$ & $_{-0.024}^{+0.074}$ &
   $_{-0.077}^{+0.046}$ & $_{-0.085}^{+0.071}$ &
   $_{-0.113}^{+0.109}$ \\
\hline
$10^2 \times {\cal A}^{\rm mix, \, \perp}_{\rm CP}$
   & $-0.273$ & $_{-0.426}^{+0.318}$ &
   $_{-0.192}^{+0.148}$ & $_{-0.066}^{+0.185}$ &
   $_{-0.000}^{+0.031}$ & $_{-0.343}^{+0.195}$ &
   $_{-0.141}^{+0.060}$ & $_{-0.010}^{+0.014}$ &
   $_{-0.024}^{+0.022}$ \\
\hline
$10^4 \times \left ( A^{\perp}_{\Delta \Gamma} - 1 \right )$
   & $-1.820$ & $_{-0.802}^{+0.456}$ &
   $_{-0.108}^{+0.039}$ & $_{-0.671}^{+0.338}$ &
   $_{-0.234}^{+0.117}$ & $_{-0.015}^{+0.000}$ &
   $_{-0.197}^{+0.100}$ & $_{-0.160}^{+0.129}$ &
   $_{-0.228}^{+0.205}$ \\
\hline
\hline
\end{tabular}
}
\caption{Theory predictions of the CP-averaged branching fraction,
the two polarization fractions and the six CP-violating observables
for $B_s \to \gamma \gamma$ in the presence of the neutral-meson mixing
with the total uncertainties obtained by adding all separate uncertainties
in quadrature, where the numerically important individual uncertainties
for each physical quantity are further displayed for completeness.}
\label{Results of Bs to mu mu gamma}
\end{table}

Remarkably, including the subleading power soft corrections to the time-integrated
decay rates  will enlarge the combined theory uncertainties  by almost a factor of two
for both exclusive radiative decay processes with the central values of $\lambda_{B_q}$,
which can be attributed to the even more pronounced sensitivity to the
shape parameters $\widehat{\sigma}_{B_d}^{(1)}$ and $\widehat{\sigma}_{B_d}^{(2)}$
under this circumstance.
We can readily understand such interesting observation from the fact that the two
inverse-logarithmic moments will start to emerge in the leading power factorization formulae
for the helicity amplitudes by means of either the perturbative correction to
the hard-collinear matching coefficient
or the renormalization-group evolution of the inverse moment $\lambda_{B_q}$,
both of which will lead to an effective suppression factor
$\alpha_s(\mu) \, C_F / (4 \, \pi) \approx 0.04$ at $\mu = 1.5 \, {\rm GeV}$ numerically.
On the contrary, the NLP resolved photon contribution (\ref{final formula of soft correction})
estimated with the dispersion approach
depends on the functional form (particularly the small-$\omega$ behaviour)
of the leading twist $B_q$-meson distribution amplitude $\phi_B^{+}(\omega, \, \mu)$ already
at tree level (see \cite{Wang:2015vgv,DeFazio:2007hw} for earlier discussions in the context of
the heavy-to-light decay form factors).
A number of important comments on the numerical results presented in
Tables \ref{Results of Bd to mu mu gamma} and \ref{Results of Bs to mu mu gamma}
are in order.

\begin{itemize}

\item{It is instructive to derive the transparent expression of the ratio
for the two branching fractions of $B_q \to \gamma \gamma$ in the
leading-order approximation of the double expansion in powers of
$\Lambda/m_b$ and $\alpha_s$
\begin{eqnarray}
\frac{{\cal BR}(B_s \to \gamma \gamma)}{{\cal BR}(B_d \to \gamma \gamma)}
&=& {\tau_{B_s} \over \tau_{B_d}} \, \left |{ V_{ts} \over V_{td} } \right |^2 \,
\left ( {m_{B_s} \over m_{B_d}} \right )^3 \, \left ( {f_{B_s} \over f_{B_d}} \right )^2 \,
\left ( {\lambda_{B_d} \over \lambda_{B_s}} \right )^2 \,
\left ( {1- y_d^2 \over 1- y_s^2} \right )
+ {\cal O}  \left( {\Lambda \over m_b}, \alpha_s \right )
\nonumber \\
&=& 33.80 \, \left ( {\lambda_{B_d} \over \lambda_{B_s}} \right )^2
+ {\cal O}  \left( {\Lambda \over m_b}, \alpha_s \right )  \,.
\label{BR law}
\end{eqnarray}
We have verified explicitly that the $\lambda_{B_q}$-scaling violation
effect at leading power in the heavy quark expansion appears to be
insignificant numerically (at the level of ${\cal O} (1 \, \%)$);
while the subleading power factorizable and soft contributions
can result in approximately $(10-20) \, \%$  corrections
to the scaling behaviour (\ref{BR law}) for the default intervals of
the two inverse moments.
It is therefore of high interest to carry out the precision measurements
for such ``golden observable" ${\cal BR}(B_s \to \gamma \gamma) : {\cal BR}(B_d \to \gamma \gamma)$
at the Belle II experiment, which allow for an excellent determination
of the ratio of the two inverse moments with the systematic theory uncertainties
at the level of $(5-10) \, \%$ in the leading-power approximation.
}

\item{In comparison with the previous predictions of the exclusive  radiative $B_q \to \gamma \gamma $
decay rates from the QCD factorization approach \cite{DescotesGenon:2002ja}
\begin{eqnarray}
{\cal BR} (B_d \to \gamma \gamma) \big |_{\rm DS} \simeq 3 \times 10^{-8} \,,
\qquad
{\cal BR}(B_s \to \gamma \gamma)  \big |_{\rm DS} \simeq  10^{-6} \,,
\end{eqnarray}
our numerical results for the corresponding leading power contributions with the central values
of the input parameters turn out to be smaller by a factor of $2-3$ approximately.
The observed discrepancies in large part stem from approximating the $\overline{\rm MS}$-mass
of the $b$-quark by the bottom-meson mass for the resulting hard-scattering kernel (19) in  \cite{DescotesGenon:2002ja},
from adopting the higher value of the effective Wilson coefficient
$C_7^{\rm eff} (4.8 \, {\rm GeV}) = - 0.390$
and from  taking the lower value of the inverse moment for the $B_s$-meson
distribution amplitude $\lambda_{B_s}(\mu_0) = 350 \, {\rm MeV}$
implemented in the phenomenological analysis of \cite{DescotesGenon:2002ja}.
}

\item{Adding the factorizable NLP weak-annihilation correction on the top of
the leading-power contribution from the magnetic penguin operator $P_7$
and adopting further the central input parameters  gives rise to  \cite{Bosch:2002bv}
\begin{eqnarray}
{\cal BR}(B_d \to \gamma \gamma)  \big |_{\rm BB} = 3.11 \times 10^{-8} \,,
\qquad
{\cal BR}(B_s \to \gamma \gamma)  \big |_{\rm BB} =  1.23 \times 10^{-6} \,,
\label{BB results for BR}
\end{eqnarray}
which differ prominently  from our numerical predictions with the same theory accuracy
\begin{eqnarray}
{\cal BR}(B_d \to \gamma \gamma) = 1.78 \times 10^{-8} \,,
\qquad
{\cal BR}(B_s \to \gamma \gamma) = 3.88 \times 10^{-7} \,.
\label{SWW results for BR}
\end{eqnarray}
The dominating mechanisms accounting for the notable differences
between (\ref{BB results for BR}) and (\ref{SWW results for BR})
lie in the replacement of the $\overline {\rm MS}$-scheme bottom-quark mass
in the effective weak operator $P_7$ by the counterpart heavy-meson mass $m_{B_q}$
for deriving the leading-power transversity amplitudes  (3.2) in \cite{Bosch:2002bv},
which invokes additionally the approximate SU(3)-flavour symmetry relation
$\lambda_{B_s} = \lambda_{B_d}$ in the numerical extrapolation.
}

\end{itemize}

\begin{figure}
\begin{center}
\includegraphics[width=1.0 \columnwidth]{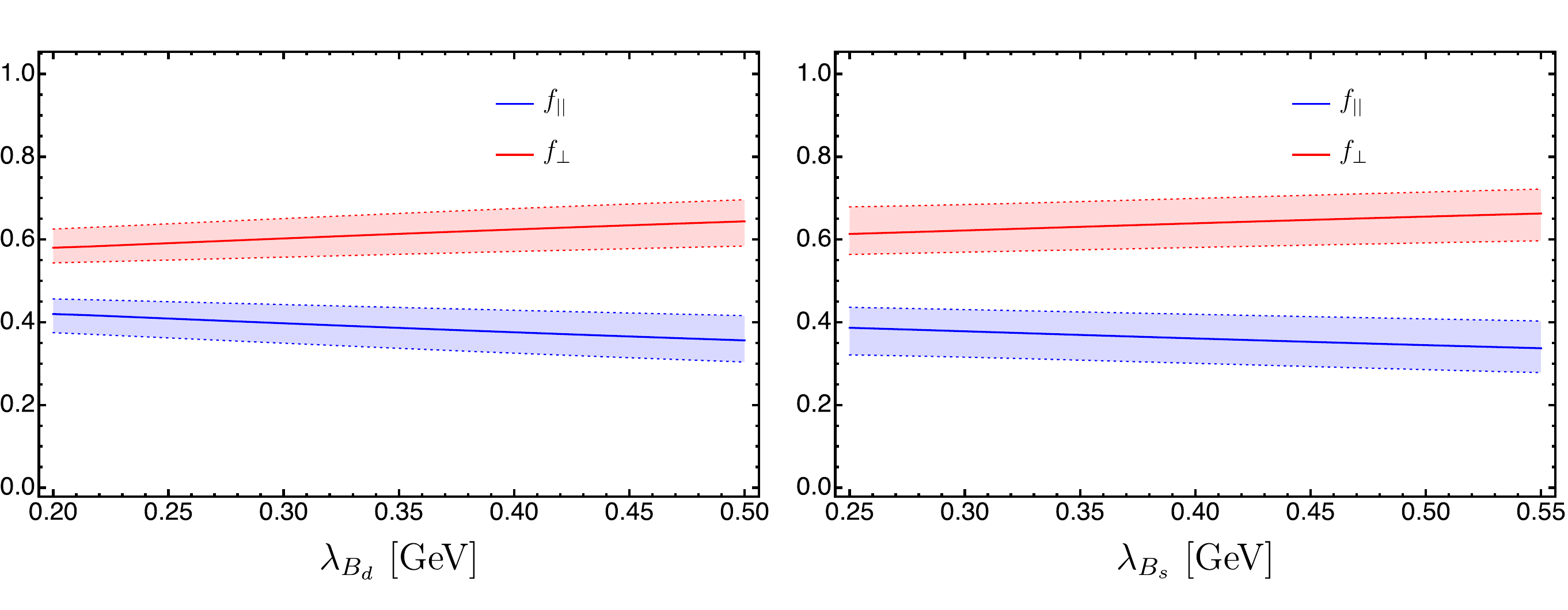}
\vspace*{0.1cm}
\caption{Theory predictions for the $\lambda_{B_q}$-dependence of the
CP-averaged polarization fractions  of  $\bar B_d \to \gamma \gamma$ [left panel]
and $\bar B_s \to \gamma \gamma$ [right  panel].}
\label{fig: polarization fractions results with errors}
\end{center}
\end{figure}

We proceed to present our predictions for the polarization fractions
of the double radiative $B_q$-meson decays within the default intervals
of the inverse moments in Figure
\ref{fig: polarization fractions results with errors}.
Apparently, the yielding uncertainties for such ratio observables
are substantially improved, when compared with the numerical results of
the CP-averaged  branching fractions displayed in (\ref{fig: BR results with errors}),
thanks to the striking cancellation of the errors from varying the shape parameters
of the bottom-meson distribution amplitudes.
The most significant uncertainties for the polarization fractions arise from
the variations of  QCD renormalization scale $\nu$ as shown in Tables
\ref{Results of Bd to mu mu gamma} and \ref{Results of Bs to mu mu gamma}.
It is appropriate to emphasize that the deviation of the quantity $f_{\|} : f_{\perp}$
from one evidently characterizes the typical size of the subleading power correction
in the $\Lambda_{\rm QCD}/m_b$ expansion.
In addition, the SU(3)-flavour symmetry violation effects for the polarization fractions
are estimated to be extraordinary small, around $(4 -7) \, \%$ numerically,
on account of an extra suppression factor from the heavy quark mass expansion.

\begin{figure}
\begin{center}
\includegraphics[width=1.0 \columnwidth]{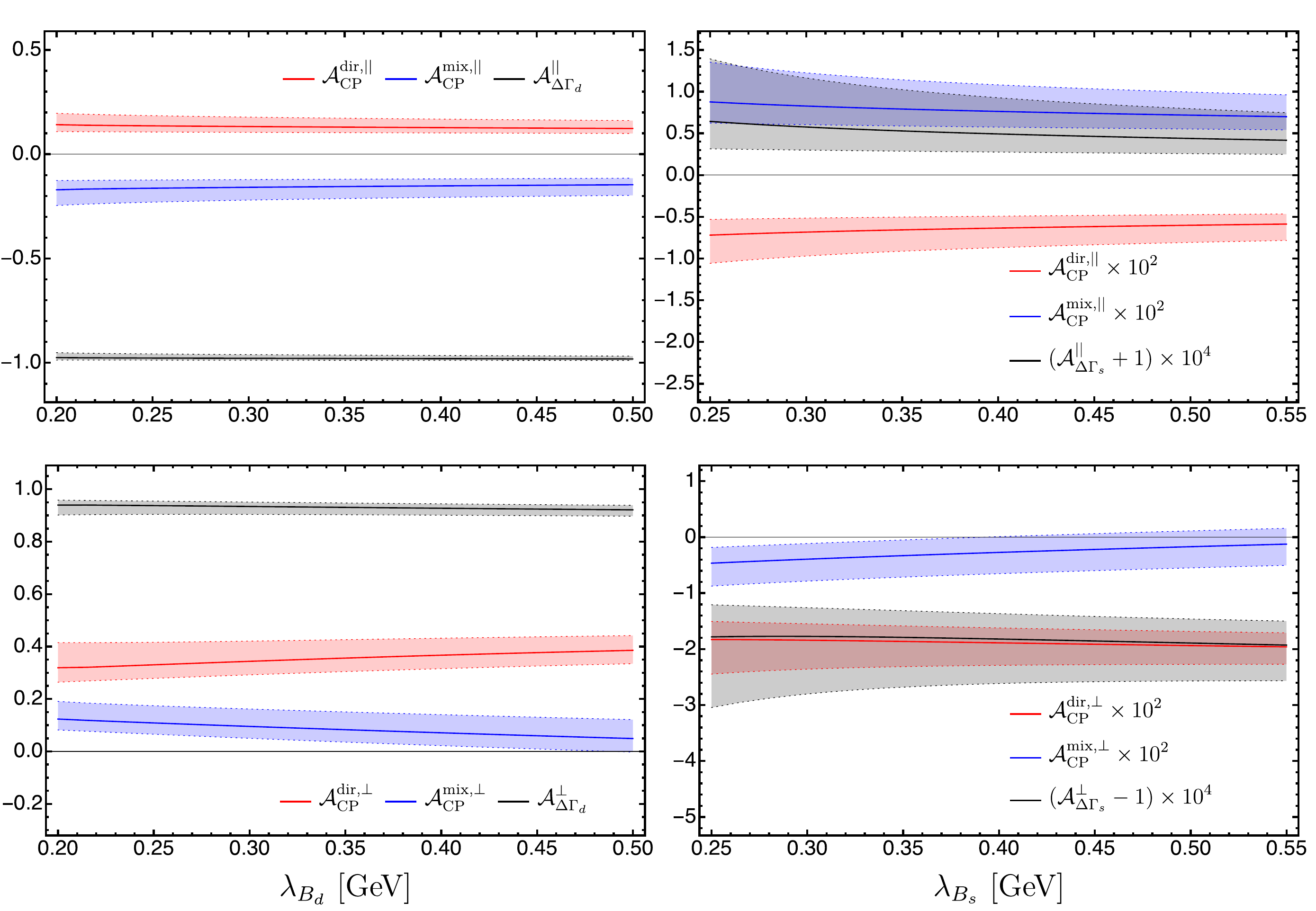}
\vspace*{0.1cm}
\caption{Theory predictions for the $\lambda_{B_q}$-dependence of the
three observables ${\cal A}_{\rm CP}^{{\rm dir}, \, \chi}$,
${\cal A}_{\rm CP}^{{\rm mix}, \, \chi}$  and ${\cal A}_{\Delta \Gamma_q}^{\chi}$
(with $\chi=\|, \perp$) determining the time-dependent CP asymmetries of
of  $\bar B_d \to \gamma \gamma$ [left panel]
and $\bar B_s \to \gamma \gamma$ [right  panel] in the approximation of
discarding the negligible CP violation in mixing.}
\label{fig: ACP results with errors}
\end{center}
\end{figure}

Finally we turn to display in Figure \ref{fig: ACP results with errors}
the theory predictions for the three observables determining the time-dependent
CP asymmetries of $B_q \to \gamma \gamma$ with the nonvanishing  width splittings
$\Delta \Gamma_q$.
The apparent hierarchy structure for the CKM matrix elements  contributing to
the $B_s \to  \gamma \, \gamma$ helicity amplitudes
\begin{eqnarray}
|V_{ub} \, V_{us}^{\ast}| :
|V_{cb} \, V_{cs}^{\ast}|
\approx  1 : 50 \,,
\end{eqnarray}
indicates the model-independent results of the mass-eigenstate rate asymmetries
${\cal A}_{\Delta \Gamma}^{\|} \simeq  -1$
and ${\cal A}_{\Delta \Gamma}^{\perp} \simeq + 1$
by dropping out the negligible $V_{ub} \, V_{us}^{\ast}$ terms
in (\ref{final amplitude of anti-B-meson decay}),
which are further supported by the complete NLL QCD calculations
at leading power in $\Lambda_{\rm QCD} / m_b$ with the various NLP corrections
at tree level as summarized in Table \ref{Results of Bs to mu mu gamma}.
The yielding predictions for ${\cal A}_{\rm CP}^{\rm dir, \, \|}$,
${\cal A}_{\rm CP}^{\rm mix, \, \|}$ and ${\cal A}_{\rm CP}^{\rm dir, \, \perp}$
suffer from the relatively smaller uncertainties than the counterpart results
for the CP-averaged branching fractions.
By contrast, the resulting values of ${\cal A}_{\rm CP}^{\rm mix, \, \perp}$
collected in Tables \ref{Results of Bd to mu mu gamma}
and \ref{Results of Bs to mu mu gamma}  appear to be  more uncertain,
mainly from varying the inverse moment $\lambda_{B_q}$ as well as
the perturbative hard  scale $\nu$.
According to the definition of ${\cal A}_{\rm CP}^{\rm mix, \, \perp}$
in (\ref{definition: three CP asymmetries}), we can readily identify
the emerged considerable uncertainties from the strong sensitivity
of the Wilson-coefficient  combination $\left [ C_F \, C_1(\nu) + C_2(\nu) \right ]$
entering the factorizable weak-annihilation amplitude ${\cal F}_V^{(p), \, \rm WA}$
in (\ref{1-loop WA functions}) on the renormalization scale $\nu$.
 As can be understood from the factorization formula
(\ref{factorized amplitude at NLL}), the two direct CP-violating quantities
${\cal A}_{\rm CP}^{\rm dir, \, \|}$  and ${\cal A}_{\rm CP}^{\rm dir, \, \perp}$
are expected to be identical  in the leading-power approximation.
However, the obtained numerical results displayed in  Tables \ref{Results of Bd to mu mu gamma}
and \ref{Results of Bs to mu mu gamma} tend to suggest that the subleading power contributions
to the radiative $B_q$-meson decay amplitudes in the heavy quark expansion can numerically
generate the substantial corrections to the above-mentioned asymptotic scaling.
This pattern can be attributed to the fact that the higher order terms in the heavy quark mass
expansion will generally lead to the enormous impacts on the perturbative strong phases
for the exclusive $B_q$-decay amplitudes due to the absence of the $\alpha_s$ suppression.



\section{Conclusions}
\label{section:conclusions}

In the present paper  we have carried out the improved calculations of
the double radiative $B_q$-meson decay amplitudes beyond the leading-power
accuracy by taking advantage of the perturbative factorization technique
and the dispersion approach.
The NLO hard matching coefficients from the two-loop QCD matrix elements
of $\langle q \, \gamma | H_{\rm eff}| b\rangle$ including the QCD penguin operators $P_{3,...,6}$
were taken into account for the first time based upon the  analytical calculations
performed in \cite{Buras:2002tp}, which are apparently of importance for exploring
the phenomenological observables of  $B_d \to \gamma \, \gamma$.
A complete NLL resummation for the enhanced logarithms of $m_b / \Lambda_{\rm QCD}$
appearing in the SCET factorization formula for the radiative $B_q \to \gamma \, \gamma$
decay amplitude has been accomplished with the renormalization group formalism in momentum space.
Subsequently, we constructed the factorized expressions for the NLP contributions
from the non-local hard-collinear interaction due to the energetic photon
radiation off the light-flavour quark of the bottom meson,
from the nonvanishing light-quark mass,
from the subleading power heavy-to-light current $J^{(\rm A2)}$ in SCET,
from the two-particle and three-particle $B_q$-meson distribution amplitudes
up to the twist-six accuracy, from the (anti)-collinear photon emission from the bottom quark
and from the one-particle irreducible weak-annihilation diagrams at tree level in detail
in Section \ref{section:NLP}.
It is remarkable to mention that all the above-mentioned non-local NLP corrections other than
the light-quark mass effect (\ref{T7 NLP mq}) can be expressed in terms of the convolution
integrals of the hard-collinear functions and the HQET distribution amplitudes
free of the potential rapidity divergences,
which would invalidate the resulting soft-collinear factorization formulae otherwise.
We further evaluated the NLP resolved photon contribution to the two helicity form
factors with the OPE-controlled dispersion technique at ${\cal O} (\alpha_s)$.
The large-recoil symmetry violating  effect between the two transversity amplitudes
has proven to be solely generated by the local NLP contribution by means of
the weak-annihilation mechanism.

Adopting the three-parameter ansatz for the essential $B_q$-meson distribution amplitudes
\cite{Beneke:2018wjp} fulfilling the classical equation-of-motion constraints
and the appropriate asymptotic behaviors at small quark and gluon momenta,
we have investigated the numerical implications of the newly derived theory ingredients
on the $B_q \to \gamma \gamma$  decay phenomenology systematically
in Section \ref{section:numerics}.
The attractive advantage of implementing this nonperturbative LCDA model
for the nonlocal NLP contributions consists in the complete independence of the yielding
factorized results on the very profile function $f(\omega)$
parameterizing the leading-twist $B_q$-meson distribution amplitude
as presented in (\ref{NLP integral relation I}), (\ref{NLP integral relation II}),
(\ref{NLP integral relation III}), (\ref{NLP integral relation IV})
and (\ref{NLP integral relation V}).
The NLL QCD corrections to the leading power predictions
of the helicity amplitudes ${\rm Re} \, \bar {\cal A}_L (\bar B_q \to \gamma \gamma)$
have been estimated to be  in the range of  $(2-10) \%$ for
$\lambda_{B_d} = (350 \pm 50) \, {\rm MeV}$ and $\lambda_{B_s} = (400 \pm 50) \, {\rm MeV}$.
On the other hand, the factorizable NLP contributions to
${\rm Re} \, \bar {\cal A}_L (\bar B_q \to \gamma \gamma)$
appeared to flip the sign numerically with the growing value of the inverse moment $\lambda_{B_q}$
due to the intrinsic competitions between the two different sectors of
$\{ \bar {\cal A}_{L}^{A2, \, {\rm NLP}}, \,\,  \bar{\cal A}_{L}^{e_b,  \, {\rm NLP}}, \,\, \bar{\cal A}_{L}^{\rm WA, \, NLP} \}$
and $\{ \bar {\cal A}_{L}^{\rm hc, \, NLP}, \,\,  \bar{\cal A}_{L}^{m_q,  \, {\rm NLP}}, \,\, \bar{\cal A}_{L}^{\rm HT, \, NLP} \}$;
and the resulting correction factors could reach an amount of ${\cal O} (15 \, \%)$  in magnitude maximally.
In particular, the factorizable NLP corrections to the imaginary part of the helicity amplitude
${\rm Im} \, \bar {\cal A}_L (\bar B_d \to \gamma \gamma)$ turned out to be even more pronounced:
as large as $(26 - 35) \, \%$ numerically.
Moreover, it has been demonstrated that the subleading power nonfactorizable contribution
from the ``hadronic" photon component  develops the strong sensitivity on the precise shape of
the leading twist $B_q$-meson distribution amplitude.

Having at our disposal the desired expressions for the transversity amplitudes $\bar {\cal A}_{\|}$
and $\bar {\cal A}_{\perp}$, we further provided the theory predictions for the CP-averaged branching fractions,
the polarization fractions and the CP-violating observables for $B_q \to \gamma \gamma$
in the presence of the neutral-meson mixing.
The major phenomenological results for confronting the forthcoming Belle II measurements
have been collected in Tables \ref{Results of Bd to mu mu gamma}
and \ref{Results of Bs to mu mu gamma}
as well as Figures \ref{fig: BR results with errors},
\ref{fig: polarization fractions results with errors}
and \ref{fig: ACP results with errors}.
The yielding  results for the CP-averaged branching fractions read
\begin{eqnarray}
{\cal BR}(B_d \to \gamma \gamma) = \left ( 1.352^{+1.242}_{-0.745} \right ) \times 10^{-8} \,,
\qquad
{\cal BR}(B_s \to \gamma \gamma) = \left ( 2.964^{+1.800}_{-1.614} \right ) \times 10^{-7} \,,
\end{eqnarray}
with the dominating theory uncertainties from varying the inverse-logarithmic
moments $\lambda_{B_q}$, $\widehat{\sigma}_{B_q}^{(1)}$, $\widehat{\sigma}_{B_q}^{(2)}$
and the QCD renormalization scale $\nu$.
It is worth noticing that precision measurements of the ratio for the two branching
fractions ${\cal BR}(B_s \to \gamma \gamma) : {\cal BR}(B_d \to \gamma \gamma)$
have been proven to allow for determining the quantity $\lambda_{B_d} : \lambda_{B_s}$
with the estimated systematic uncertainties not worse than  $(5-10) \%$
(see also \cite{Khodjamirian:2020hob} for an interesting calculation from the HQET sum rules).
It has been also verified that the resulting uncertainties for the  CP violation observables
${\cal A}_{\rm CP}^{\rm dir, \, \|}$, ${\cal A}_{\rm CP}^{\rm mix, \, \|}$
and ${\cal A}_{\rm CP}^{\rm dir, \, \perp}$ are reduced by approximately
a factor of two, in comparison with the counterpart results for the CP-averaged branching fractions.

Further developments of the precision QCD calculations for the double radiative
$B_q$-meson decays beyond the current work can be pursued forward in distinct directions.
First, implementing the systematic ${\rm QCD} \to {\rm SCET_{I}} \to {\rm SCET_{II}}$
matching for the correlation functions (\ref{QCD correlator: 7})
and (\ref{QCD correlator: 1-6 and 8}) at NLP in the heavy quark mass expansion
and constructing the appropriate SCET factorization formulae beyond the ${\cal O}(\alpha_s^0)$ accuracy
will  be in high demand for enhancing the predictive power of the perturbative factorization method
and for promoting our understanding towards the emerged $B$-physics anomalies.
To this end, it will be of utmost importance to develop the technical  framework
for tackling the soft-collinear convolution integrals in the presence
of the rapidity divergences with  effective filed theories
(see \cite{Liu:2019oav,Wang:2019mym} for further discussions in the different context).
Second,  improving the current theory constraints on the shape parameters
of the twist-two  HQET distribution amplitude will be indispensable for
pinning down the yet considerable uncertainties particularly for
the CP-averaged branching fractions.
In this respect, it remains to be demonstrated evidently that
whether the newly proposed method of accessing the nonperturbative light-front
quantities with the corresponding equal-time correlation functions
calculable on a Euclidean lattice and the standard OPE technique
will be able to provide us with  phenomenological encouraging results \cite{Wang:2019msf}.
Third, extending the established strategies of evaluating the factorizable NLP corrections
to the factorization analysis for a wide range of  exclusive $B$-meson decays
$B \to \gamma \, \ell \,   \bar \nu_{\ell}$,
$B \to \ell \, \bar \ell \, \ell^{\prime } \, \bar \nu_{\ell^{\prime}}$ \cite{Wang:2020}
and $B \to M \, \ell \,   \bar \nu_{\ell}$
(with $M=\pi, \,  \rho, \, \omega, \,  K^{(\ast)},  \, D^{(\ast)}$)
will be of high interest for probing the delicate strong interaction mechanisms
governing a variety of  heavy quark decays.

\subsection*{Acknowledgements}

We are grateful to Yao Ji for illuminating discussions.
Y.M.W acknowledges support from the National Youth Thousand Talents Program,
the Youth Hundred Academic Leaders Program of Nankai University,
the  National Natural Science Foundation of China  with
Grant No. 11675082 and 11735010, and  the Natural Science Foundation of Tianjin
with Grant No. 19JCJQJC61100.


\appendix

\section{The NLO coefficient functions $F_{i, 7}^{(p)}$}
\label{appendix: hard functions}

In this appendix, we will  collect the explicit expressions of the perturbative
coefficients $F_{i, 7}^{(p)}$  entering the NLO hard functions (\ref{combined hard function at NLO})
by taking advantage of the obtained results of \cite{Buras:2002tp}.
\begin{eqnarray}
F_{1, 7}^{(p)} &=& - {208 \over 243} \, \ln \left ( {m_b \over \nu} \right )
+ {833 \over 729} - {1 \over 3} \, \left [ a(z_p) + b(z_p) \right ]
+ {40 \over 243} \, i \, \pi  \,,  \nonumber \\
F_{2, 7}^{(p)} &=&  {416 \over 81} \, \ln \left ( {m_b \over \nu} \right )
- {1666 \over 243} + 2 \, \left [ a(z_p) + b(z_p) \right ]
- {80 \over 81} \, i \, \pi  \,,  \nonumber \\
F_{3, 7} &=&  - {176 \over 81} \, \ln \left ( {m_b \over \nu} \right )
+ {2392 \over 243} + {8 \pi \over 3 \sqrt{3}} + {32 \over 9} \, X_b
- a(1) + 2 \, b(1) +  {56 \over 81} \, i \, \pi  \,,  \nonumber \\
F_{4, 7} &=&  - {152 \over 243} \, \ln \left ( {m_b \over \nu} \right )
- {761 \over 729} - {4 \pi \over 9 \sqrt{3}} - {16 \over 27} \, X_b
+ {1 \over 6} a(1) + {5 \over 3} \, b(1) + 2 \, b(z_c)
-  {148 \over 243} \, i \, \pi  \,,  \nonumber \\
F_{5, 7} &=&  - {6272 \over 81} \, \ln \left ( {m_b \over \nu} \right )
+ {56680 \over 243} + {32 \pi \over 3 \sqrt{3}} + {128 \over 9} \, X_b
- 16  a(1) + 32 \, b(1)  +  {896 \over 81} \, i \, \pi  \,,  \nonumber \\
F_{6, 7} &=&   {4624 \over 243} \, \ln \left ( {m_b \over \nu} \right )
+ {5710 \over 729} - {16 \pi \over 9 \sqrt{3}} - {64 \over 27} \, X_b
- {10 \over 3}  a(1) + {44 \over 3} \, b(1) + 12 \, a(z_c) + 20 b(z_c) \nonumber \\
&& - {2296 \over 243} \, i \, \pi  \,,  \nonumber \\
F_{8, 7} &=&   -{32 \over 9} \, \ln \left ( {m_b \over \nu} \right )
+ {44 \over 9} - {8  \over 27} \pi^2  + {8 \over 9} \, i \, \pi \,,
\end{eqnarray}
with the dimensionless quantity $z_p=m_p^2/m_b^2$.
The integral representations of the constant $X_b$
and the two functions $a(z_p)$ and $b(z_p)$ are given by
\begin{eqnarray}
X_b &=& \int_0^1 dx \,  \int_0^1 dy \,  \int_0^1 dv \,
x \, y \, \ln [v + x (1-x) (1-v) (1 - v + vy)] \,, \nonumber \\
a(z_p) &=& {8 \over 9}  \int_0^1 dx \,  \int_0^1 dy \,  \int_0^1 dv \,
\big \{ \left [2- v + x y \, (2 v -3) \right ] \,
\ln \left [ v z_p + x (1-x) (1-v) (1-v+vy) \right ]  \nonumber \\
&&  +  \left [ 1- v + x y \, (2 v -1)  \right ] \,
\ln [z_p - x (1-x) y v - i \epsilon]
+ {43 \over 9}  +  {4 \over 9} \,  i \, \pi \big \}  \,,  \nonumber \\
b(z_p) &=&  {4 \over 81}  \, \ln z_p + {16 \over 27} \, z_p^2  +  {224 \over 81} z_p
- {92 \over 243}  +  {4 \over 81} i \, \pi
+ {-48 z_p^2 - 64 z_p + 4 \over 81} \, \sqrt{1-4 z_p} \, f(z_p) \nonumber \\
&& - {8 \over 9}   z_p^2  \, \left ( {2 \over 3}  z_p -1  \right ) \,
\left [ f(z_p) \right ] ^2  - {8 \over 9} \, \int_0^1 dx \,  \int_0^1 dy \,
{1 \over (1-y)^2 }  \nonumber \\
&& \times \, \left [ {1 \over 2} y^2 (y^2 -1)  x  (1-x)
+  (2 - y) \, u_1 \, \ln u_1 + (2 y^2 - 2 y -1) \, u_2 \, \ln u_2 \right ] \,,
\end{eqnarray}
where we have introduced the following conventions
\begin{eqnarray}
u_k &=& y^k \, x (1-x) + (1-y) z_p \,, \nonumber \\
f(z_p) &=& \left [ \ln{ 1 + \sqrt{1- 4 z_p}  \over 1 - \sqrt{1- 4 z_p}} - i \, \pi  \right ]  \theta(1-4 \, z_p)
- 2 i \, {\rm arctan} {1 \over \sqrt{4 z_p -1}}  \, \theta(4 z_p -1) \,.
\end{eqnarray}

\section{Renormalization-group evolution functions}
\label{appendix: B-meson LCDA}

Here we will summarize the expanded expressions for the evolution functions
in the SCET factorization formula (\ref{factorized amplitude at NLL})
at the NLL accuracy in perturbation theory.
As already explained in Section \ref{section:LP}, the approximate expressions
for the evolution kernels $\hat{U}_1(m_b, \mu_{\rm h}, \mu)$ and
$\hat{U}_2(m_b, \mu_{\rm h}, \mu)$ at ${\cal O}(\alpha_s)$ can be derived from
the  expanded solution  of $U_1(E_{\gamma}, \mu_{\rm h}, \mu)$
as displayed in \cite{Beneke:2011nf}.
Defining the soft-collinear terms in the last three lines of (\ref{factorized amplitude at NLL})
as the function ${\cal R}(m_b, \mu_0, \mu)$,
the corresponding approximate expression from perturbative expansions
of the anomalous dimensions $\Gamma_{\rm cusp}(\alpha_s)$ and $\gamma_{\eta}(\alpha_s)$
as well as the QCD beta function reads
\begin{eqnarray}
{\cal R}(m_b, \mu_0, \mu) &=& \hat{U}_3^{(0)}(\bar{\omega}, \mu_{0}, \mu)
\, {\rm exp} [2 \, \gamma_E \, a_{\Gamma}^{(0)}] \,\,
\frac{\Gamma  (1+a_{\Gamma}^{(0)})}{\Gamma  (1-a_{\Gamma}^{(0)}  )} \,
\hat{\mathcal{J}}^{(0)}\left ({\partial \over \partial \eta}, \mu \right )
\,\,  \tilde{\phi}_B^{+}(a_{\Gamma}^{(0)}, \mu_0) \bigg |_{\eta=0} \nonumber \\
&& \bigg [ 1 +  {\alpha_s(\mu_0) \over 4 \pi} \,
\bigg \{ \bigg [ {\cal D}(\xi=0, \mu_0, \mu) -  \ln \left ( {\mu^2 \over m_{b} \, \bar{\omega}} \right ) \bigg ]
\, a_{\Gamma}^{(1)}  +  a_{G}^{(1)} +  \hat{U}_3^{(1)}(\bar{\omega}, \mu_{0}, \mu)
\nonumber \\
&& \hspace{0.2 cm}
+ \, \left [ {\alpha_s(\mu) \over \alpha_s(\mu_0)} \right ] \,
\hat{\mathcal{J}}^{(1)} \left ({\partial \over \partial \eta}, \mu \right ) \,
{\rm exp}  \left [ \int_0^{\eta} \, d \xi \,\,  {\cal D}(\xi, \mu_0, \mu) \right ] \bigg |_{\eta=0} \bigg \} \bigg ]  \,,
\label{R function: RGE}
\end{eqnarray}
where we have introduced the following conventions for brevity
\begin{eqnarray}
\hat{U}_3(\bar{\omega}, \mu_{0}, \mu) &=& \hat{U}_3^{(0)}(\bar{\omega}, \mu_{0}, \mu) \,
\left [ 1 + \sum_{n=1}^{\infty} \,  \left (  {\alpha_s(\mu_0) \over 4 \pi} \right )^{n}
\hat{U}_3^{(n)}(\bar{\omega}, \mu_{0}, \mu) \right ] \,,
\nonumber \\
\hat{\mathcal{J}} \left ({\partial \over \partial \eta}, \mu \right ) \,
&=&  \hat{\mathcal{J}}^{(0)} \left ({\partial \over \partial \eta}, \mu \right )  \,
\left [ 1 + \sum_{n=1}^{\infty} \,  \left (  {\alpha_s(\mu) \over 4 \pi} \right )^{n}
\hat{\mathcal{J}}^{(n)} \left ({\partial \over \partial \eta}, \mu \right )  \right ]  \,, \nonumber \\
a_{\Gamma}(\mu_{0}, \mu) &=&  \sum_{n=0}^{\infty} \,\left (  {\alpha_s(\mu_0) \over 4 \pi} \right )^{n}
\, a_{\Gamma}^{(n)}(\mu_{0}, \mu) \,,  \nonumber \\
a_{G}(\mu_{0}, \mu) &=&  \int_{\alpha_s(\mu_0)}^{\alpha_s(\mu)} \, {d \alpha \over \beta(\alpha)}  \,
\mathcal{G}(a_{\Gamma}(\mu_{\alpha}, \mu), \alpha)
= \sum_{n=1}^{\infty} \,\left (  {\alpha_s(\mu_0) \over 4 \pi} \right )^{n}
\, a_{G}^{(n)}(\mu_{0}, \mu)  \,.
\end{eqnarray}
The manifest expression of the evolution factor $\hat{U}_3(\bar{\omega}, \mu_{0}, \mu)$
can be constructed from  $U_1(E_{\gamma}, \mu_{\rm h}, \mu)$ \cite{Beneke:2011nf} with the
necessary replacement rules
\begin{eqnarray}
\gamma(\alpha_s) \to  - \gamma_{\eta}(\alpha_s), \qquad
E_{\gamma} \to \bar{\omega}/2, \qquad
\mu_{\rm h} \to \mu_0.
\end{eqnarray}
For completeness, we further collect the expansion coefficients  in demand
\begin{eqnarray}
a_{\Gamma}^{(0)}(\mu_{0}, \mu) &=& {\Gamma_{\rm cusp}^{(0)}  \over 2 \, \beta_0} \, \ln r \,,
\qquad
a_{\Gamma}^{(1)}(\mu_{0}, \mu) = {\Gamma_{\rm cusp}^{(0)}  \over 2 \, \beta_0} \,
\left (  {\Gamma_{\rm cusp}^{(1)}  \over \Gamma_{\rm cusp}^{(0)}}
- {\beta_1 \over \beta_0} \right )  \,  (r-1) \,,  \nonumber \\
a_{G}^{(1)}(\mu_{0}, \mu)  &=&  {2 \, C_F \over \beta_0} \,
\int_0^1 \, d y  \,\,  {h(y)  \over 1-y}  \,
\frac{r^{-{2 \, C_F \over \beta_0}\, \ln y}  - r}
{1+ {2 \, C_F \over \beta_0} \, \ln y}  \,, \qquad
r =  {\alpha_s(\mu) \over \alpha_s(\mu_0)}    \,.
\end{eqnarray}
In addition, the newly defined function ${\cal D}(\xi, \mu_0, \mu)$
is given by
\begin{eqnarray}
{\cal D}(\xi, \mu_0, \mu) &=& \ln \left ( {\mu^2 \over m_{b} \, \bar{\omega}} \right )
+ \psi(1 + \xi + a_{\Gamma}^{(0)}) + \psi(1 - \xi - a_{\Gamma}^{(0)})
- \psi(1 + \xi) - \psi(1 - \xi)  \nonumber \\
&& +  \, \frac{\tilde{\phi}_B^{+ \prime}(\xi + a_{\Gamma}^{(0)}, \mu_0)}
{\tilde{\phi}_B^{+}(\xi + a_{\Gamma}^{(0)}, \mu_0)} \,,
\end{eqnarray}
with the customary definition for the logarithmic derivative of the Gamma function
\begin{eqnarray}
\psi(x) = {d \over d x} \, \ln \Gamma(x)
= \int_0^1 d t \,\, {1 - t^{x-1} \over 1-t}  - \gamma_{E}\,.
\end{eqnarray}



\begin{thebibliography}{99}



\bibitem{Beneke:2001at}
  M.~Beneke, T.~Feldmann and D.~Seidel,
  {\it Systematic approach to exclusive $B \to  V  \ell \ell$, $V \gamma$ decays,}
  Nucl.\ Phys.\ B {\bf 612} (2001) 25
  [hep-ph/0106067].





\bibitem{Beneke:2004dp}
  M.~Beneke, T.~Feldmann and D.~Seidel,
  {\it Exclusive radiative and electroweak $b \to  d$ and $b \to  s$ penguin decays at NLO,}
  Eur.\ Phys.\ J.\ C {\bf 41} (2005) 173
  [hep-ph/0412400].




\bibitem{Bosch:2001gv}
  S.~W.~Bosch and G.~Buchalla,
  {\it The Radiative decays $B \to  V \gamma$ at next-to-leading order in QCD,}
  Nucl.\ Phys.\ B {\bf 621} (2002) 459
  [hep-ph/0106081].



\bibitem{DescotesGenon:2004hd}
  S.~Descotes-Genon and C.~T.~Sachrajda,
  {\it Spectator interactions in $B \to V \gamma$ decays and QCD factorization,}
  Nucl.\ Phys.\ B {\bf 693} (2004) 103
  [hep-ph/0403277].


\bibitem{Ali:2004hn}
  A.~Ali, E.~Lunghi and A.~Y.~Parkhomenko,
  {\it Implication of the $B \to (\rho, \omega) \, \gamma$ branching ratios for the CKM phenomenology,}
  Phys.\ Lett.\ B {\bf 595} (2004) 323
  [hep-ph/0405075].






\bibitem{Becher:2005fg}
  T.~Becher, R.~J.~Hill and M.~Neubert,
  {\it Factorization in $B \to V \gamma$ decays,}
  Phys.\ Rev.\ D {\bf 72} (2005) 094017
  [hep-ph/0503263].



\bibitem{Ali:2007sj}
  A.~Ali, B.~D.~Pecjak and C.~Greub,
  {\it $B \to V \gamma$ Decays at NNLO in SCET,}
  Eur.\ Phys.\ J.\ C {\bf 55} (2008) 577
  [arXiv:0709.4422 [hep-ph]].





\bibitem{Ball:2006eu}
  P.~Ball, G.~W.~Jones and R.~Zwicky,
  {\it $B \to  V \gamma$ beyond QCD factorisation,}
  Phys.\ Rev.\ D {\bf 75} (2007) 054004
  [hep-ph/0612081].




\bibitem{Khodjamirian:2010vf}
  A.~Khodjamirian, T.~Mannel, A.~A.~Pivovarov and Y.-M.~Wang,
  {\it Charm-loop effect in $B \to K^{(\ast)} \ell^{+} \ell^{-}$ and $B\to K^{\ast} \gamma$,}
  JHEP {\bf 1009} (2010) 089
  [arXiv:1006.4945 [hep-ph]].



\bibitem{Wang:2017ijn}
  Y.~M.~Wang and Y.~L.~Shen,
  {\it Subleading power corrections to the pion-photon transition form factor in QCD,}
  JHEP {\bf 1712} (2017) 037
  [arXiv:1706.05680 [hep-ph]].



\bibitem{Li:2013xna}
  H.~N.~Li, Y.~L.~Shen and Y.~M.~Wang,
  {\it Joint resummation for pion wave function and pion transition form factor,}
  JHEP {\bf 1401} (2014) 004
  [arXiv:1310.3672 [hep-ph]].





\bibitem{Wang:2018wfj}
  Y.~M.~Wang and Y.~L.~Shen,
  {\it Subleading-power corrections to the radiative leptonic $B \to \gamma \ell \nu$ decay in QCD,}
  JHEP {\bf 1805} (2018) 184
  [arXiv:1803.06667 [hep-ph]].



\bibitem{Li:2020rcg}
  H.~D.~Li, C.~D.~L\"{u}, C.~Wang, Y.~M.~Wang and Y.~B.~Wei,
  {\it QCD calculations of radiative heavy meson decays with subleading power corrections,}
  JHEP {\bf 2004} (2020) 023
  [arXiv:2002.03825 [hep-ph]].




\bibitem{Bosch:2002bv}
  S.~W.~Bosch and G.~Buchalla,
  {\it The Double radiative decays $B \to \gamma \gamma$ in the heavy quark limit,}
  JHEP {\bf 0208} (2002) 054
  [hep-ph/0208202].




\bibitem{Kou:2018nap}
  E.~Kou {\it et al.} [Belle-II Collaboration],
  {\it The Belle II Physics Book,}
  PTEP {\bf 2019} (2019)12,  123C01;
   Erratum: [PTEP {\bf 2020} (2020) 2,  029201]
  [arXiv:1808.10567 [hep-ex]].



\bibitem{DescotesGenon:2002ja}
  S.~Descotes-Genon and C.~T.~Sachrajda,
  {\it Universality of nonperturbative QCD effects in radiative B decays,}
  Phys.\ Lett.\ B {\bf 557} (2003) 213
  [hep-ph/0212162].



\bibitem{Buras:2002tp}
  A.~J.~Buras, A.~Czarnecki, M.~Misiak and J.~Urban,
 {\it Completing the NLO QCD calculation of $B \to X_s \gamma$,}
  Nucl.\ Phys.\ B {\bf 631} (2002) 219
  [hep-ph/0203135].




\bibitem{Asatrian:2004et}
  H.~M.~Asatrian, H.~H.~Asatryan and A.~Hovhannisyan,
  {\it Rare decays $\bar B \to X_{s(d)} \gamma$  at the NLO,}
  Phys.\ Lett.\ B {\bf 585} (2004) 263
  [hep-ph/0401038].








\bibitem{Lin:1989vj}
  G.~L.~Lin, J.~Liu and Y.~P.~Yao,
  {\it Flavor Changing Two Photon Decay,}
  Phys.\ Rev.\ Lett.\  {\bf 64} (1990) 1498.



\bibitem{Lin:1990kw}
  G.~L.~Lin, J.~Liu and Y.~P.~Yao,
 {\it Top quark mass dependence of the decay $B_s \to \gamma \gamma$ in the standard electroweak model,}
  Phys.\ Rev.\ D {\bf 42} (1990) 2314.


\bibitem{Reina:1997my}
  L.~Reina, G.~Ricciardi and A.~Soni,
  {\it QCD corrections to $b \to s \gamma \gamma$ induced decays: $B \to X_s \gamma \gamma$ and $B_s \to \gamma \gamma$},
  Phys.\ Rev.\ D {\bf 56} (1997) 5805
  [hep-ph/9706253].


\bibitem{Herrlich:1991bq}
  S.~Herrlich and J.~Kalinowski,
 {\it Direct CP violation in $K, B \to \gamma \gamma$ with heavy top quark,}
  Nucl.\ Phys.\ B {\bf 381} (1992) 501.






\bibitem{Chang:1997fs}
  C.~H.~V.~Chang, G.~L.~Lin and Y.~P.~Yao,
  {\it QCD corrections to $b \to s \gamma \gamma$ and exclusive $B_s \to \gamma\gamma$ decay,}
  Phys.\ Lett.\ B {\bf 415} (1997) 395
  [hep-ph/9705345].



\bibitem{Devidze:1996np}
  G.~G.~Devidze, G.~R.~Jibuti and A.~G.~Liparteliani,
 {\it On the double radiative decays of the $B_s$ meson and mu atom: $B_s \to \gamma \gamma $, $\mu^{+} e^{-} \to \gamma \gamma$,}
  Nucl.\ Phys.\ B {\bf 468} (1996) 241.




\bibitem{Dincer:2001hu}
  Y.~Dincer and L.~M.~Sehgal,
  {\it Charge asymmetry and photon energy spectrum in the decay $B_s \to \ell^{+} \ell^{-} \gamma$,}
  Phys.\ Lett.\ B {\bf 521} (2001) 7
  [hep-ph/0108144].



\bibitem{Choudhury:1998rb}
  D.~Choudhury and J.~R.~Ellis,
  {\it Estimates of long distance contributions to the $B_s \to \gamma \gamma$ decay,}
  Phys.\ Lett.\ B {\bf 433} (1998) 102
  [hep-ph/9804300].




\bibitem{Liu:1999qz}
  W.~Liu, B.~Zhang and H.~Q.~Zheng,
  {\it On the long distance contribution to the $B_s \to \gamma \gamma$ decay in the effective Lagrangian approach,}
  Phys.\ Lett.\ B {\bf 461} (1999) 295
  [hep-ph/9905504].



\bibitem{Hiller:1997ie}
  G.~Hiller and E.~O.~Iltan,
  {\it Leading logarithmic QCD corrections to the $B_s \to \gamma \gamma$ decay rate
   including long distance effects through $B_s \to \phi \gamma \to \gamma \gamma$,}
  Phys.\ Lett.\ B {\bf 409} (1997) 425
  [hep-ph/9704385].



\bibitem{Hiller:1997kp}
  G.~Hiller and E.~O.~Iltan,
  {\it Estimate of the long distance contribution through $b \to s \psi$ to the $B_s \to \gamma \gamma$ decay rate,}
  Mod.\ Phys.\ Lett.\ A {\bf 12} (1997) 2837
  [hep-ph/9708477].






\bibitem{Braun:2017liq}
  V.~M.~Braun, Y.~Ji and A.~N.~Manashov,
  {\it Higher-twist $B$-meson Distribution Amplitudes in HQET,}
  JHEP {\bf 1705} (2017) 022
  [arXiv:1703.02446 [hep-ph]].




\bibitem{Khodjamirian:1997tk}
  A.~Khodjamirian,
  {\it Form-factors of $\gamma^{\ast} \rho \to  \pi$ and $\gamma^{\ast} \gamma \to  \pi^{0}$ transitions and light cone sum rules,}
  Eur.\ Phys.\ J.\ C {\bf 6} (1999) 477
  [hep-ph/9712451].




\bibitem{Agaev:2010aq}
  S.~S.~Agaev, V.~M.~Braun, N.~Offen and F.~A.~Porkert,
 {\it Light Cone Sum Rules for the $\pi^{0} \gamma^{\ast} \gamma$ Form Factor Revisited,}
  Phys.\ Rev.\ D {\bf 83} (2011) 054020
  [arXiv:1012.4671 [hep-ph]].



\bibitem{Agaev:2012tm}
  S.~S.~Agaev, V.~M.~Braun, N.~Offen and F.~A.~Porkert,
  {\it BELLE Data on the $\pi^0 \gamma^{\ast} \gamma$ Form Factor: A Game Changer?,}
  Phys.\ Rev.\ D {\bf 86} (2012) 077504
  [arXiv:1206.3968 [hep-ph]].




\bibitem{Wang:2016qii}
  Y.~M.~Wang,
  {\it Factorization and dispersion relations for radiative leptonic $B$ decay,}
  JHEP {\bf 1609} (2016) 159
  [arXiv:1606.03080 [hep-ph]].



\bibitem{Beneke:2018wjp}
  M.~Beneke, V.~M.~Braun, Y.~Ji and Y.~B.~Wei,
  {\it Radiative leptonic decay $B\to \gamma \ell \nu_\ell$ with subleading power corrections,}
  JHEP {\bf 1807} (2018) 154
  [arXiv:1804.04962 [hep-ph]].




\bibitem{Politzer:1980me}
  H.~D.~Politzer,
  {\it Power Corrections at Short Distances,}
  Nucl.\ Phys.\ B {\bf 172} (1980) 349.




\bibitem{Grinstein:1990tj}
  B.~Grinstein, R.~P.~Springer and M.~B.~Wise,
  {\it Strong Interaction Effects in Weak Radiative $\bar{B}$ Meson Decay,}
  Nucl.\ Phys.\ B {\bf 339} (1990) 269.



\bibitem{Chetyrkin:1996vx}
  K.~G.~Chetyrkin, M.~Misiak and M.~Munz,
 {\it Weak radiative B meson decay beyond leading logarithms,}
  Phys.\ Lett.\ B {\bf 400} (1997) 206;
   Erratum: [Phys.\ Lett.\ B {\bf 425} (1998) 414]
  [hep-ph/9612313].



\bibitem{Buchalla:1995vs}
  G.~Buchalla, A.~J.~Buras and M.~E.~Lautenbacher,
  {\it Weak decays beyond leading logarithms,}
  Rev.\ Mod.\ Phys.\  {\bf 68} (1996) 1125
  [hep-ph/9512380].




\bibitem{Gambino:2003zm}
  P.~Gambino, M.~Gorbahn and U.~Haisch,
  {\it Anomalous dimension matrix for radiative and rare semileptonic B decays up to three loops,}
  Nucl.\ Phys.\ B {\bf 673} (2003) 238
  [hep-ph/0306079].




\bibitem{Beneke:2020}
  M.~Beneke, C.~Bobeth and Y.~M.~Wang,
  {\it $B_{d,s} \to \gamma \ell \bar{\ell}$ decay with an energetic photon,}
  arXiv:2008.12494 [hep-ph].




\bibitem{Beneke:2006hg}
  M.~Beneke, J.~Rohrer and D.~Yang,
  {\it Branching fractions, polarisation and asymmetries of $B \to VV$ decays,}
  Nucl.\ Phys.\ B {\bf 774} (2007) 64
  [hep-ph/0612290].



\bibitem{Beneke:2002ph}
  M.~Beneke, A.~P.~Chapovsky, M.~Diehl and T.~Feldmann,
  {\it Soft collinear effective theory and heavy to light currents beyond leading power,}
  Nucl.\ Phys.\ B {\bf 643} (2002) 431
  [hep-ph/0206152].




\bibitem{Beneke:2002ni}
  M.~Beneke and T.~Feldmann,
  {\it Multipole expanded soft collinear effective theory with nonAbelian gauge symmetry,}
  Phys.\ Lett.\ B {\bf 553} (2003) 267
  [hep-ph/0211358].



\bibitem{Beneke:2004rc}
  M.~Beneke, Y.~Kiyo and D.~S.~Yang,
  {\it Loop corrections to subleading heavy quark currents in SCET,}
  Nucl.\ Phys.\ B {\bf 692} (2004) 232
  [hep-ph/0402241].





\bibitem{Beneke:2005gs}
  M.~Beneke and D.~Yang,
  {\it Heavy-to-light B meson form-factors at large recoil energy: Spectator-scattering corrections,}
  Nucl.\ Phys.\ B {\bf 736} (2006) 34
  [hep-ph/0508250].


\bibitem{Lunghi:2002ju}
  E.~Lunghi, D.~Pirjol and D.~Wyler,
  {\it Factorization in leptonic radiative B $\to \gamma  \ell \nu$ decays,}
  Nucl.\ Phys.\ B {\bf 649} (2003) 349
  [hep-ph/0210091].




\bibitem{Grozin:1996pq}
  A.~G.~Grozin and M.~Neubert,
  {\it Asymptotics of heavy meson form-factors,}
  Phys.\ Rev.\ D {\bf 55} (1997) 272
  [hep-ph/9607366].



\bibitem{Beneke:2000wa}
  M.~Beneke and T.~Feldmann,
  {\it Symmetry breaking corrections to heavy to light B meson form-factors at large recoil,}
  Nucl.\ Phys.\ B {\bf 592} (2001) 3
  [hep-ph/0008255].




\bibitem{Bosch:2003fc}
  S.~W.~Bosch, R.~J.~Hill, B.~O.~Lange and M.~Neubert,
  {\it Factorization and Sudakov resummation in leptonic radiative B decay,}
  Phys.\ Rev.\ D {\bf 67} (2003) 094014
  [hep-ph/0301123].



\bibitem{Liu:2020ydl}
  Z.~L.~Liu and M.~Neubert,
  {\it Two-Loop Radiative Jet Function for Exclusive $B$-Meson and Higgs Decays,}
  JHEP {\bf 2006} (2020) 060
  [arXiv:2003.03393 [hep-ph]].




\bibitem{Beneke:2011nf}
  M.~Beneke and J.~Rohrwild,
  {\it B meson distribution amplitude from $B \to \gamma l \nu$,}
  Eur.\ Phys.\ J.\ C {\bf 71} (2011) 1818
  [arXiv:1110.3228 [hep-ph]].



\bibitem{Braun:2019wyx}
  V.~M.~Braun, Y.~Ji and A.~N.~Manashov,
  {Two-loop evolution equation for the $B$-meson distribution amplitude,}
  Phys.\ Rev.\ D {\bf 100} (2019) 014023
  [arXiv:1905.04498 [hep-ph]].







\bibitem{Galda:2020epp}
  A.~M.~Galda and M.~Neubert,
  {Tomography of the $B$-Meson Light-Cone Distribution Amplitude,}
  arXiv:2006.05428 [hep-ph].



\bibitem{Baikov:2016tgj}
  P.~A.~Baikov, K.~G.~Chetyrkin and J.~H.~K¨¹hn,
  {\it Five-Loop Running of the QCD coupling constant,}
  Phys.\ Rev.\ Lett.\  {\bf 118} (2017)   082002
  [arXiv:1606.08659 [hep-ph]].



\bibitem{Braun:2019zhp}
  V.~M.~Braun, Y.~Ji and A.~N.~Manashov,
  {\it Scale-dependence of the $B$-meson LCDA beyond leading order from conformal symmetry,}
  arXiv:1912.03210 [hep-ph].




\bibitem{Lange:2003ff}
  B.~O.~Lange and M.~Neubert,
  {\it Renormalization group evolution of the B meson light cone distribution amplitude,}
  Phys.\ Rev.\ Lett.\  {\bf 91} (2003) 102001
  [hep-ph/0303082].





\bibitem{Bell:2013tfa}
  G.~Bell, T.~Feldmann, Y.~M.~Wang and M.~W.~Y.~Yip,
  {\it Light-Cone Distribution Amplitudes for Heavy-Quark Hadrons,}
  JHEP {\bf 1311} (2013) 191
  [arXiv:1308.6114 [hep-ph]].



\bibitem{Braun:2014owa}
  V.~M.~Braun and A.~N.~Manashov,
  {\it Conformal symmetry of the Lange-Neubert evolution equation,}
  Phys.\ Lett.\ B {\bf 731} (2014) 316
  [arXiv:1402.5822 [hep-ph]].













\bibitem{Becher:2014oda}
  T.~Becher, A.~Broggio and A.~Ferroglia,
 {\it Introduction to Soft-Collinear Effective Theory,}
  Lect.\ Notes Phys.\  {\bf 896} (2015) pp.1-206,
  [arXiv:1410.1892 [hep-ph]].



\bibitem{Kawamura:2001jm}
  H.~Kawamura, J.~Kodaira, C.~F.~Qiao and K.~Tanaka,
  {\it B-meson light cone distribution amplitudes in the heavy quark limit,}
  Phys.\ Lett.\ B {\bf 523} (2001) 111;
   Erratum: [Phys.\ Lett.\ B {\bf 536} (2002) 344]
  [hep-ph/0109181].



\bibitem{Kawamura:2001bp}
  H.~Kawamura, J.~Kodaira, C.~F.~Qiao and K.~Tanaka,
  {\it B meson light cone distribution amplitudes and heavy quark symmetry,}
  Int.\ J.\ Mod.\ Phys.\ A {\bf 18} (2003) 1433
  [hep-ph/0112146].




\bibitem{Manohar:2000dt}
  A.~V.~Manohar and M.~B.~Wise,
  {\it Heavy quark physics,}
  Camb.\ Monogr.\ Part.\ Phys.\ Nucl.\ Phys.\ Cosmol.\  {\bf 10} (2000) 1.





\bibitem{Beneke:2003zv}
  M.~Beneke and M.~Neubert,
 {\it QCD factorization for $B \to PP$ and $B \to PV$ decays,}
  Nucl.\ Phys.\ B {\bf 675} (2003) 333
  [hep-ph/0308039].




\bibitem{Beneke:2007zz}
  M.~Beneke,
  {\it Hadronic $B$ decays,}
  eConf C {\bf 0610161} (2006) 030
   [Nucl.\ Phys.\ Proc.\ Suppl.\  {\bf 170} (2007) 57]
  [hep-ph/0612353].





\bibitem{Mannel:2020fts}
  T.~Mannel, D.~Moreno and A.~Pivovarov,
  {\it Heavy Quark Expansion for Heavy Hadron Lifetimes: Completing the $1/m_b^3$ Corrections,}
  arXiv:2004.09485 [hep-ph].




\bibitem{Falk:1993dh}
  A.~F.~Falk, M.~E.~Luke and M.~J.~Savage,
  {\it Nonperturbative contributions to the inclusive rare decays $B \to  X_s \gamma$ and $B \to  X_s l^+ l^-$,}
  Phys.\ Rev.\ D {\bf 49} (1994) 3367
  [hep-ph/9308288].




\bibitem{Ball:1998sk}
  P.~Ball, V.~M.~Braun, Y.~Koike and K.~Tanaka,
 {\it Higher twist distribution amplitudes of vector mesons in QCD: Formalism and twist-three distributions,}
  Nucl.\ Phys.\ B {\bf 529} (1998) 323
  [hep-ph/9802299].





\bibitem{Ball:1998ff}
  P.~Ball and V.~M.~Braun,
  {\it Higher twist distribution amplitudes of vector mesons in QCD: Twist-4 distributions and meson mass corrections,}
  Nucl.\ Phys.\ B {\bf 543} (1999) 201
  [hep-ph/9810475].




\bibitem{Wandzura:1977qf}
  S.~Wandzura and F.~Wilczek,
  {\it Sum Rules for Spin Dependent Electroproduction: Test of Relativistic Constituent Quarks,}
  Phys.\ Lett.\  {\bf 72B} (1977) 195.





\bibitem{Lu:2018cfc}
  C.~D.~L\"{u}, Y.~L.~Shen, Y.~M.~Wang and Y.~B.~Wei,
 {\it QCD calculations of $B \to \pi, K$ form factors with higher-twist corrections,}
  JHEP {\bf 1901} (2019) 024
  [arXiv:1810.00819 [hep-ph]].




\bibitem{Balitsky:1987bk}
  I.~I.~Balitsky and V.~M.~Braun,
  {\it Evolution Equations for QCD String Operators,}
  Nucl.\ Phys.\ B {\bf 311} (1989) 541.






\bibitem{Khodjamirian:2012rm}
  A.~Khodjamirian, T.~Mannel and Y.~M.~Wang,
  {\it $B \to K \ell^{+}\ell^{-}$ decay at large hadronic recoil,}
  JHEP {\bf 1302} (2013) 010
  [arXiv:1211.0234 [hep-ph]].




\bibitem{Gao:2019lta}
  J.~Gao, C.~D.~L\"{u}, Y.~L.~Shen, Y.~M.~Wang and Y.~B.~Wei,
  {\it Precision calculations of $B \to V$ form factors from soft-collinear effective theory sum rules on the light-cone,}
  Phys.\ Rev.\ D {\bf 101} (2020)  074035
  [arXiv:1907.11092 [hep-ph]].




\bibitem{Straub:2015ica}
  A.~Bharucha, D.~M.~Straub and R.~Zwicky,
  {\it $B\to V\ell^+\ell^-$ in the Standard Model from light-cone sum rules,}
  JHEP {\bf 1608} (2016) 098
  [arXiv:1503.05534 [hep-ph]].



\bibitem{Beneke:2003pa}
  M.~Beneke and T.~Feldmann,
  {\it Factorization of heavy to light form-factors in soft collinear effective theory,}
  Nucl.\ Phys.\ B {\bf 685} (2004) 249
  [hep-ph/0311335].



\bibitem{Ball:2002ps}
  P.~Ball, V.~M.~Braun and N.~Kivel,
  {\it Photon distribution amplitudes in QCD,}
  Nucl.\ Phys.\ B {\bf 649} (2003) 263
  [hep-ph/0207307].




\bibitem{Tanabashi:2018oca}
P.~A.~Zyla {\it et al.} [Particle Data Group],
  {\it Review of Particle Physics,}
  PTEP {\bf 2020} (2020)   083C01.








\bibitem{Beneke:2014pta}
  M.~Beneke, A.~Maier, J.~Piclum and T.~Rauh,
 {\it The bottom-quark mass from non-relativistic sum rules at NNNLO,}
  Nucl.\ Phys.\ B {\bf 891} (2015) 42
  [arXiv:1411.3132 [hep-ph]].



\bibitem{Aoki:2019cca}
  S.~Aoki {\it et al.} [Flavour Lattice Averaging Group],
  {\it FLAG Review 2019: Flavour Lattice Averaging Group (FLAG),}
  Eur.\ Phys.\ J.\ C {\bf 80} (2020)   113
  [arXiv:1902.08191 [hep-lat]].





\bibitem{King:2019lal}
  D.~King, A.~Lenz and T.~Rauh,
  {\it $B_{s}$ mixing observables and $|V_{td}/V_{ts}|$ from sum rules,}
  JHEP {\bf 1905} (2019) 034
  [arXiv:1904.00940 [hep-ph]].





\bibitem{Bobeth:1999mk}
  C.~Bobeth, M.~Misiak and J.~Urban,
  {\it Photonic penguins at two loops and $m_t$ dependence of $BR[B \to  X_s l^+ l^-]$,}
  Nucl.\ Phys.\ B {\bf 574} (2000) 291
  [hep-ph/9910220].



\bibitem{Bobeth:2003at}
  C.~Bobeth, P.~Gambino, M.~Gorbahn and U.~Haisch,
  {\it Complete NNLO QCD analysis of $\bar B \to X_s \ell^{+} \ell^{-}$ and higher order electroweak effects,}
  JHEP {\bf 0404} (2004) 071
  [hep-ph/0312090].




\bibitem{Huber:2005ig}
  T.~Huber, E.~Lunghi, M.~Misiak and D.~Wyler,
  {\it Electromagnetic logarithms in $\bar B \to  X_s \ell^+ \ell^-$,}
  Nucl.\ Phys.\ B {\bf 740} (2006) 105
  [hep-ph/0512066].



\bibitem{Beneke:1998rk}
  M.~Beneke,
  {\it A Quark mass definition adequate for threshold problems,}
  Phys.\ Lett.\ B {\bf 434} (1998) 115
  [hep-ph/9804241].



\bibitem{Hoang:2008yj}
  A.~H.~Hoang, A.~Jain, I.~Scimemi and I.~W.~Stewart,
  {\it Infrared Renormalization Group Flow for Heavy Quark Masses,}
  Phys.\ Rev.\ Lett.\  {\bf 101} (2008) 151602
  [arXiv:0803.4214 [hep-ph]].




\bibitem{Bazavov:2017lyh}
  A.~Bazavov {\it et al.},
  {\it $B$- and $D$-meson leptonic decay constants from four-flavor lattice QCD,}
  Phys.\ Rev.\ D {\bf 98} (2018)  074512
  [arXiv:1712.09262 [hep-lat]].



\bibitem{Carrasco:2015xwa}
  N.~Carrasco, V.~Lubicz, G.~Martinelli, C.~T.~Sachrajda, N.~Tantalo, C.~Tarantino and M.~Testa,
  {\it QED Corrections to Hadronic Processes in Lattice QCD,}
  Phys.\ Rev.\ D {\bf 91} (2015)  074506
  [arXiv:1502.00257 [hep-lat]].





\bibitem{Beneke:2019slt}
  M.~Beneke, C.~Bobeth and R.~Szafron,
 {\it Power-enhanced leading-logarithmic QED corrections to $B_q \to \mu^+\mu^-$,}
  JHEP {\bf 1910} (2019) 232
  [arXiv:1908.07011 [hep-ph]].


\bibitem{Grozin:1996hk}
  A.~G.~Grozin and M.~Neubert,
  {\it Hybrid renormalization of penguins and five-dimension heavy light operators,}
  Nucl.\ Phys.\ B {\bf 495} (1997) 81
  [hep-ph/9701262].




\bibitem{Nishikawa:2011qk}
  T.~Nishikawa and K.~Tanaka,
  {\it QCD Sum Rules for Quark-Gluon Three-Body Components in the $B$ Meson,}
  Nucl.\ Phys.\ B {\bf 879} (2014) 110
  [arXiv:1109.6786 [hep-ph]].




\bibitem{Braun:2003wx}
  V.~M.~Braun, D.~Y.~Ivanov and G.~P.~Korchemsky,
 {\it The $B$-meson distribution amplitude in QCD,}
  Phys.\ Rev.\ D {\bf 69} (2004) 034014
  [hep-ph/0309330].




\bibitem{Wang:2015vgv}
  Y.~M.~Wang and Y.~L.~Shen,
  {\it QCD corrections to $B \to \pi$ form factors from light-cone sum rules,}
  Nucl.\ Phys.\ B {\bf 898} (2015) 563
  [arXiv:1506.00667 [hep-ph]].




\bibitem{Wang:2017jow}
  Y.~M.~Wang, Y.~B.~Wei, Y.~L.~Shen and C.~D.~L\"{u},
  {\it Perturbative corrections to $B \to D$ form factors in QCD,}
  JHEP {\bf 1706} (2017) 062
  [arXiv:1701.06810 [hep-ph]].




\bibitem{Li:2012nk}
  H.~n.~Li, Y.~L.~Shen and Y.~M.~Wang,
  {\it Next-to-leading-order corrections to $B \to \pi$ form factors in $k_T$ factorization,}
  Phys.\ Rev.\ D {\bf 85} (2012) 074004
  [arXiv:1201.5066 [hep-ph]].





\bibitem{Li:2012md}
  H.~N.~Li, Y.~L.~Shen and Y.~M.~Wang,
  {\it Resummation of rapidity logarithms in $B$-meson wave functions,}
  JHEP {\bf 1302} (2013) 008
  [arXiv:1210.2978 [hep-ph]].


\bibitem{Khodjamirian:2020hob}
  A.~Khodjamirian, R.~Mandal and T.~Mannel,
  {\it Inverse moment of the $B_s$-meson distribution amplitude from QCD sum rule,}
  arXiv:2008.03935 [hep-ph].








\bibitem{Lee:2005gza}
  S.~J.~Lee and M.~Neubert,
  {\it Model-independent properties of the $B$-meson distribution amplitude,}
  Phys.\ Rev.\ D {\bf 72} (2005) 094028
  [hep-ph/0509350].





\bibitem{Feldmann:2014ika}
  T.~Feldmann, B.~O.~Lange and Y.~M.~Wang,
  {\it $B$-meson light-cone distribution amplitude: Perturbative constraints and asymptotic behavior in dual space,}
  Phys.\ Rev.\ D {\bf 89} (2014)   114001
  [arXiv:1404.1343 [hep-ph]].



\bibitem{Charles:2004jd}
  J.~Charles {\it et al.} [CKMfitter Group],
  {\it CP violation and the CKM matrix: Assessing the impact of the asymmetric $B$ factories,}
  Eur.\ Phys.\ J.\ C {\bf 41} (2005)  1
  [hep-ph/0406184].




\bibitem{Ball:1996tb}
  P.~Ball and V.~M.~Braun,
  {\it The $\rho$ meson light cone distribution amplitudes of leading twist revisited,}
  Phys.\ Rev.\ D {\bf 54} (1996) 2182
  [hep-ph/9602323].




\bibitem{Nierste:2009wg}
  U.~Nierste,
  {\it Three Lectures on Meson Mixing and CKM phenomenology,}
  arXiv:0904.1869 [hep-ph].



\bibitem{Dunietz:1986vi}
  I.~Dunietz and J.~L.~Rosner,
  {\it Time Dependent CP Violation Effects in $B_0-\bar B_0$ Systems,}
  Phys.\ Rev.\ D {\bf 34} (1986) 1404.



\bibitem{Dunietz:2000cr}
  I.~Dunietz, R.~Fleischer and U.~Nierste,
  {\it In pursuit of new physics with $B_s$ decays,}
  Phys.\ Rev.\ D {\bf 63} (2001) 114015
  [hep-ph/0012219].



\bibitem{Amhis:2019ckw}
  Y.~S.~Amhis {\it et al.} [HFLAV Collaboration],
  {\it Averages of $b$-hadron, $c$-hadron, and $\tau$-lepton properties as of 2018,}
  arXiv:1909.12524 [hep-ex].





\bibitem{Anikeev:2001rk}
  K.~Anikeev {\it et al.},
  {\it $B$ physics at the Tevatron: Run II and beyond,}
  hep-ph/0201071.



\bibitem{Artuso:2015swg}
  M.~Artuso, G.~Borissov and A.~Lenz,
  {\it CP violation in the $B_s^0$ system,}
  Rev.\ Mod.\ Phys.\  {\bf 88} (2016)  045002;
  Addendum: Rev.\ Mod.\ Phys.\  {\bf 91} (2019)  049901
  [arXiv:1511.09466 [hep-ph]].




\bibitem{Jubb:2016mvq}
  T.~Jubb, M.~Kirk, A.~Lenz and G.~Tetlalmatzi-Xolocotzi,
  {\it On the ultimate precision of meson mixing observables,}
  Nucl. Phys. B \textbf{915} (2017) 431
  [arXiv:1603.07770 [hep-ph]].




\bibitem{Glashow:1970gm}
  S.~L.~Glashow, J.~Iliopoulos and L.~Maiani,
  {\it Weak Interactions with Lepton-Hadron Symmetry,}
  Phys.\ Rev.\ D {\bf 2} (1970) 1285.





\bibitem{Inami:1980fz}
  T.~Inami and C.~S.~Lim,
  {\it Effects of Superheavy Quarks and Leptons in Low-Energy Weak Processes $K_L \to \mu \, \bar \mu$,
  $K^{+} \to \pi^{+} \, \nu \, \bar \nu$ and $K^0 \leftrightarrow \bar K^0$}
  Prog.\ Theor.\ Phys.\  {\bf 65} (1981) 297;
   Erratum: [Prog.\ Theor.\ Phys.\  {\bf 65} (1981) 1772].



\bibitem{DescotesGenon:2011pb}
  S.~Descotes-Genon, J.~Matias and J.~Virto,
  {\it An analysis of $B_{d, \, s}$ mixing angles in presence of New Physics and an update of $B_s \to \bar {K}^{0 \ast} K^{0 \ast}$,}
  Phys.\ Rev.\ D {\bf 85} (2012) 034010
  [arXiv:1111.4882 [hep-ph]].




\bibitem{Harrison:1998yr}
  D.~Boutigny {\it et al.} [BaBar Collaboration],
  {\it The BABAR physics book: Physics at an asymmetric $B$ factory,}
  edited by P.~F.~Harrison and H.~R.~Quinn.





\bibitem{DeBruyn:2012wj}
  K.~De Bruyn, R.~Fleischer, R.~Knegjens, P.~Koppenburg, M.~Merk and N.~Tuning,
  {\it Branching Ratio Measurements of $B_s$ Decays,}
  Phys.\ Rev.\ D {\bf 86} (2012) 014027
  [arXiv:1204.1735 [hep-ph]].




\bibitem{DeFazio:2007hw}
  F.~De Fazio, T.~Feldmann and T.~Hurth,
  {\it SCET sum rules for $B \to P$ and $B \to V$ transition form factors,}
  JHEP {\bf 0802} (2008) 031
  [arXiv:0711.3999 [hep-ph]].





\bibitem{Liu:2019oav}
  Z.~L.~Liu and M.~Neubert,
  {\it Factorization at subleading power and endpoint-divergent convolutions in $h \to \gamma \gamma$ decay,}
  JHEP {\bf 2004} (2020) 033
  [arXiv:1912.08818 [hep-ph]].



\bibitem{Wang:2019mym}
  J.~Wang,
  {\it Resummation of double logarithms in loop-induced processes with effective field theory,}
  arXiv:1912.09920 [hep-ph].





\bibitem{Wang:2019msf}
  W.~Wang, Y.~M.~Wang, J.~Xu and S.~Zhao,
  {\it $B$-meson light-cone distribution amplitude from Euclidean quantities,}
  Phys.\ Rev.\ D {\bf 102} (2020)   011502
  [arXiv:1908.09933 [hep-ph]].



\bibitem{Wang:2020}
  C.~Wang, Y.~M.~Wang  and  Y.~B.~Wei,
  {\it QCD factorization for the four-body leptonic $B$-meson decays,}
  in prepartion.







\end{thebibliography}
\end{document}